\documentclass[%
aps,
prd,
preprintnumbers,
showpacs,
nofootinbib,
superscriptaddress,
amsmath,
amssymb,
floatfix,
notitlepage
]{revtex4-1}

\usepackage[english]{babel}
\usepackage[applemac]{inputenc}

\usepackage{hyperref}
\usepackage{amsfonts, amsmath, amssymb}
\usepackage{graphicx}
\usepackage{booktabs, multirow}
\usepackage{siunitx}
\usepackage{hyperxmp}
\usepackage{color}
\usepackage{simplewick}

\usepackage{relsize}
\newcommand{\babar}{\mbox{\slshape B\kern-0.1em{\smaller A}\kern-0.1em B\kern-0.1em{\smaller A\kern-0.2em R}}}
    
\newcommand{\belle}{Belle}

\newcommand{\VMD}{\mathrm{VMD}}
\newcommand{\LMD}{\mathrm{LMD}}
\newcommand{\LMDV}{\mathrm{LMD+V}}
\newcommand{\dof}{\mathrm{d.o.f.}}

\newcommand{\fm}{\mathrm{fm}}
\newcommand{\MeV}{\mathrm{MeV}}
\newcommand{\GeV}{\mathrm{GeV}}

\newcommand{\psib}{\overline{\psi}}	

\newcommand{\<}{\langle}
\renewcommand{\>}{\rangle}

\newcommand{\circq}{q^{^{\!\!\!\circ}}}
\newcommand{\circp}{p^{^{\!\!\!\circ}}}

\newcommand{\FF}{{\cal F}_{\pi^0\gamma^*\gamma^*}}

\newcommand{\amu}{a_\mu^{\mathrm{HLbL}; \pi^0}}

\newcommand{\be}{\begin{equation}}
\newcommand{\ee}{\end{equation}}
\newcommand{\bea}{\begin{eqnarray}}
\newcommand{\eea}{\end{eqnarray}}
\newcommand{\bdm}{\begin{displaymath}}
\newcommand{\edm}{\end{displaymath}}

\newcommand\mc[1]{\multicolumn{1}{c}{#1}}

\makeatother

\begin{document}
%
\title{Lattice calculation of the pion transition form factor $\pi^0 \to \gamma^* \gamma^*$}
\author{Antoine G\'erardin}
\email{gerardin@kph.uni-mainz.de}
\affiliation{PRISMA Cluster of Excellence and Institut f\"ur Kernphysik, Johannes Gutenberg-Universit\"at Mainz, 55099 Mainz, Germany}
\author{Harvey B. Meyer}
\email{meyerh@uni-mainz.de}
\affiliation{PRISMA Cluster of Excellence and Institut f\"ur Kernphysik, Johannes Gutenberg-Universit\"at Mainz, 55099 Mainz, Germany}
\affiliation{Helmholtz Institute Mainz, Johannes Gutenberg-Universit\"at Mainz, 55099 Mainz, Germany}
\author{Andreas Nyf\/feler}
\email{nyffeler@kph.uni-mainz.de}
\affiliation{PRISMA Cluster of Excellence and Institut f\"ur Kernphysik, Johannes Gutenberg-Universit\"at Mainz, 55099 Mainz, Germany}
\preprint{MITP/16-079}

\begin{abstract} 
We calculate the $\pi^0\to \gamma^*\gamma^*$ transition form factor
${\cal F}_{\pi^0\gamma^*\gamma^*}(q_1^2,q_2^2)$ in lattice QCD with two flavors of
quarks.  Our main motivation is to provide the input to calculate the
$\pi^0$-pole contribution to hadronic light-by-light scattering in the muon $(g-2)$,
$a_\mu^{\rm HLbL;\pi^0}$.
We therefore focus on the region where both photons are spacelike up to
virtualities of about $1.5~\GeV^2$, which has so far not been
experimentally accessible.  Results are obtained in the continuum at the physical pion mass
by a combined extrapolation. We reproduce the prediction of the
chiral anomaly for real photons with an accuracy of about $8-9\%$.
We also compare to various recently proposed
models and find reasonable agreement for the parameters of some of these models
with their phenomenological values.
Finally, we use the parametrization of our lattice data by these models
to calculate $a_\mu^{\rm HLbL;\pi^0}$.
\end{abstract}

\maketitle

\section{Introduction} 

\begin{figure}[b!]
\center
\includegraphics*[height=3cm]{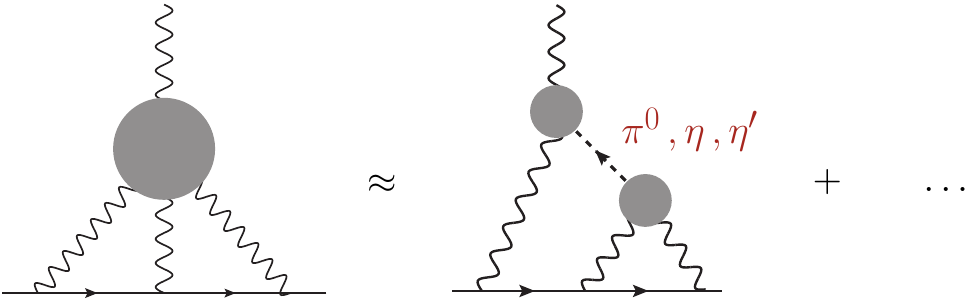}	
\caption{Pseudoscalar-pole contribution to hadronic light-by-light
  scattering in the muon $g-2$. The blobs on the right-hand side
  represent the ${\rm P} \to \gamma^* \gamma^*$ transition form
  factors with ${\rm P} = \pi^0, \eta, \eta^\prime$.}
\label{fig:hlbl}
\end{figure}

The anomalous magnetic moment of the muon provides one of the most
precise tests of the Standard Model of particle
physics~\cite{Jegerlehner:2009ry,Miller:2012opa}. It is known to
comparable precision in experiment~\cite{Bennett:2006fi} and theory
but the results disagree by about $3-4$ standard
deviations~\cite{Agashe:2014kda} depending on the theoretical
estimate. To interpret this tension as a sign of new physics,
improving the accuracy is of primary importance. On the experimental
side, new experiments at Fermilab and J-PARC are expected to reduce
the error by a factor of four~\cite{future_g-2_exp}. Therefore, a
corresponding theoretical effort is necessary to fully benefit from
the increased experimental precision. The theory error of $(g-2)_\mu$
is dominated by hadronic contributions: the hadronic vacuum
polarization (HVP) and hadronic light-by-light scattering (HLbL). The
first contribution can be related to the cross section $e^+ e^- \to$
hadrons using a dispersion relation such that the estimate can, in
principle, be improved by accumulating more data. Also, in recent years, more and more precise
lattice QCD calculations of the HVP have become available but are not yet competitive
with the dispersive approach \cite{DellaMorte:2011aa, Boyle:2011hu, Burger:2013jya, Chakraborty:2016mwy}.
However, the HLbL contribution to the muon $g-2$ cannot fully be related to direct
experimental information and current determinations usually rely on
model assumptions where systematic errors are difficult to
estimate~\cite{Prades:2009tw,Jegerlehner:2009ry,Bijnens:2015jqa}. However,
recently a dispersive approach was proposed~\cite{HLbL_DR} which
relates the, presumably, numerically dominant pseudoscalar-pole
contribution, as depicted in Fig.~\ref{fig:hlbl}, and the pion loop in
HLbL with on-shell intermediate pseudoscalar states to
measurable form factors and cross sections with off-shell photons:
$\gamma^* \gamma^* \to \pi^0, \eta, \eta^\prime$ and $\gamma^*
\gamma^* \to \pi^+ \pi^-, \pi^0 \pi^0$. Furthermore, increasingly
realistic lattice calculations of the HLbL contribution to the muon
$g-2$ have been carried out recently \cite{Blum:2014oka,Blum:2015gfa}.
Also, the hadronic light-by-light scattering amplitude \emph{per se}
has been calculated on the lattice in \cite{Green:2015sra}.

Within the dispersive framework, the pseudoscalar-pole contribution
requires as hadronic input the transition form factor ${\cal F}_{{\rm
    P}\gamma^*\gamma^*}(q_1^2,q_2^2)$ describing the interaction of an
on-shell pseudoscalar meson, ${\rm P} = \pi^0, \eta, \eta^\prime$,
with two off-shell photons with virtualities $q_1^2$ and $q_2^2$. The
HLbL contribution is then obtained by integrating some weight
functions times the product of a single-virtual and a double-virtual
transition form factor for spacelike
momenta~\cite{Jegerlehner:2009ry}. For the pion, the weight functions
turn out to be peaked at low momenta such that the main contribution
to $a_{\mu}^{{\rm HLbL}; \pi^0}$ arises from photon virtualities below
$1~\GeV^2$~\cite{KN_02,Nyffeler:2016gnb}, a kinematical range
accessible on the lattice.

The single-virtual transition form factor for the pion ${\cal
  F}_{\pi^0\gamma^*\gamma^*}(-Q^2,0)$ in the spacelike region has been
measured experimentally by several collaborations
\cite{Behrend:1990sr, Gronberg:1997fj,Aubert:2009mc, Uehara:2012ag} in
a wide kinematic range, although only for $Q^2 \geq 0.5~\GeV^2$. More
precise data down to $0.3~\GeV^2$ are expected soon from
BESIII~\cite{Denig:2014mma}. There are currently no data available for
the double-virtual transition form factor, a first measurement is
planned at BESIII~\cite{BESIII_double_virtual}. The
double-virtual form factor has also been addressed on the lattice in
\cite{Lin:2013im}. Finally, the authors of \cite{Feng:2012ck}
also considered the double-virtual form factor at a single lattice spacing but focused their study
on the pion decay $\pi^0 \to \gamma\gamma$, i.e.\ they were interested
mostly in the behavior of the form factor at very low momenta.
Transition form factors of mesons were first addressed in the context of the $\eta_c$ in~\cite{Dudek:2006ut,Chen:2016yau}.

Here we compute the transition form factor on the lattice in the
kinematical region relevant to hadronic light-by-light scattering in
the $(g-2)_\mu$.  Several lattice spacings and pion masses are used to
extrapolate our results to the physical point. Our calculation
involves several technical improvements over previous calculations.

This paper is structured as follows. In Sec.~\ref{sec:pion_ff}, we
give the precise definition of the transition form factor, describe
its phenomenology and theoretical constraints from QCD and introduce
the models whose functional form we will use to parametrize our
lattice data. In Sec.~\ref{sec:methodology}, we describe the
methodology of the lattice calculation, including the analytic
continuation, the required Wick contractions and the kinematic setup
that we choose.  In Sec.~\ref{sec:lattice_simulation}, the lattice
calculation itself is presented, with the final result for the
transition form factor presented in
Sec.~\ref{subsec:final_res}. Sec.~\ref{sec:pheno}
compares our fits to the lattice data with the available experimental
and theoretical information on the pion transition form factor and in
Sec.~\ref{subsec:amu} the pion-pole contribution to HLbL in the
muon $g-2$ is evaluated with the form factor determined on the
lattice. The paper ends with a summary of what has been achieved and
an outlook on possible future improvements. Several appendixes contain
some derivations and further discussions of some technical aspects, as
well as tables with detailed results of the fits.

\section{The pion transition form factor \label{sec:pion_ff}} 

In Minkowski spacetime, the transition form factor describing the
interaction between a neutral pion and two off-shell photons is
defined via the following matrix element\footnote{Equivalently, the
  form factor is given by 
\[ 
-2 \, [q_1^2 \, q_2^2 -(q_1\cdot q_2)^2]\, 
\FF(q_1^2, q_2^2) = \epsilon_{\mu\nu\alpha\beta} \, q_1^{\alpha} \,
q_2^{\beta} M^{\mu\nu} \,. 
\label{eq:form_factor}
\] 
}
\be 
M_{\mu\nu}(p,q_1)  = i \int \mathrm{d}^4 x \, e^{i q_1 x} \, \langle
\Omega | T \{ J_{\mu}(x) J_{\nu}(0) \} | \pi^0(p) \rangle =
\epsilon_{\mu\nu\alpha\beta} \, q_1^{\alpha} \, q_2^{\beta} \,
\FF(q_1^2, q_2^2) \,, 
\label{eq:MFF}
\ee 
where $q_1$ and $q_2$ are the photon momenta, $p = q_1 + q_2$ the
on-shell pion momentum, $p^2 = m_\pi^2$, $J_{\mu} = \sum_f Q_f \,
\psib_f \gamma_{\mu} \psi_f$ is the hadronic component of the
electromagnetic current and where we use the relativistic
normalization of states $\langle \pi^0(p) | \pi^0(p^{\prime}) \rangle
= (2\pi)^3\, 2 E_{\pi}(\vec{p}) \ \delta^{(3)}(\vec{p}-\vec{p}^{\
  \prime})$.  We use the mostly minus metric, $\epsilon^{0123}=+1$,
the axial current is given by $A_\mu^a= \psib\gamma_\mu\gamma^5
\frac{\tau^a}{2}\psi$ with $\tau^a$ a Pauli matrix, and the phase of
the one-pion state is fixed by $\<0|A_\mu^a(x)|\pi^b(p)\>= i F_\pi
p_\mu \delta^{ab}\,e^{-ipx}$ with $F_\pi = 92.4~\MeV$.  In the chiral
limit and at low energy, the form factor is constrained by the
Adler-Bell-Jackiw (ABJ) anomaly \cite{Adler:1969gk, Bell:1969ts}. At
the physical pion mass, there are corrections due to quark mass
effects which can be captured to a large extent by replacing the pion
decay constant in the chiral limit by the pion decay constant $F_{\pi}
= 92.4~\MeV$ obtained from charged pion
decay~\cite{Agashe:2014kda}. This leads to the following theoretical
normalization of the form factor:
\be
\vspace{-0.03cm} 
\FF(0, 0) = \frac{1}{4\pi^2 F_{\pi}} \,. 
\label{eq:ABJ}
\ee 
At leading order in QED, one gets for the decay rate 
\be 
\Gamma(\pi^0 \to \gamma\gamma) = \frac{\pi \alpha_e^2 m_{\pi}^3}{4}
\mathcal{F}^2_{\pi^0\gamma^{*}\gamma^{*}}(0, 0) \,, 
\label{eq:pion_decay}
\ee 
where $\alpha_e$ is the fine structure constant. Together with
Eq.~(\ref{eq:ABJ}) this reproduces quite well the measured decay width
$\Gamma(\pi^0 \to \gamma\gamma) = 7.73(16)$~eV~
\cite{Agashe:2014kda}. The PDG average is dominated by the PrimEx
experiment~\cite{Larin:2010kq} where a precision of 2.8\% has already
been achieved and a further reduction of the error by a factor of two
is expected soon. For a detailed comparison of theory and experiment
at the level of a few percent, higher order quark mass and radiative
corrections need to be taken into account, using chiral perturbation
theory ($\chi PT$) together with some form of resonance estimates of
the relevant low-energy constants \cite{pi0_gamma_gamma_theory,Moussallam:1994xp}. 

On the other hand, at large Euclidean (spacelike) momentum, the
single-virtual form factor has been computed in the framework of
factorization in QCD (operator-product expansion (OPE) on the
light cone) with a perturbatively calculable hard-scattering part and
a nonperturbative pion distribution amplitude. At leading order in
$\alpha_s$, one finds the Brodsky-Lepage behavior~\cite{BL_3_papers}
\be 
\FF(-Q^2, 0) \xrightarrow[Q^2 \to \infty]{}
\frac{2 F_{\pi}}{Q^2} \,.  
\label{eq:BL}
\ee 
In this formula, the prefactor should be taken with caution since its
value actually depends on the shape of the pion distribution amplitude
used in the calculation, which is usually modeled. When we impose
below the Brodsky-Lepage behavior according to Eq.~(\ref{eq:BL}), we
will only demand a $1/Q^2$ falloff of the form factor, without
insisting that the prefactor be reproduced exactly. On the
experimental side, the single-virtual form factor has been measured
for spacelike momenta in the range $[0.7-2.2]~\GeV^2$ by
CELLO~\cite{Behrend:1990sr} and for $[1.6-8.0]~\GeV^2$ by
CLEO~\cite{Gronberg:1997fj}.  Later \babar~\cite{Aubert:2009mc} and
\belle~\cite{Uehara:2012ag} obtained results at larger momentum
transfers both in the range $[4-40]~\GeV^2$. However their results
differ significantly at large momenta: the results of \babar{} showed an
unexpected slower falloff of the single-virtual form factor, while
the \belle{} data are compatible with a Brodsky-Lepage behavior.  In any
case, however, the data suggest that the asymptotic behavior is
approached only at a momentum transfer above $Q^2 = 10~\GeV^2$,
outside the kinematical range considered in this paper.  An analysis
by BESIII~\cite{Denig:2014mma} should be released soon which will
cover the low-momentum region $[0.3-3.1]~\GeV^2$ more relevant for the
muon $g-2$.

Finally, the double-virtual form factor where both momenta become
simultaneously large has been computed using the OPE at short
distances. In the chiral limit the result
reads~\cite{Nesterenko:1982dn, Novikov:1983jt}
\be 
\FF(-Q^2, -Q^2) \xrightarrow[Q^2 \to \infty]{} \frac{2
  F_{\pi}}{3} \left[ \frac{1}{Q^2} - \frac{8}{9} \frac{\delta^2}{Q^4} +
  \mathcal{O}\left(\frac{1}{Q^6}\right) \right] \,, 
\label{eq:OPE}
\ee 
where order $\alpha_s$ corrections are neglected and the quantity
$\delta^2 = (0.20 \pm 0.02)~\GeV^2$ parametrizes the higher-twist
matrix element in the OPE and was estimated in
Ref.~\cite{Novikov:1983jt} using QCD sum rules. In the double virtual
case, no experimental data exist yet but some results from the BESIII
experiment are expected in the coming years in the range $Q_{1,2}^2
\in [0.3-3]~\GeV^2$ \cite{BESIII_double_virtual,
  Nyffeler:2016gnb}. Therefore, the dependence of the double-virtual
form factor in the kinematical range of interest $[0-1]~\GeV^2$ for
the computation of the hadronic light-by-light contribution to the
muon $g-2$ is still unknown and the available estimates all rely on
phenomenological models \cite{Jegerlehner:2009ry,
  Bijnens:2015jqa}. The model parameters are either fixed using
theoretical and experimental constraints from various sources or by
fitting the experimental data of the single-virtual form factor and
then extrapolating to the double-virtual case, i.e.\ by assuming a
factorization of the form factor $\FF(-Q_1^2,-Q_2^2) = f(Q_1^2) \times f(Q_2^2)$. However, this method might be
unreliable and a model-independent theoretical estimate of the
transition form factor from lattice QCD is highly desirable.  Another
a priori model-independent approach is the use of a dispersion
relation for the form factor~\cite{Hoferichter:2012pm, DR_pion_TFF}, which is based on
general properties of analyticity and unitarity. For the practical
implementation, however, some assumptions and approximations need to
be made.

Different phenomenological models have been proposed in the literature
to describe the form factor in the whole kinematical range, see Ref.~\cite{FF_reviews} and references therein. The
simplest model is the vector meson dominance (VMD) model, where the
form factor is given by
\be 
\FF^{\VMD}(q_1^2, q_2^2) = \frac{ \alpha
  M_V^4}{(M_V^2-q_1^2)(M_V^2-q_2^2)} \,,  
\label{eq:VMD_model}
\ee 
where $\alpha = \alpha_{\mathrm{th}} = 1/(4\pi^2F_{\pi}) =
0.274~\GeV^{-1}$ to reproduce the anomaly constraint (\ref{eq:ABJ})
and with $M_V$ usually set to the $\rho$ meson mass. We will, however,
treat $\alpha$ and $M_V$ as free model parameters in our fits to the
lattice data below. The VMD model is compatible with the
Brodsky-Lepage behavior (\ref{eq:BL}) in the single-virtual
case. However, it behaves as $1/Q^4$ when both photons carry large
virtualities and falls off faster than the OPE
prediction~(\ref{eq:OPE}). The second model considered in this paper
is the lowest meson dominance (LMD)
model~\cite{Moussallam:1994xp,Knecht:1999gb}, within the large-$N_C$
approximation to QCD, which can be parametrized as
\be 
\FF^{\LMD}(q_1^2, q_2^2) = \frac{ \alpha M_V^4+ \beta(q_1^2
  + q_2^2)}{(M_V^2-q_1^2)(M_V^2-q_2^2)} \,. 
\label{eq:LMD_model}
\ee 
Again, one can set $\alpha = 1/(4\pi^2F_{\pi})$ to recover the anomaly
constraint. The form factor behaves as $1/Q^2$ in the double-virtual
case and for $\beta = \beta^{\mathrm{OPE}} = -F_\pi/3 = -0.0308~\GeV$
reproduces the leading OPE prediction, which is imposed in the
original LMD model by construction. On the other hand, the model does
not reproduce the Brodsky-Lepage behavior for the single-virtual form
factor (\ref{eq:BL}) but tends to a constant at large Euclidean
momentum for the off-shell photon. The original LMD model has no free
parameters, but we will treat $\alpha, \beta$ and $M_V$ as free
parameters in our fits below. 

Finally, in Ref.~\cite{Knecht:2001xc} the LMD+V model has been
proposed as a refinement of the LMD model where a second vector
resonance ($\rho^{\prime}$) is considered, see
Ref.~\cite{Nyffeler:2016gnb} for a recent brief review of the
model. The LMD+V model can simultaneously fulfill the Brodsky-Lepage
and the leading OPE behavior. Using a slightly different
parametrization from Ref.~\cite{Knecht:2001xc}, it can be written as
\be 
\FF^{\LMDV}(q_1^2, q_2^2) = \frac{\widetilde{h}_0\, q_1^2
  q_2^2 (q_1^2 + q_2^2) + \widetilde{h}_1 (q_1^2+q_2^2)^2  +
  \widetilde{h}_2\, q_1^2 q_2^2 + \widetilde{h}_5\, M_{V_1}^2
  M_{V_2}^2\,  (q_1^2+q_2^2) +
  \alpha\, M_{V_1}^4
  M_{V_2}^4}{(M_{V_1}^2-q_1^2)(M_{V_2}^2-q_1^2)
  (M_{V_1}^2-q_2^2)(M_{V_2}^2-q_2^2)} \,. 
\label{eq:LMDV_model}
\ee 
We have the relation $\widetilde{h}_{1} = -(F_\pi/3) h_{1}$,
$\widetilde{h}_{2} = -(F_\pi/3) {\bar h}_{2}$ and $\widetilde{h}_{5} =
-(F_\pi/(3M_{V_1}^2 M_{V_2}^2)) {\bar h}_{5}$ between the above
parametrization and the original model parameters $h_i$ (defined in
the chiral limit) and ${\bar h}_i$ (the latter parameters include
corrections proportional to powers of the pion mass). In the LMD+V
model proposed in Ref.~\cite{Knecht:2001xc} only the parameters $h_i$
(or ${\bar h}_i$) are treated as free parameters while the masses
$M_{V_1}$ and $M_{V_2}$ are set equal to the physical masses of the
$\rho$ and $\rho^\prime$ mesons. Furthermore the anomaly constraint is
imposed, $\alpha = 1/(4\pi^2F_{\pi})$, as is the Brodsky-Lepage
behavior which leads to $\widetilde{h}_1=0$. The form factor also has
by construction the correct leading OPE behavior in the double-virtual
case when both photons carry large Euclidean momenta by setting
$\widetilde{h}_0 = \widetilde{h}_0^{\mathrm{OPE}} = -F_\pi / 3$. As
pointed out in Ref.~\cite{MV_04}, the parameter ${\bar h}_2$ can be
fixed by comparing with the subleading term in the OPE in
Eq.~(\ref{eq:OPE}). Finally the parameter ${\bar h}_5$ has been
determined in Ref.~\cite{Knecht:2001xc} by a fit to the CLEO
data~\cite{Gronberg:1997fj} for the single-virtual form factor. One
then obtains the model parameters
\begin{eqnarray}
{\widetilde h}_2 & = & 0.327~\mbox{GeV}^3, \qquad 
[{\bar h}_2 = -4 (M_{V_1}^2 + M_{V_2}^2) + (16/9) \delta^2
= -10.63~\mbox{GeV}^2], \label{h2} \\   
{\widetilde h}_5 & = & (-0.166 \pm 0.006)~\mbox{GeV}, \qquad 
[{\bar h}_5 = (6.93 \pm 0.26)~\mbox{GeV}^4]. \label{h5}    
\end{eqnarray}
Following Ref.~\cite{Moussallam:1994xp}, information on ${\bar h}_5$
can also be obtained from the decay $\rho^+ \to \pi^+ \gamma$
(assuming octet symmetry) which leads to the less precise
determination ${\bar h}_5 = (6.3 \pm
0.9)~\mbox{GeV}^4$~\cite{Knecht:2001xc}. In our fits below, we will in
principle treat the parameters $\alpha, \widetilde{h}_i$ and the
masses $M_{V_1}$ and $M_{V_2}$ as free parameters.  The additional factors $M_{V_1}^2 M_{V_2}^2$ 
in the term with $\widetilde{h}_5$ in the numerator in Eq.~(\ref{eq:LMDV_model}) will lead to more stable fits later.

A summary of the different asymptotic limits for each model and
from the theory is given in Table.~\ref{tab:fit_properties}.
\renewcommand{\arraystretch}{1.2}
\begin{table}[t!]
\caption{Asymptotic behavior of the form factor for the different
  models (for LMD+V, $\widetilde{h}_1=0$ is assumed). The last line
  corresponds to the theoretical constraints discussed in the text.} 
\begin{center}
\begin{tabular}{lc@{\quad}c@{\quad}c}
\hline 
      & $\FF(0, 0)$	& $\FF(-Q^2, 0)$ & $\FF(-Q^2,-Q^2)$	\\   
\hline 
VMD   &	$\alpha$	& $\alpha M_V^2/Q^2$ & $\alpha M_V^4 / Q^4$  \\ 
LMD   &	$\alpha$	& $-\beta/M_V^2$     & $-2 \beta/Q^2$	 \\  
LMD+V &	$\alpha$	& $-\widetilde{h}_5/ Q^2$ &
$-2\widetilde{h}_0/Q^2$	 \\  
\hline
Eqs.~(\ref{eq:ABJ}) (\ref{eq:BL}) (\ref{eq:OPE}) &	
$1/(4 \pi^2 F_{\pi})$	& $2 F_{\pi}/Q^2$	 & $2 F_{\pi}/(3Q^2)$ \\ 
\hline 
\end{tabular}
\label{tab:fit_properties}
\end{center}
\end{table}

\section{Methodology \label{sec:methodology}}

From this section on, we use Euclidean notation by default. In particular, time evolution is governed by $e^{-H\tau}$ 
rather than $e^{-iHt}$, and $(J_\mu)_{\rm Minkowski} = (J_0,-iJ_k)_{\rm Euclid}$. However the four-vectors $q_1$ and $q_2$
are always understood to be Minkowskian, i.e.\ $q_1^2 = (q_1^0)^2 - \sum_{k=1}^3 (q_1^k)^2$.

\subsection{Extraction of the form factor} 

Using the method introduced in \cite{Ji:2001wha,Ji:2001nf}, and first implemented on the lattice in \cite{Dudek:2006ut}, 
one can show that the matrix element of Eq.~(\ref{eq:MFF}) can be written in Euclidean spacetime as \cite{Feng:2012ck} 
\begin{equation}
M_{\mu\nu}  =  (i^{n_0}) M_{\mu\nu}^{\rm E}, \quad 
M_{\mu\nu}^{\rm E}  \equiv  - \int \mathrm{d} \tau \, e^{\omega_1 \tau}  \int \mathrm{d}^3 z \, e^{-i \vec{q}_1 \vec{z}} \, 
\langle 0 | T \left\{ J_{\mu}(\vec{z}, \tau) J_{\nu}(\vec{0}, 0) \right\} | \pi(p) \rangle  \,,
\label{eq:Mminkowski}
\end{equation}
where $\omega_1$ is a real free parameter such that $q_1 = (\omega_1, \vec{q}_1)$ and $n_0$ denotes the number of temporal indices carried by the two vector currents. To obtain this formula, it is important to assume that $q_{1,2}^2 < M^2_V = {\rm min}(M^2_{\rho}, 4m_\pi^2)$ so that the integration contour does not encounter a singularity, where one of the photons can mix with an on-shell particle. Therefore, one is led to consider the following three-point correlation function on the lattice
\begin{align}
C^{(3)}_{\mu\nu}(\tau,t_{\pi}) = a^6\sum_{\vec{x}, \vec{z}} \, \big\langle  T \left\{ J_{\mu}(\vec{z}, t_i) J_{\nu}(\vec{0}, t_f)  P^{\dag}(\vec{x},t_0) \right\} \big\rangle \, e^{i \vec{p}\, \vec{x}} \, e^{-i \vec{q}_1 \vec{z}} \,,
\label{eq:lat_cor}
\end{align}
where 
\be
\tau=t_i-t_f
\ee
 is the time separation between the two vector currents and 
\be
t_{\pi}={\rm min}(t_f-t_0,t_i-t_0)
\ee
 is the minimal time separation between the pion interpolating operator and the two vector currents. Inserting a complete set of eigenstates, we obtain the following asymptotic behavior 
\begin{align}
\tau>0: \quad&\quad C^{(3)}_{\mu\nu}(\tau,t_{\pi})    \xrightarrow[t_{\pi} \to \infty]{}  -\frac{Z_{\pi}\,a^3}{2E_{\pi}}  \sum_{\vec{z}} \, \langle 0 | J_{\mu}(\vec{z}, \tau) J_{\nu}(\vec{0}, 0) | \pi(p) \rangle \, e^{-i \vec{q}_1 \vec{z}} \, e^{-E_{\pi} t_{\pi} } \,, \\
\tau<0:  \quad&\quad C^{(3)}_{\mu\nu}(\tau,t_{\pi})     \xrightarrow[t_{\pi} \to \infty]{}  - \frac{Z_{\pi}\,a^3}{2E_{\pi}} \sum_{\vec{z}} \, \langle 0 | J_{\nu}(\vec{0}, -\tau) J_{\mu}(\vec{z}, 0) | \pi(p) \rangle \, e^{-i \vec{q}_1 \vec{z}} \, e^{-E_{\pi} t_{\pi} } \,,
\end{align}
where $\langle0 | P(\vec{x},t) |  \pi(p) \rangle = Z_{\pi} e^{-Et+i \vec{p} \vec{x}} $ is the overlap factor of our interpolating operator with the pion state\footnote{With the choice of $P=\psib\gamma_5\tau^3\psi$ made below in Eq.\ (\ref{eq:Pdef}), the overlap is given by the partially conserved axial current (PCAC) relation, $Z_\pi = -i F_\pi m_\pi^2 / m$, where $m$ is the average $(u,d)$ quark mass.} and the factor $2E_{\pi}$ in the denominator comes from the relativistic normalization of states. The large time behavior of the three-point correlation function (\ref{eq:lat_cor}) ensures that the pion is on shell and that the excited states contribution in the pseudoscalar channel is small.  Finally, the overlap $Z_{\pi}$ and the pion mass are extracted from the two-point correlation function 
\begin{align}
C^{(2)}(t) = a^3 \sum_{\vec{x}} \, \big\langle  P(\vec{x},t)  P^{\dag}(\vec{0},0)  \big\rangle \, e^{-i \vec{p} \vec{x}}   \xrightarrow[t \to \infty]{}   \frac{|Z_{\pi}|^2}{2E_{\pi}} \left( e^{-E_{\pi} t} + e^{-E_{\pi}(T-t)} \right) \,,
\label{eq:cor2pt}
\end{align}
where $T$ is the temporal extent of the lattice. 
It is convenient to remove the explicit pion energy time dependence in the three-point correlation function and to define 
\begin{align}
A_{\mu\nu}(\tau) = \lim_{t_{\pi} \rightarrow +\infty} C^{(3)}_{\mu\nu}(\tau,t_{\pi}) \, e^{E_{\pi}t_{\pi}}  \,.
\label{eq:Amunu}
\end{align}
Then, from Eq.~(\ref{eq:Mminkowski}), $M_{\mu\nu}$ can be obtained via
\begin{gather}
 M_{\mu\nu}^{\rm E} =  \frac{2 E_{\pi}}{ Z_{\pi} }  \left( \int_{-\infty}^{0} \, \mathrm{d}\tau \, e^{\omega_1 \tau} \, A_{\mu\nu}(\tau) \,  e^{-E_{\pi} \tau} +  \int_{0}^{\infty} \, \mathrm{d}\tau  \, e^{\omega_1 \tau} \, A_{\mu\nu}(\tau)  \right)  
=   \frac{2 E_{\pi}}{ Z_{\pi} }  \int_{-\infty}^{\infty} \, \mathrm{d}\tau \, e^{\omega_1 \tau} \, \widetilde{A}_{\mu\nu}(\tau) \,,
\label{eq:lat_M}
\\ 
\widetilde{A}_{\mu\nu}(\tau) = \lim_{t_{\pi} \rightarrow + \infty} e^{E_\pi(t_f-t_0)} C^{(3)}_{\mu\nu}(\tau,t_{\pi})
= \left\{\begin{array}{l@{~~~}l}  A_{\mu\nu}(\tau) & \tau >0 \\  A_{\mu\nu}(\tau) \,  e^{-E_{\pi} \tau} & \tau<0 \end{array} \right.\;. 
\label{eq:lat_A_Atilde}
\end{gather}
The integral (\ref{eq:lat_M}) is convergent as long as\footnote{The bound applies in infinite volume. In finite volume, the threshold
can be at a slightly different energy than $\sqrt{M_V^2+\vec q_{1,2}^{\, 2}}$.} $q_{1,2}^2<M^2_V$: 
the three-point correlation function falls off with a factor $e^{-E_V |\tau|}$, with $E_V$ the energy of a vector state.
We point out that it is 
$\widetilde{A}_{\mu\nu}(\tau)$, 
rather than $A_{\mu\nu}(\tau)$ which is most directly related to the matrix element of interest $M_{\mu\nu}$; 
see Appendix \ref{app:A} for more details.

\subsection{Kinematic setups} 

\begin{figure}[t]
	\begin{minipage}[c]{0.49\linewidth}
	\centering 
	\includegraphics*[width=0.95\linewidth]{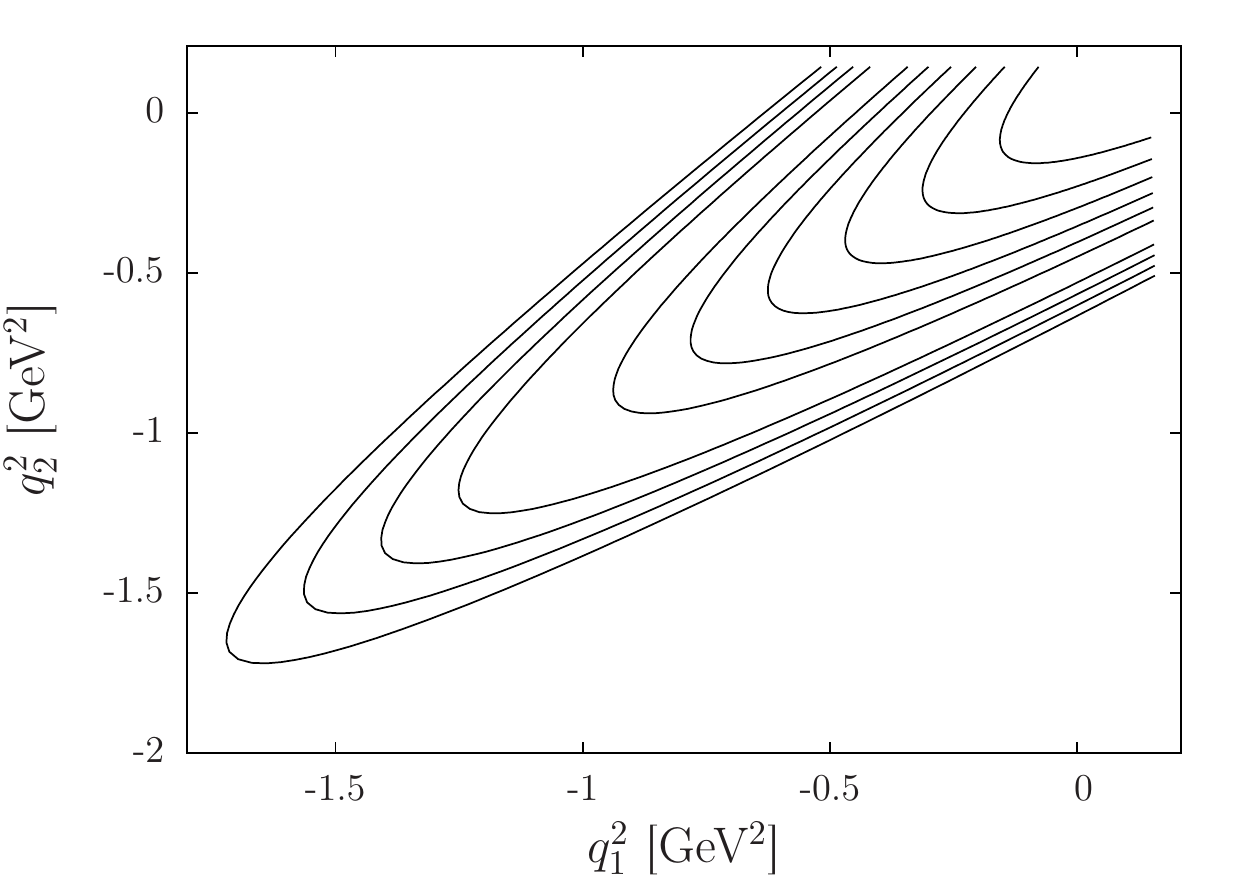}
	\end{minipage}
	\begin{minipage}[c]{0.49\linewidth}
	\centering 
	\includegraphics*[width=0.95\linewidth]{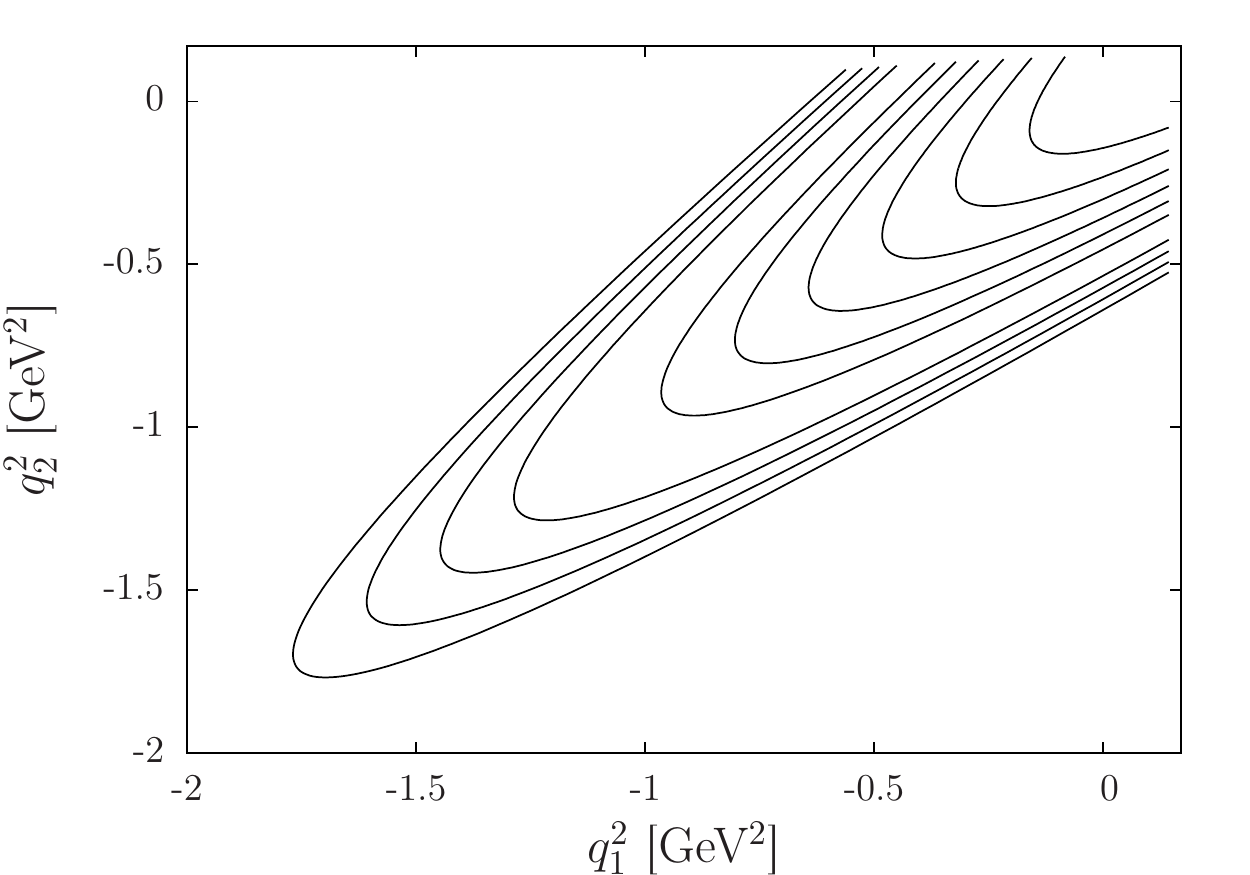}
	\end{minipage}
	\caption{Kinematic reach in the photon virtualities ($q_1^2,q_2^2$) in our setup with the pion at rest, for the lattice resolution $48^3\times96$ at $a=0.065\,{\rm fm}$ (left) and for the lattice resolution $64^3\times128$ at $a=0.048\,{\rm fm}$ (right). Each curve corresponds to a different value of the spatial momentum $|\vec{q}_1|^2$.}	
	\label{fig:kin}
\end{figure}

On the lattice, the momentum of the pion is set explicitly through the pseudoscalar interpolating operator used in Eq.~(\ref{eq:lat_cor}) and its energy $E_{\pi}$ is then imposed by the on-shell condition. We are also free to choose one vector current spatial momentum (e.g. $\vec{q}_1$), $\vec{q}_2$ being determined by the momentum conservation $\vec{p} = \vec{q}_1 + \vec{q}_2$. Finally, in Eq.~(\ref{eq:lat_M}), we can vary continuously $\omega_1$, with $\omega_2$ determined by the energy conservation $E_{\pi} = \omega_1 + \omega_2$. Therefore, the kinematical range accessible on the lattice can be parametrized by
\begin{align}
\nonumber q_1^2 &= \omega_1^2 - \vec{q}_1^{\, 2} \,, \\
q_2^2 &= (E_{\pi} - \omega_1)^2 - (\vec{p}-\vec{q}_1)^2 \,.
\label{eq:kin}
\end{align}
Choosing the pion reference frame where $\vec{p}=0$, both photons have back-to-back spatial momenta ($\vec{q}_2 = - \vec{q}_1$) and $E_{\pi} = m_{\pi}$. The kinematic range corresponding to different choices of $|\vec{q}_1|$ is plotted in Fig.~\ref{fig:kin} for two different lattice resolutions. As explained below, from a numerical point of view, different momenta $\vec{q}_1$ can be obtained without any new inversion of the Dirac operator. Therefore, this setup is adapted to study the form factors at large $q_{1,2}^2$ and, for each ensemble, the three-point correlation function has been computed up to momenta $|q_{1,2}^2| \approx 1.5~\GeV^2$.   Using the Lorentz structure of the form factor 
(see Eqs.~(\ref{eq:MFF}), (\ref{eq:Mminkowski}) and (\ref{eq:lat_M})), $A_{\mu\nu}$ with one or more temporal indices vanishes
and the spatial components can be written in the form
\be\label{eq:Ascalar}
A_{kl}(\tau) = -i q_{kl} \, A(\tau)\,, \qquad q_{kl}\equiv \epsilon_{kl\alpha\beta} \, q_1^{\alpha} \, q_2^{\beta} 
= m_\pi\,\epsilon_{kli}\,q_1^i\,,
\ee
where $A(\tau)$ is a scalar under the spatial rotation group. 
From $\widetilde A_{kl}(\tau)$ we define $\widetilde A(\tau)$ in the same way.
Averaging over equivalent momenta through the cubic group, the
statistic can be significantly increased. The total number of
equivalent contributions for each value of $|\vec{q}_1|^2$ is
summarized in Table~\ref{tab:nb_config}.

\renewcommand{\arraystretch}{1.1}
\begin{table}[t]
	\caption{Number of equivalent contributions to $A(\tau)$ for different values of $|\vec{q}_1|^2$.}
	\begin{center}
	\begin{tabular}{c@{\hskip 0.1in}c@{\hskip 0.1in}c@{\hskip 0.1in}c@{\hskip 0.1in}c@{\hskip 0.1in}c@{\hskip 0.1in}c@{\hskip 0.1in}c@{\hskip 0.1in}c@{\hskip 0.1in}c@{\hskip 0.1in}c@{\hskip 0.1in}c@{\hskip 0.1in}c@{\hskip 0.1in}c}
	\hline
$(|\vec{q}_1| \times L/(2\pi))^2$		& $1$ & 	$2$ & $3$ & $4$ & $5$ &	$6$ & $8$ & $9$ & $10$ & $11$ & $12$ & $13$ & $14$	\\ 
	\hline 
	Number 			& $12$ &	$48$ & $48$ &	$12$ & $96$ &	$144$ & $48$ & $156$ & $96$ & $144$ & $48$ & $96$ &$288$	\\ 
	\hline
	\end{tabular}	
	\label{tab:nb_config}
	\end{center}	
\end{table}

\subsection{Correlation functions \label{sec:cor_fun}} 

\begin{figure}[b]
	\centering 
	\includegraphics*[width=8.0cm]{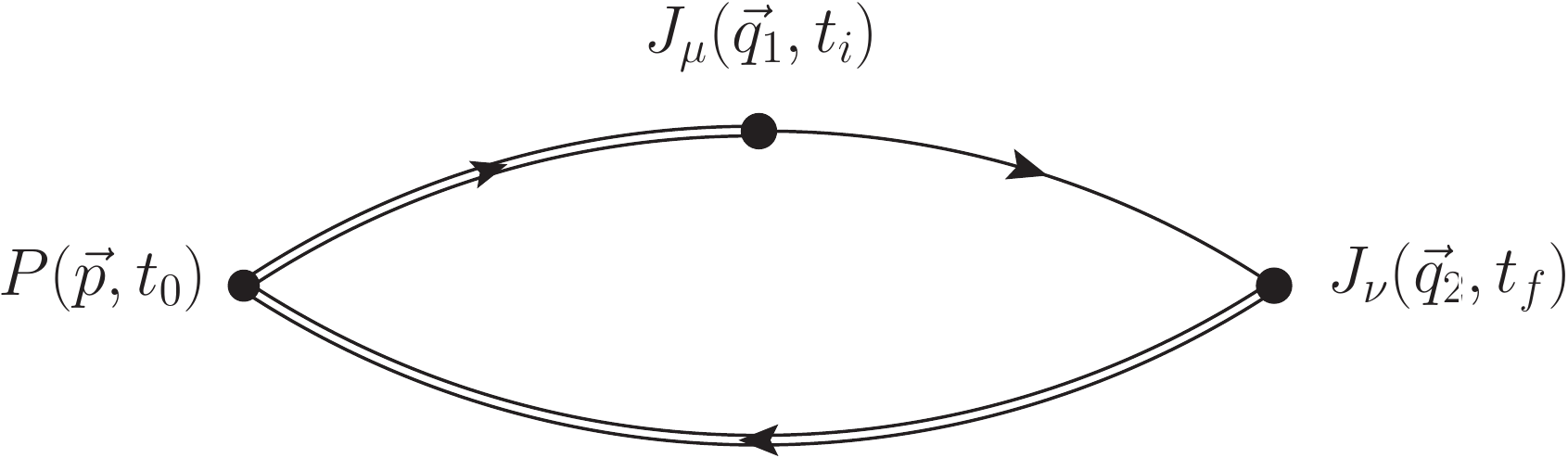}
	\caption{The connected contribution computed using point sources. The two double lines taken together correspond
                 to $\widetilde{G}(y,z;t_0;\vec{p})$.}
	\label{fig:connected}
\end{figure}  

We use the following (anti-Hermitian) interpolating operator for the neutral pion $\pi^{0}$,
\begin{equation}\label{eq:Pdef}
P(x) = \overline{u}(x) \gamma_{5} u(x) - \overline{d}(x) \gamma_{5} d(x) = \overline{\psi}(x) \gamma_5 \tau^3 \psi(x) \,.
\end{equation}
At the quark level, the three-point correlation function receives three contributions,
\be
C^{(3)}_{\mu\nu}(\tau,t_{\pi}) = C_{\mu\nu}^{\mathrm{conn}}(\tau,t_{\pi}) + C_{\mu\nu}^{\mathrm{disc1}}(\tau, t_{\pi})
+ C_{\mu\nu}^{\mathrm{disc2}}(\tau, t_{\pi}).
\ee
Let $x=(\vec x,t_0)$, $y=(\vec y,t_i)$, $z=(\vec 0,t_f)$, and $Q_u = +2/3$ and $Q_{d} = -1/3$ are the electromagnetic charges. Only up and down quark contributions are considered in this paper.  If one uses two ``local'' vector currents,
\be
J_{\mu}^{l}(x) = \sum_f Q_f \ \overline{\psi}_f(x) \gamma_{\mu} \psi_f(x) \,, 
\ee
the connected contribution to the three-point correlation function reads
\begin{align}
\nonumber C_{\mu\nu}^{\mathrm{conn}}(\tau,t_{\pi}) &=  a^6 \sum_{\vec{x}, \vec{y}} \big\langle  J_{\nu}(\vec{0},t_f) J_{\mu}(\vec{y},t_i) P^{\dag}(\vec{x},t_0)  \big\rangle \ e^{-i\vec{q}_1 \vec{y}} \ e^{i\vec{p} \vec{x}} \\
%
\contraction[2ex]{= -  \mathrm{tr}\left[ \tau^3 Q^2 \right] \ a^6 \sum_{\vec{x}, \vec{y}} \big\langle \  }{   \psib   }{   _f(x) \gamma_{\nu} \psi_f(x)  \psib_f(0) \gamma_{\mu} \psi_f(0)  \psib_f(y) \gamma_{5}    }{   \psi   }
\contraction[1.5ex]{= -  \mathrm{tr}\left[ \tau^3 Q^2 \right] \ a^6 \sum_{\vec{x}, \vec{y}} \big\langle \ \psib_f(x) \gamma_{\nu}    }{    \psi   }{   _f(x)   }{  \psib  }
\contraction[1.5ex]{= -   \mathrm{tr}\left[ \tau^3 Q^2 \right] \ a^6 \sum_{\vec{x}, \vec{y}} \big\langle \ \psib_f(x) \gamma_{\nu} \psi(x)_f  \psib_f(z) \gamma_{\mu}    }{   \psi   }{   _f(z)   }{   \psib   }
\nonumber &= - \mathrm{tr}\left[ \tau^3 Q^2 \right] \ a^6 \sum_{\vec{x}, \vec{y}} \big\langle \ \psib_f(z) \gamma_{\nu} \psi_f(z)  \psib_f(y) \gamma_{\mu} \psi_f(y)  \psib_f(x) \gamma_{5} \psi_f(x) \big\rangle  \ e^{-i\vec{q}_1 \vec{y}} \ e^{i\vec{p} \vec{x}} \\
%
\contraction[3ex]{ \quad- \mathrm{tr}\left[ \tau^3 Q^2 \right] \ a^6 \sum_{\vec{x}, \vec{y}} \big\langle \  }{  \psib  }{ _f(z) \gamma_{\nu} \psi_f(z)  \psib_f(y) \gamma_{\mu}  }{ \psi }
\contraction[1.5ex]{\quad -  \mathrm{tr}\left[ \tau^3 Q^2 \right] \ a^6 \sum_{\vec{x}, \vec{y}} \big\langle \ \psib_f(z) \gamma_{\nu} }{ \psi }{ _f(z)  \psib_f(y) \gamma_{\mu} \psi_f(y)  }{ \psi }
\contraction[2.5ex]{ \quad-  \mathrm{tr}\left[ \tau^3 Q^2 \right] \ a^6 \sum_{\vec{x}, \vec{y}} \big\langle \ \psib_f(z) \gamma_{\nu} \psi_f(z)  }{ \psi }{ _f(y) \gamma_{\mu} \psi_f(y)  \psib_f(x) \gamma_{5} }{ \psi }
\nonumber &\quad - \mathrm{tr}\left[ \tau^3 Q^2 \right] \ a^6 \sum_{\vec{x}, \vec{y}} \big\langle \ \psib_f(z) \gamma_{\nu} \psi_f(z)  \psib_f(y) \gamma_{\mu} \psi_f(y)  \psib_f(x) \gamma_{5} \psi_f(x) \big\rangle\ e^{-i\vec{q}_1 \vec{y}} \ e^{i\vec{p} \vec{x}} \\
\nonumber &= 2\, \mathrm{tr}\left[ \tau^3 Q^2 \right] \ \big\langle a^6 \sum_{\vec{x}, \vec{y}} \mathrm{Re} \, \mathrm{Tr} \left[ G(x,z) \gamma_{\nu} G(z,y) \gamma_{\mu} G(y,x) \gamma_{5}  \right]   \ e^{-i\vec{q}_1 \vec{y}} \ e^{i\vec{p} \vec{x}} \ \big\rangle_U \\
&= 2 \, \mathrm{tr}\left[ \tau^3 Q^2 \right] \ \big\langle a^3 \sum_{\vec{y}} \mathrm{Re} \, \mathrm{Tr} \left[ \gamma_{\nu} \gamma_5 G^{\dag}(y,z) \gamma_5 \gamma_{\mu} \widetilde{G}(y,z;t_0;\vec{p})  \right]  \ e^{-i\vec{q}_1 \vec{y}}  \big\rangle_U \,,
\label{eq:connected_contrib}
\end{align}
where Tr is the trace over spinor and color indices; $\tau^3$ is the Pauli matrix; $Q = \mathrm{diag}(2/3, -1/3)$ is the charge matrix; tr the trace over flavors only; $G(x,y)$ denotes the light quark propagator; and $\widetilde{G}(y,z;t_0;\vec{p})$ is a sequential propagator defined below. The correlation function depicted in Fig.~\ref{fig:connected} is computed in two steps: a first inversion on a point source $\eta(z)$ leads to the solution vector $\psi(x)= \sum_{z} G(x,z)\eta(z)$. This solution vector is projected against the pion momenta $\vec{p}$ and, when restricted to a given time slice $t_0$, is used as a secondary source to obtain the sequential propagator (represented by a double line in Fig.~\ref{fig:connected})
\begin{equation}
\widetilde{G}(y,z;t_0;\vec{p}) = a^3 \sum_{\vec{x}} G(y,x) \gamma_5 G(x,z) e^{i\vec{p}\,\vec{x}}  \,.
\end{equation}
In particular, the sequential propagator satisfies the equation
\begin{equation}
a \sum_{y} D(w,y) \widetilde{G}(y,z;t_0,\vec{p}) = \delta_{t_w,t_0}  \gamma_5 G(w,z)  e^{i\vec{p}\,\vec{w}} \equiv \widetilde{\eta}(w) \,,
\end{equation}
where $\widetilde{\eta}$ is the sequential source and $D$ is the lattice Dirac operator. It is clear that a new sequential inversion would be required for each pion momentum $\vec{p}$ and each value of $t_0$. However, the momentum $\vec{q}_1$ and the indices $\mu$ and $\nu$ can be chosen freely without any new inversion of the Dirac operator. This allows us to increase the statistics (see Table~\ref{tab:nb_config}). In practice, we used ten sources per gauge configuration, randomly distributed on the lattice. 

For the results presented in this paper, the connected three-point correlation function is computed using one local and one `point-split'  current. The latter is given by
\begin{align}
J_{\mu}^{c}(x) &= \sum_f \frac{Q_f}{2} \left( \overline{\psi}_f(x+a\hat{\mu})(1+\gamma_{\mu}) U^{\dag}_{\mu}(x) \psi_f(x) - \overline{\psi}_f(x) (1-\gamma_{\mu} ) U_{\mu}(x) \psi_f(x+a\hat{\mu}) \right) \,.
\end{align}  
The Wick contraction is then only slightly modified.
The point-split vector current satisfies the Ward identity and does not need any renormalization factor, contrary to the local vector current. 
In the $\mathcal{O}(a)$-improved theory, the renormalized currents read
\begin{equation}
J_{\mu}^{\alpha,R}(x) = Z^{\alpha}_V (1+b_V^\alpha(g_0) am_q) \left(J_{\mu}^{\alpha}(x) + ac_V^\alpha \partial_{\nu} T_{\mu\nu} \right) \,,
\end{equation}
where the label  $\alpha$ stands for local or conserved and for isospin $I=0$ or $I=1$,
$b_V^\alpha$ and $c_V^\alpha$ are improvement coefficients and $T_{\mu\nu}(x) = -\frac{1}{2}\, \overline{\psi}(x) [\gamma_{\mu}, \gamma_{\nu}] \frac{\tau^3}{2} \psi(x)$ is the tensor density (written here for the improvement of the isovector part of the electromagnetic current). In particular, $Z_V^{c,I} = 1$ and $b_V^{c,I} = 0$, while the renormalization constant $Z_V^{l,I=1}$ has been computed nonperturbatively in \cite{DellaMorteRD,Fritzsch:2012wq} with a relative error below the percent level. In this paper we use the latter values both for the $I=0$ and $I=1$ currents.
The improvement coefficients $c_V^{\alpha}$ have been evaluated in~\cite{Harris:2015vfa}, however 
in this study, we neglect the contribution from the tensor density as well as  the improvement coefficient $b_V$.
Thus $\mathcal{O}(a)$-improvement is only partially implemented. 

\begin{figure}[b]
	\centering 
	\begin{minipage}[c]{0.49\linewidth}
	\centering 
	\includegraphics*[width=0.8\linewidth]{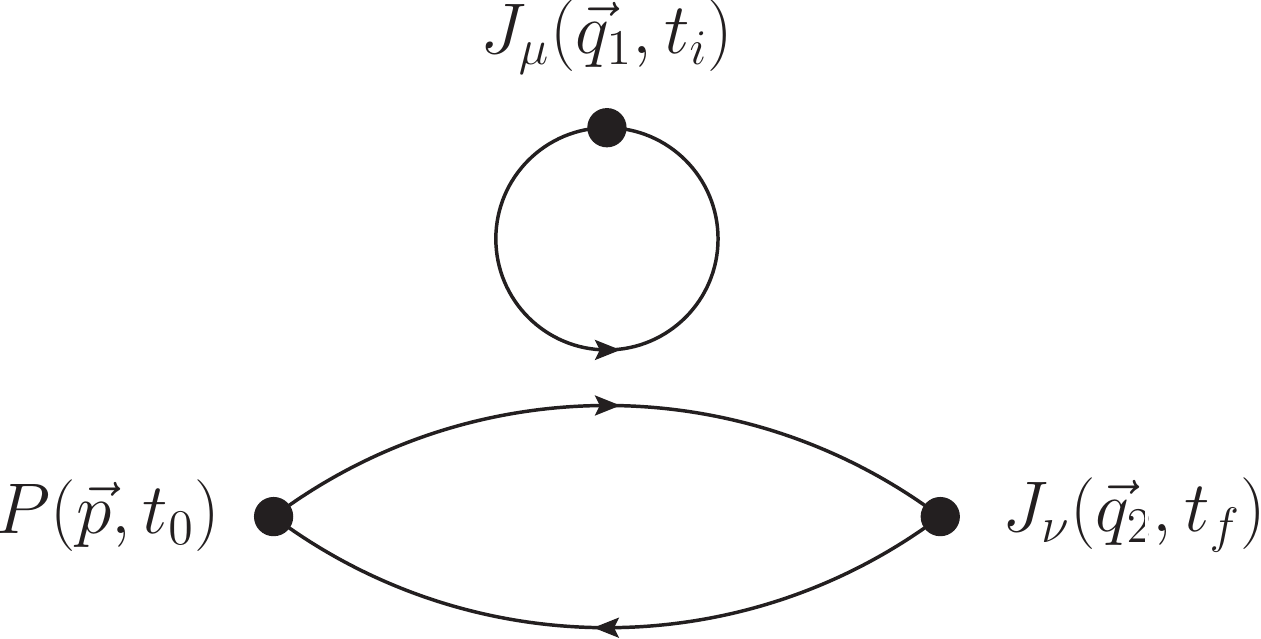}
	\end{minipage}
	\begin{minipage}[c]{0.49\linewidth}
	\centering 
	\includegraphics*[width=0.8\linewidth]{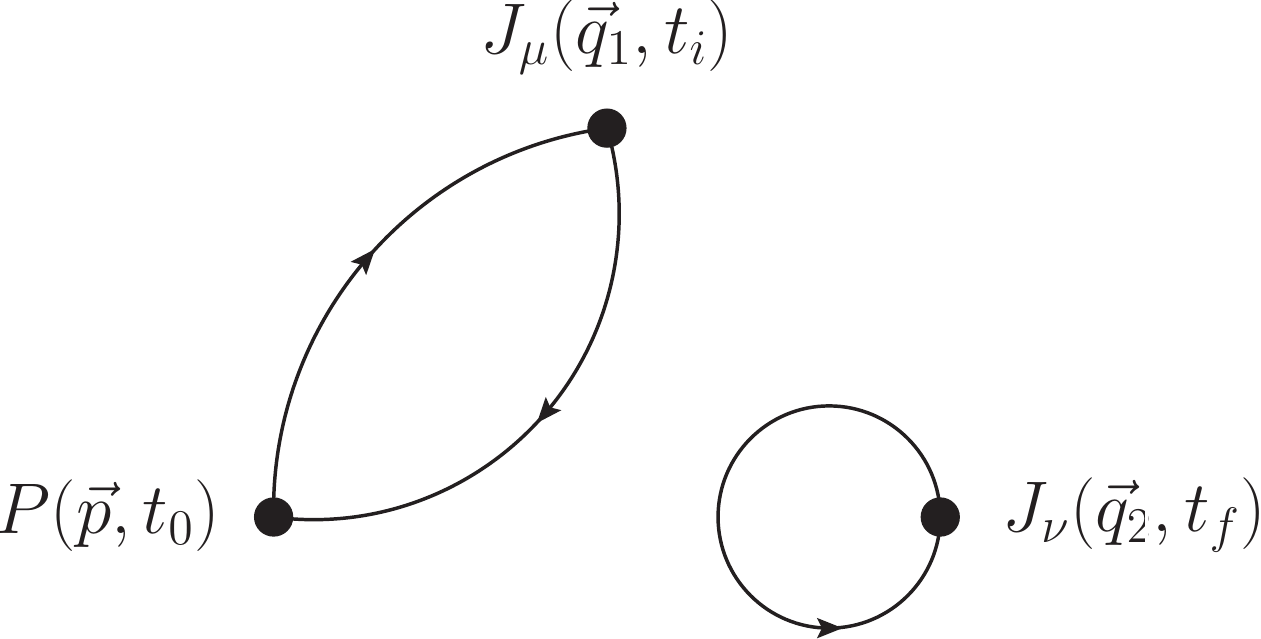}
	\end{minipage}
	\caption{The two disconnected diagrams contributing to the $\pi^0 \to \gamma^* \gamma^*$ form factor.}
		\label{fig:disconnected}
\end{figure}

For the disconnected contributions, we use two local vector currents.
Wick contractions involving only the pion do not contribute since the $u$ and $d$ contributions exactly compensate each other. Therefore, one vector current must be contracted with the pion which leads to the two diagrams depicted in Fig.~\ref{fig:disconnected}. The first diagram in the $n_f=2$ theory corresponds to the following contraction 
\begin{align}
\contraction[2.4ex]{C_{\mu\nu}^{\mathrm{disc1}}(\tau, t_{\pi}) = - \mathrm{tr}\left[ Q \right] \ \frac{a^9}{V}  \mathrm{tr}\left[ \tau^3 Q \right]  \ \sum_{\vec{x}, \vec{y}, \vec{z}} \big\langle \  }{  \psib  }{   _f(z) \gamma_{\nu} \psi_f(z)  \psib_f(y) \gamma_{\mu} \psi_f(y)  \psib_f(x) \gamma_{5}    }{   \psi   }
\contraction[1.2ex]{C_{\mu\nu}^{\mathrm{disc1}}(\tau, t_{\pi}) = - \mathrm{tr}\left[ Q \right] \ \frac{a^9}{V}  \mathrm{tr}\left[ \tau^3 Q \right]  \ \sum_{\vec{x}, \vec{y}, \vec{z}} \big\langle \ \psib_f(z) \gamma_{\nu} \psi_f(z)    }{   \psib   }{   _f(y) \gamma_{\mu}  }{  \psi   }
\contraction[2.2ex]{C_{\mu\nu}^{\mathrm{disc1}}(\tau, t_{\pi}) = - \mathrm{tr}\left[  Q \right] \ \frac{a^9}{V}  \mathrm{tr}\left[ \tau^3 Q \right]  \ \sum_{\vec{x}, \vec{y}, \vec{z}} \big\langle \ \psib_f(z) \gamma_{\nu}   }{   \psi  }{   _f(z)  \psib_f(y) \gamma_{\mu} \psi_f(y)  }{ \psib  }
\nonumber C_{\mu\nu}^{\mathrm{disc1}}&(\tau, t_{\pi}) = - \mathrm{tr}\left[ Q \right]  \ \mathrm{tr}\left[ \tau^3 Q \right]  \ \frac{a^9}{V}  \sum_{\vec{x}, \vec{y}, \vec{z}} \big\langle \ \psib_f(z) \gamma_{\nu} \psi_f(z)  \psib_f(y) \gamma_{\mu} \psi_f(y)  \psib_f(x) \gamma_{5} \psi_f(x) \big\rangle_U  \ e^{-i\vec{q}_2 \vec{z}} \ e^{-i\vec{q}_1 \vec{y}} \ e^{i\vec{p} \vec{x}} \\
&= -  \mathrm{tr}\left[ Q \right]   \  \mathrm{tr}\left[ \tau^3 Q \right]  \ \big\langle a^3 \sum_{\vec{y}} \mathrm{Tr} \left[  G(y,y) \gamma_{\mu}  \right]  e^{-i\vec{q}_1 \vec{y}} \ \frac{a^6}{V} \sum_{\vec{x},\vec{z}}  \mathrm{Tr}  \left[ G(z,x) \gamma_{5} G(x,z) \gamma_{\nu}  \right]   \ e^{-i\vec{q}_2 \vec{z}} \ e^{i\vec{p} \vec{x}} \ \big\rangle_U \,,
\label{eq:C3disc1}
\end{align}
and the second diagram reads
\begin{align}
\contraction[2ex]{C_{\mu\nu}^{\mathrm{disc2}}(\tau, t_{\pi})= - \mathrm{tr}\left[  Q \right] \  \mathrm{tr}\left[ \tau^3 Q \right] \ \frac{a^9}{V} \sum_{\vec{x}, \vec{y}, \vec{z}} \big\langle \  }{   \psib   }{   _f(z) \gamma_{\nu}  }{   \psi   }
\contraction[2ex]{ C_{\mu\nu}^{\mathrm{disc2}}(\tau, t_{\pi}) = - \mathrm{tr}\left[  Q \right] \  \mathrm{tr}\left[ \tau^3 Q \right]  \ \frac{a^9}{V} \sum_{\vec{x}, \vec{y}, \vec{z}} \big\langle \ \psib_f(z) \gamma_{\nu} \psi_f(z)    }{   \psib   }{   _f(y) \gamma_{\mu} \psi_f(y)  \psib_f(x) \gamma_{5}   }{  \psi   }
\contraction[1.5ex]{C_{\mu\nu}^{\mathrm{disc2}}(\tau, t_{\pi}) =- \mathrm{tr}\left[  Q \right] \ \frac{a^9}{V}   \mathrm{tr}\left[ \tau^3 Q \right] \ \sum_{\vec{x}, \vec{y}, \vec{z}} \big\langle \ \psib_f(z) \gamma_{\nu} \psi_f(z)  \psib_f(y) \gamma_{\mu}    }{   \psi   }{   _f(y)   }{   \psib   }
\nonumber C_{\mu\nu}^{\mathrm{disc2}}&(\tau, t_{\pi}) = -  \mathrm{tr}\left[ Q \right]  \ \mathrm{tr}\left[ \tau^3 Q \right]  \ \frac{a^9}{V} \sum_{\vec{x}, \vec{y}, \vec{z}} \big\langle \ \psib_f(z) \gamma_{\nu} \psi_f(z)  \psib_f(y) \gamma_{\mu} \psi_f(y)  \psib_f(x) \gamma_{5} \psi_f(x) \big\rangle_U \ e^{-i\vec{q}_2 \vec{z}} \ e^{-i\vec{q}_1 \vec{y}} \ e^{i\vec{p} \vec{x}} \\
&= - \mathrm{tr}\left[ Q \right] \  \mathrm{tr}\left[ \tau^3 Q \right]  \ \big\langle a^3 \sum_{\vec{z}} \mathrm{Tr} \left[  G(z,z) \gamma_{\nu}  \right]  e^{-i\vec{q}_2 \vec{z}} \ \frac{a^6}{V} \sum_{\vec{x},\vec{y}}  \mathrm{Tr}  \left[  G(y,x) \gamma_5 G(x,y) \gamma_{\mu}  \right]  \ e^{-i\vec{q}_1 \vec{y}} \ e^{i\vec{p} \vec{x}} \ \big\rangle_U \,.
\label{eq:C3disc2}
\end{align}
More details about the numerical evaluation of the disconnected contribution are given in Section~\ref{subsub:disc}.

\section{Lattice computation \label{sec:lattice_simulation}}

\begin{table}[t]
\caption{Parameters of the simulations: the bare coupling $\beta = 6/g_0^2$, the lattice resolution, the hopping parameter $\kappa$, the lattice spacing $a$ in physical units extracted from \cite{Fritzsch:2012wq}, the pion mass $m_{\pi}$, the pion decay constant $F_{\pi}$ extracted from \cite{Engel:2014eea} and the number of gauge configurations.}
\vskip 0.1in
\begin{tabular}{lcl@{\hskip 02em}l@{\hskip 01em}ccccccc}
	\hline
	CLS	&	$\quad\beta\quad$	&	$L^3\times T$ 		&	$\kappa$		&	$a~[\fm]$	&	$m_{\pi}~[\MeV]$	&	$F_{\pi}~[\MeV]$	& $m_{\pi}L$	&	$\#$confs \\
	\hline
	A5	&		$5.2$		&	$32^3\times64$	& 	$0.13594$	& 	$0.0749(8)$  	& 	$334(4)$ &	$106.0(6)$ &4.0&400 \\  
	B6	&					&	$48^3\times96$	& 	$0.13597$	& 		  		& 	$281(3)$ &	$102.3(5)$ &5.2& 400\\  
	\hline
	E5	&		$5.3$		&	$32^3\times64$	& 	$0.13625$	& 	$0.0652(6)$  	& 	$437(4)$ &	$115.2(6)$ &4.7 &400\\  
	F6	&		 	 		& 	$48^3\times96$	&	$0.13635$	& 				& 	$314(3)$ &	$105.3(6)$ &5.0&300 \\      
	F7	&		 	 		& 	$48^3\times96$	&	$0.13638$	& 				& 	$270(3)$ &	$100.9(4)$ &4.3&350 \\      
	G8	&		 	 		& 	$64^3\times128$	&	$0.136417$	& 				& 	$194(2)$ &	$95.8(4)$  &4.1&300\\    
	\hline  
	N6	&		$5.5$ 		& 	$48^3\times96$	&	$0.13667$	& 	$0.0483(4)$	& 	$342(3)$ &	$105.8(5)$ &4.0 &450\\      
	O7	&					& 	$64^3\times128$	&	$0.13671$	& 				& 	$268(3)$ &	$101.2(4)$ &4.2 & 150 \\      
	\hline
 \end{tabular} 
\label{tabsim}
\end{table}

This work is based on a subset of the $n_f=2$ Coordinated Lattice Simulations (CLS) ensembles generated using either the DD-HMC algorithm \cite{LuscherQA, LuscherRX, LuscherES, Luscherweb} or the MP-HMC algorithm \cite{MarinkovicEG}. They used the nonperturbatively $\mathcal{O}(a)$-improved Wilson-Clover action for the fermions \cite{SheikholeslamiIJ,LuscherUG} and the plaquette gauge action for  gluons \cite{WilsonSK}. The simulation parameters for each lattice ensemble are summarized in Table~\ref{tabsim}. Three lattice spacings in the range [0.05-0.075]~fm are considered with pion masses down to 193~MeV. The lattice spacings are extracted from \cite{Fritzsch:2012wq} where the kaon decay constant is used to set the scale. Finally, all ensembles satisfy the condition $Lm_{\pi}>4$ such that volume effects are expected to be negligible \cite{Meyer:2013dxa}.  For more details on the ensembles, see~\cite{Fritzsch:2012wq}.

\subsection{Two-point pion correlation function}

The pion mass and its overlap $Z_{\pi}$ with our interpolating operator are estimated using both a single and a double exponential fit. The results are summarized in Table~\ref{tab:spectrum}. As a cross-check, the effective mass 
\begin{equation}
m_{\pi}^{\rm eff}(t) = \log \left( \frac{C^{(2)}(t)}{C^{(2)}(t+1)} \right) \,,
\end{equation}
is also computed from the two-point correlator and fitted to a constant in the plateau region. The results for the single and double exponential fits are in perfect agreement within statistical errors, indicating that the contribution of excited states is under control.

\renewcommand{\arraystretch}{1.1}
\begin{table}[t]
\caption{Ground state energy $E_{\pi}$ and overlap factors $Z_{\pi}$ extracted from a single or double exponential fit of the pseudoscalar two-point correlation function for each lattice ensemble.}
\begin{center}
	\begin{tabular}{l@{\qquad}cc@{\qquad}cc@{\qquad}c}
	\hline
				& 	\multicolumn{2}{c}{Single exponential fit}		&	\multicolumn{2}{c}{Double exponential fit}	&	$m_{\pi}^{\rm eff}$ \\
	\cline{2-3}\cline{4-5}\cline{6-6} 
	CLS			& 	$iZ_{\pi}$		&	$E_{\pi}$		& 	$iZ_{\pi}$		&	$E_{\pi}$ 		&	$E_{\pi}$	\\ 
	\hline 
	A5    			& 	0.1874(18) 	&	 0.1267(9)		& 	0.1894(18)	& 	0.1274(8) 		& 	0.1274(8) \\
	\hline	
	B6    			& 	0.1778(14) 	&	0.1066(5) 		& 	0.1776(14) 	& 	0.1066(5) 		& 	0.1067(5)   \\
	\hline
	E5			&	0.1410(15)	&	0.1445(6)		&	0.1410(15)	&	0.1445(6)		&	0.1451(6) \\ 
	\hline
	F6			&	0.1259(8)		&	0.1038(4)		&	0.1258(8)		&	0.1037(4)	&	0.1039(4) \\
	\hline
	F7   			&  	0.1228(8)		& 	0.0891(4) 		&	0.1228(8)		& 	0.0890(4) 		&	 0.0893(4)\\
	\hline
	G8    		& 	0.1164(10)	& 	0.0642(4)		& 	0.1166(13)	& 	0.0643(5)		& 	0.0645(4)  \\
	\hline
	N6    		& 	0.0670(6) 		& 	0.0839(3) 		& 	0.0671(11)	&	 0.0839(6) 	&	0.0841(3) \\
	\hline
	O7    		& 	0.0613(6)	 	& 	0.0655(3)		& 	0.0617(16)	&	 0.0657(8) 	&	0.0660(3)  \\
	\hline 
	\end{tabular}
\label{tab:spectrum}
\end{center}
\end{table}

\subsection{Extraction of the form factor}

\subsubsection{Finite-time extent corrections \label{subsec:fte}}

\begin{figure}[t!]

	\begin{minipage}[c]{0.49\linewidth}
	\centering 
	\includegraphics*[width=0.9\linewidth]{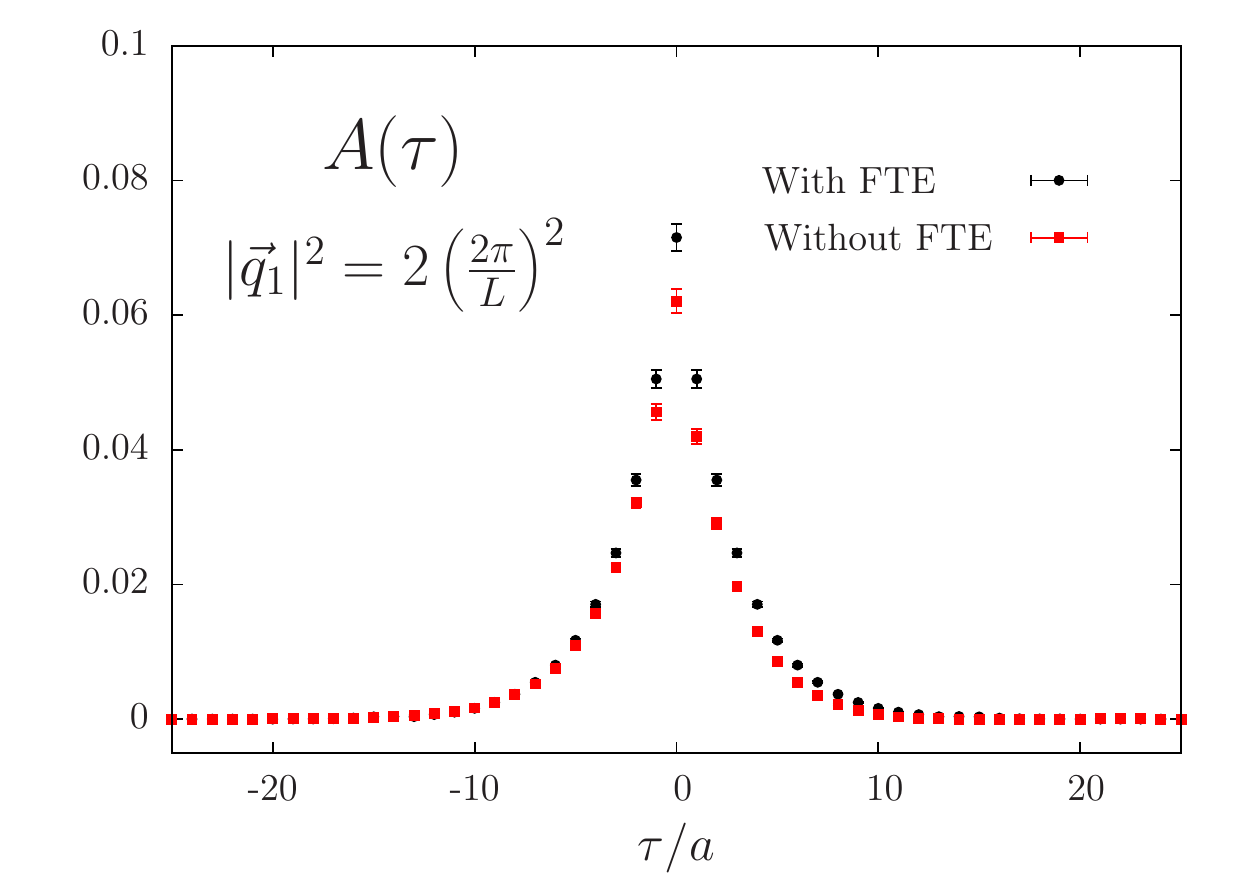}
	\end{minipage}
	\begin{minipage}[c]{0.49\linewidth}
	\centering 
	\includegraphics*[width=0.9\linewidth]{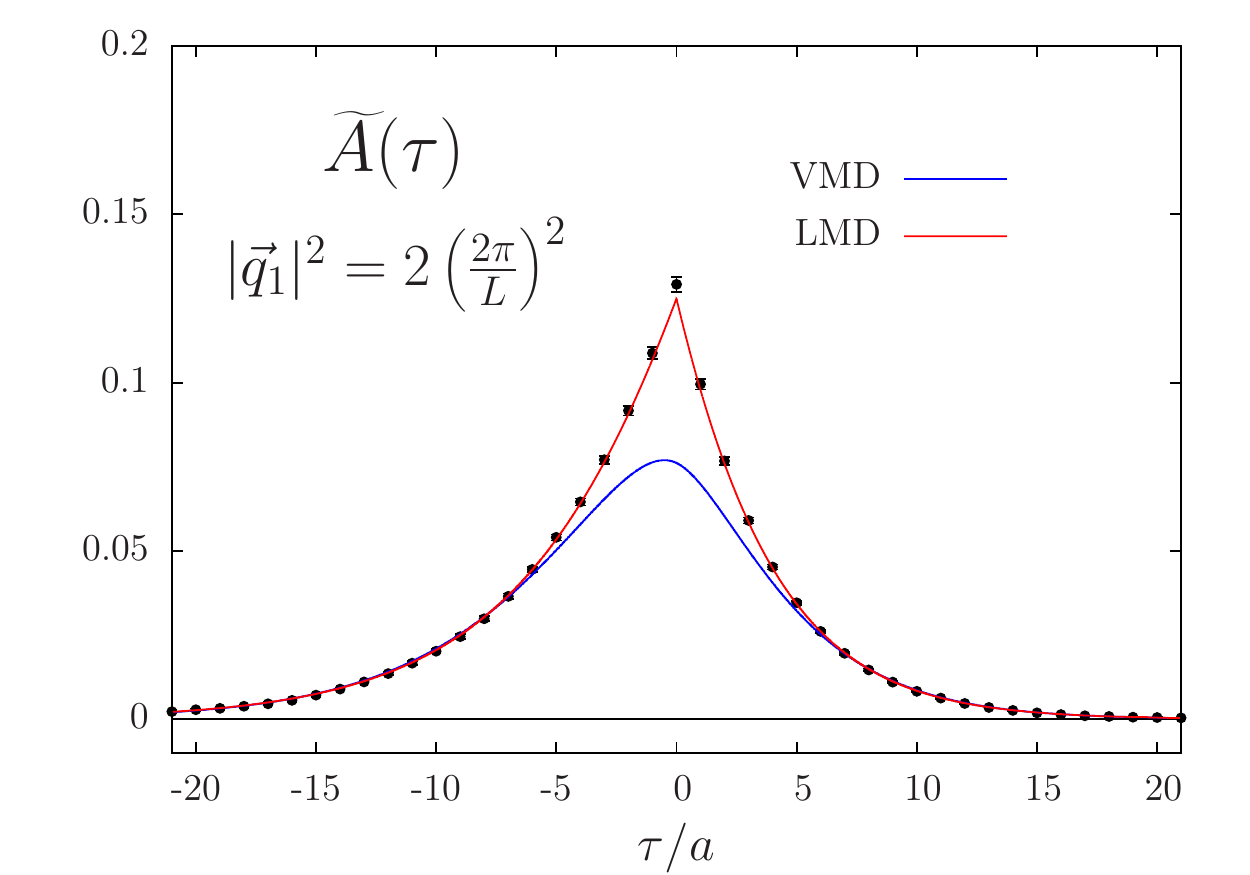}
	\end{minipage}
	
	\caption{Left: Finite-time extent corrections for $A(\tau)$ defined in Eq.~(\ref{eq:Ascalar}) at $t/a=(t_f-t_0)/a=25$ for the lattice ensemble E5. Right: The function $\widetilde{A}(\tau)$ (black points) and the VMD (blue line) and LMD (red line) fits used to describe the tail of the function at large $\tau$ for the lattice ensemble F7 (This is a global fit but only results for the spatial momentum $|\vec{q}_1|^2 = 2 (2\pi/L)^2$ are shown for clarity).}	
	\label{fig:A}
\end{figure}

Due to the finite-time extent of the lattice, backward propagating pions may contribute to the three-point correlation function. Indeed, taking into account the finite size of the box, the asymptotic behavior of the three-point correlation function now reads
\begin{align}
 C^{(3)}_{\mu\nu}(\tau,t_{\pi})    \xrightarrow[t_{\pi} \to \infty]{\tau>0}   \frac{Z_{\pi} a^3}{2E_{\pi}} \left[    \right. & \left.    \sum_{\vec{z}} \, \langle 0 | J_{\mu}(\vec{z}, \tau)  J_{\nu}(\vec{0}, 0) | \pi(p) \rangle \, e^{-i \vec{q}_1 \vec{z}} \, e^{-E_{\pi} t_{\pi} }  \right. \\
\nonumber & + \left.   \sum_{\vec{z}} \, \langle 0 | J_{\nu}(\vec{0}, \tau) J_{\mu}(\vec{z}, 0 ) | \pi(p) \rangle \, e^{-i \vec{q}_1 \vec{z}} \, e^{-E_{\pi} (T-t_{\pi}-\tau) } \right] \,,
\end{align}
such that 
\begin{gather}
A^{\rm lat}_{\mu\nu}(\tau>0) = A_{\mu\nu}(\tau) + A_{\nu\mu}(\tau) e^{-E_{\pi} (T-\tau-t_{\pi}) } = A_{\mu\nu}(\tau) \left( 1 - e^{-E_{\pi} (T-2t_{\pi}-\tau) } \right) \,,
\end{gather}
and similarly for $\tau<0$. In particular, for $A(\tau)$ defined in Eq.~(\ref{eq:Ascalar}), one has
\begin{gather}
A^{\rm lat}(\tau>0) = A(\tau) \left( 1 - e^{-E_{\pi} (T-2t_{\pi}-\tau) } \right) \,.
\end{gather}
Therefore, for values of $t_{\pi}$ close to $T/2$, one expects large corrections which tend to lower the real value. This effect is shown in the left panel of Fig.~\ref{fig:A} for the lattice ensemble E5 at our largest time separation $t/a=(t_f-t_0)/a=25$. After a finite-time extent correction, the function $A(\tau)$ is indeed symmetric within error bars. For lattice ensembles with larger resolutions ($T/a=96,128$) these effects are exponentially suppressed and completely negligible at our level of precision. \\

In Eq.~(\ref{eq:lat_M}), integration bounds are $\pm \infty$. The function $\widetilde{A}(\tau)$ decreases exponentially fast at large $|\tau|$ but the exponential factor $\exp(\omega_1 \tau)$ in Eq.~(\ref{eq:lat_M}) tends to probe the tail of the function $\widetilde{A}(\tau)$ at large $|\tau|$, making the numerical integration difficult for two reasons: first, the finite-time extent of the lattice obviously limits the range of integration. Secondly, the signal-to-noise ratio decreases when $|\tau|$ increases. To circumvent these problems, we take advantage of the idea that the VMD model is expected to work well in the large~$|\tau|$ limit where the excited states' contribution in the vector channel is small, and we fit the lattice data at large $|\tau|$ using 
\begin{multline}
A^{\VMD}_{kl}(\tau) = -i q_{kl} A^{\VMD}(\tau) =  \frac{ Z_{\pi} \, q_{kl}  }{ 4 m_{\pi}} \left[  \frac{\alpha \, M_V^4}{ m_{\pi} \sqrt{M_V^2+|\vec{q}_1|^2} \left( 2 \sqrt{M_V^2+|\vec{q}_1|^2 } - m_{\pi} \right)  } \, e^{- \sqrt{M_V^2+|\vec{q}_1|^2} \, |\tau|}   \right. \\
\left. - \frac{\alpha \,M_V^4}{  m_{\pi} \sqrt{M_V^2+|\vec{q}_1|^2}  \left(2 \sqrt{M_V^2+|\vec{q}_1|^2} + m_{\pi} \right) } \, e^{-  \left( m_{\pi}  + \sqrt{M_V^2+|\vec{q}_1|^2}  \right)  |\tau|} \right] \,,
\label{eq:A_VMD}
\end{multline}
where $\alpha$ and $M_V$ are free parameters and $q_{kl}$ is defined in Eq~(\ref{eq:Ascalar}). We have performed a global fit for each lattice ensemble where all momenta are fitted simultaneously. Then, we introduced a cutoff $\tau_c$ where the data are too noisy or not available and the VMD fit is used to perform the integration up to infinity in Eq.~(\ref{eq:lat_M}). The time $\tau_c \gtrsim 1.3~\fm$ is chosen such that it takes approximately the same value in physical units for all lattice ensembles. We will discuss the potential systematic error introduced by this method in Sec.~\ref{sec:syst_err}. A typical fit for the lattice ensemble F7 is depicted in the right panel of Fig.~\ref{fig:A} where the result using the LMD model rather that the VMD model is also shown. The main advantage of the LMD model is that it is able to describe the cusp at $\tau=0$ (Appendix {\ref{app:A}}). As shown in Appendix \ref{app:OPE}, the cusp is directly related to the behaviour of the doubly-virtual form factor predicted by the OPE in Eq.~(\ref{eq:OPE}).

\subsubsection{Fits in four-momentum space \label{subsec:fit_res}}

In this section, we propose to compare our results with the phenomenological models introduced in Sec.~\ref{sec:pion_ff}. In particular, since we are using Wilson fermions, the chiral symmetry is lost even in the chiral limit and is recovered only once the results are extrapolated to the continuum and chiral limit. It is then important to check that our results are in agreement with the ABJ anomaly.

\begin{figure}[t]
	\begin{minipage}[c]{0.49\linewidth}
	\centering 
	\includegraphics*[width=0.89\linewidth]{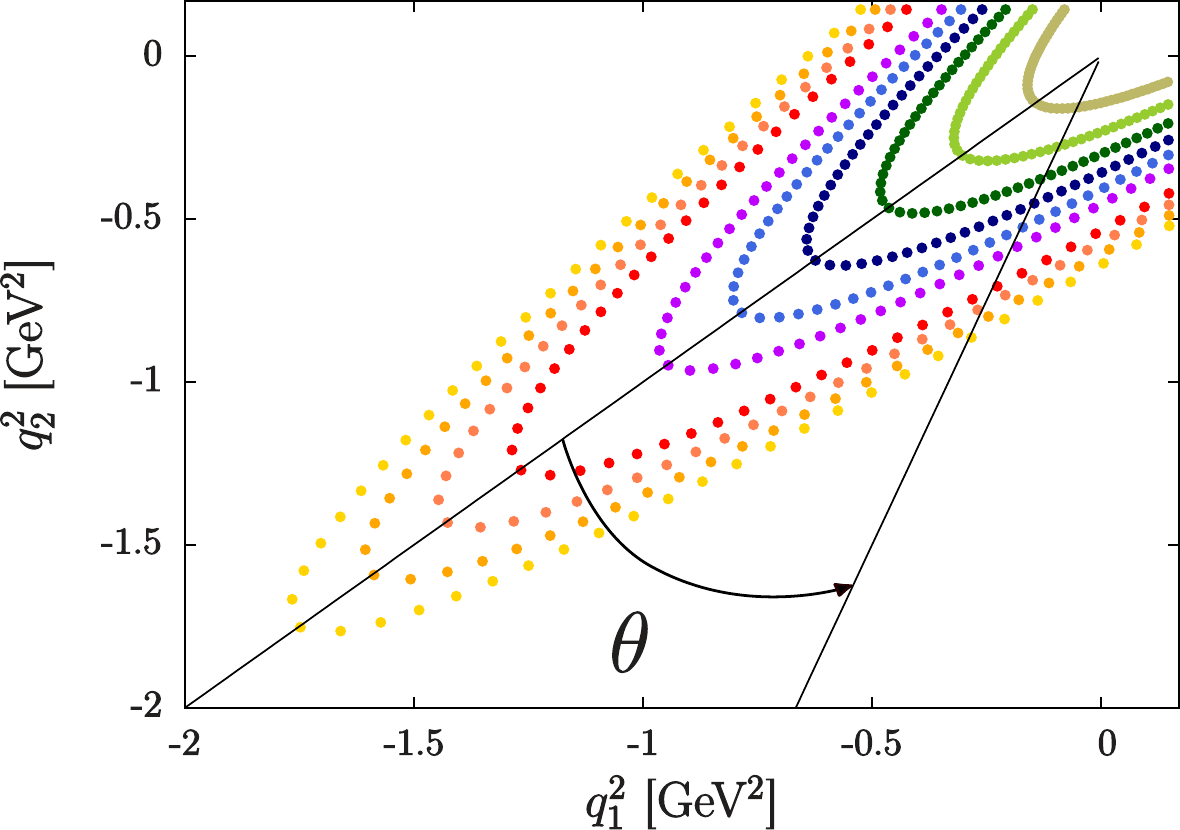}
	\vspace{-0.25cm}
	\end{minipage}
	\begin{minipage}[c]{0.49\linewidth}
	\centering 
	\includegraphics*[width=0.95\linewidth]{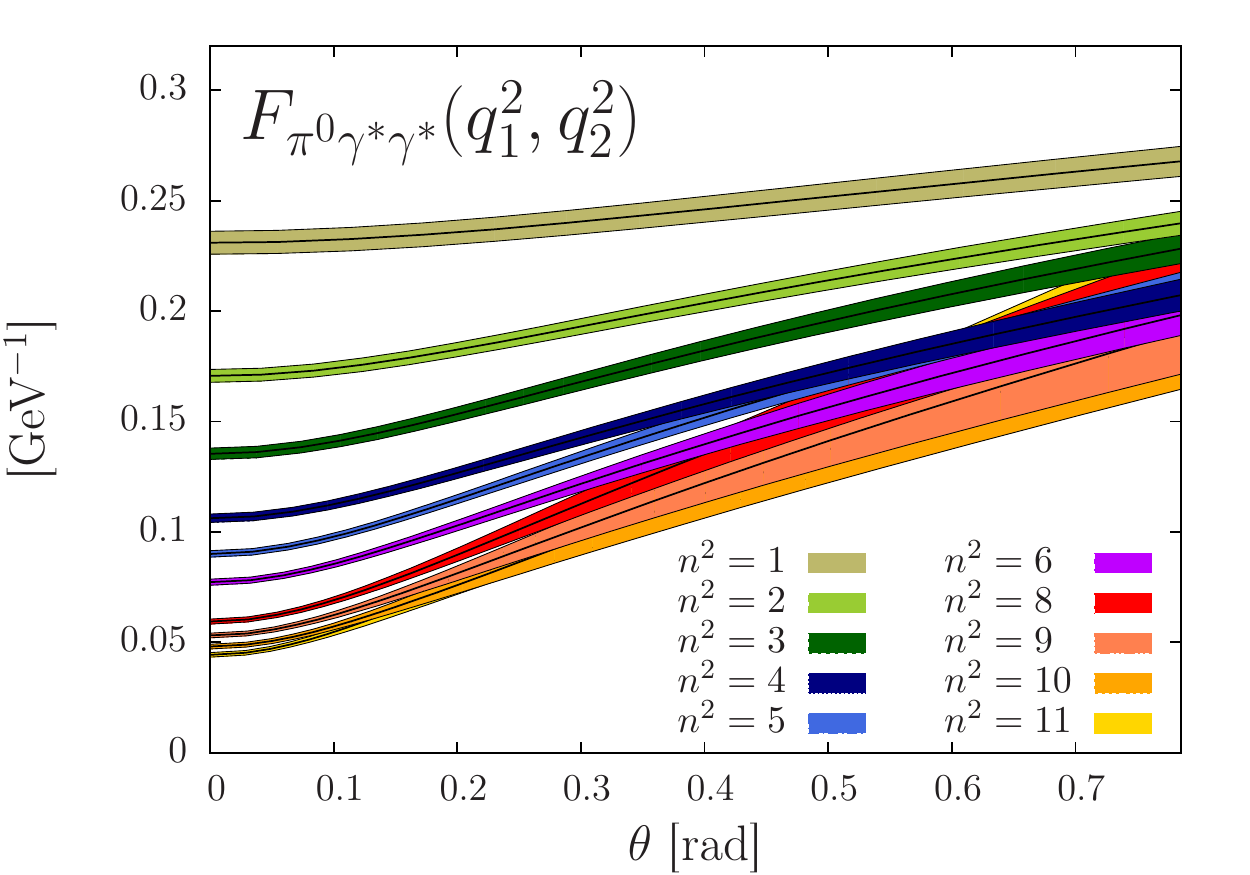}
	\end{minipage}
	
	\caption{Left: Sampling of our data in the ($q_1^2,q_2^2)$ plane. Right: The form factor for different values of $|\vec{q}_1|^2$. For each value of $|\vec{q}_1|^2 = n^2 (2\pi/L)^2 = (q_2^2-q_1^2-m_{\pi}^2)^2/(4m_{\pi}^2) - q_1^2$, one gets a curve by varying continuously the value of $\tan(\theta-\pi/4) = q_1^2 / q_2^2$. Data correspond to the lattice ensemble O7.}	
	\label{fig:FF_lat}
\end{figure}

On the lattice, the form factor is obtained as a continuous function of $\omega_1$ for each value of the discretized spatial momentum $|\vec{q}_1|^2$ and a typical example for the lattice ensemble F6 is depicted in Fig.~\ref{fig:FF_lat}. Therefore, to fit the form factor, we first have to sample our data. We have selected values of $\omega_1$ such that data points are regularly distributed along each curve in the $(q_1^2,q_2^2)$ plane as depicted in the left panel of Fig.~\ref{fig:FF_lat}. However, as discussed in Sec.~\ref{sec:syst_err}, no significant difference has been observed by using different samplings.\\

\begin{figure}[t!]
	\begin{minipage}[c]{0.32\linewidth}
	\centering 
	\includegraphics*[width=\linewidth]{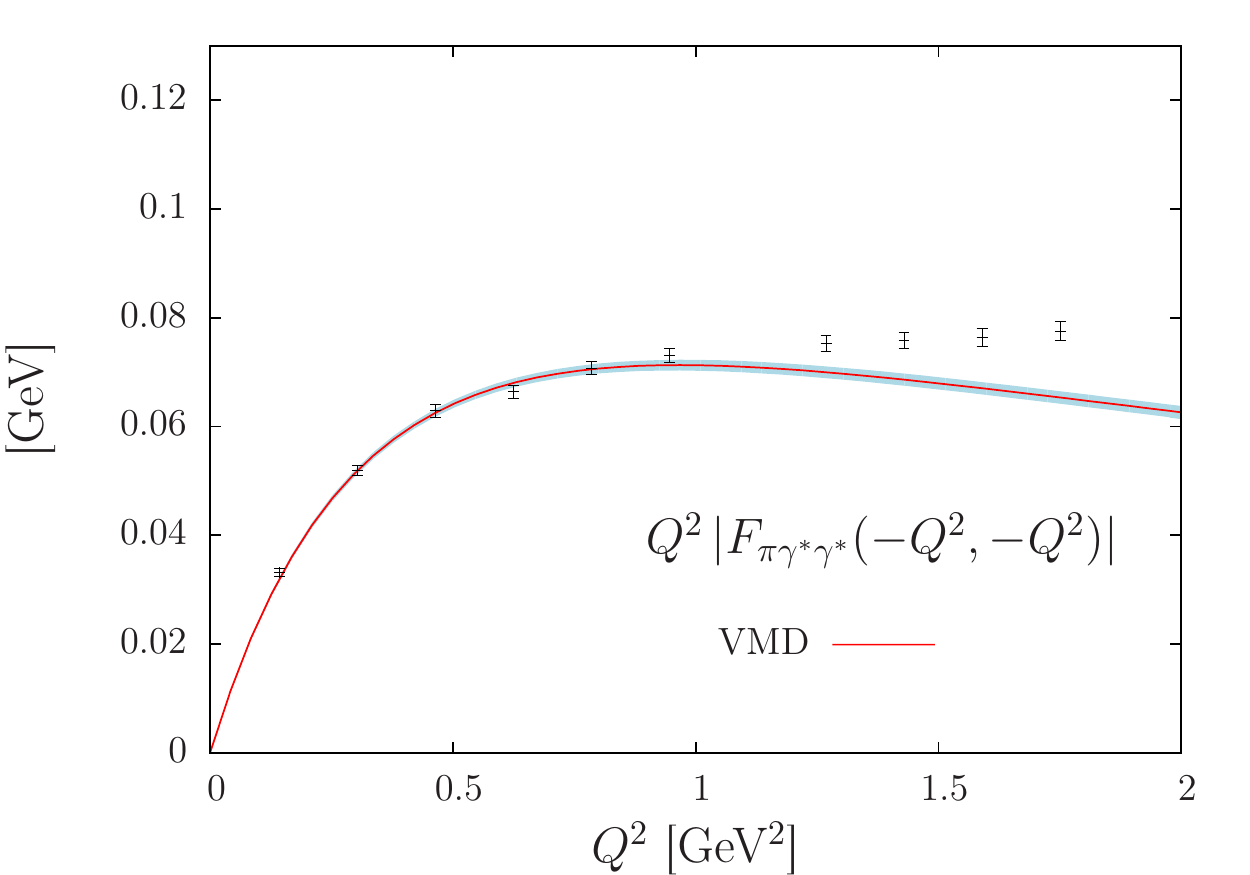}
	\end{minipage}
	\begin{minipage}[c]{0.32\linewidth}
	\centering 
	\includegraphics*[width=\linewidth]{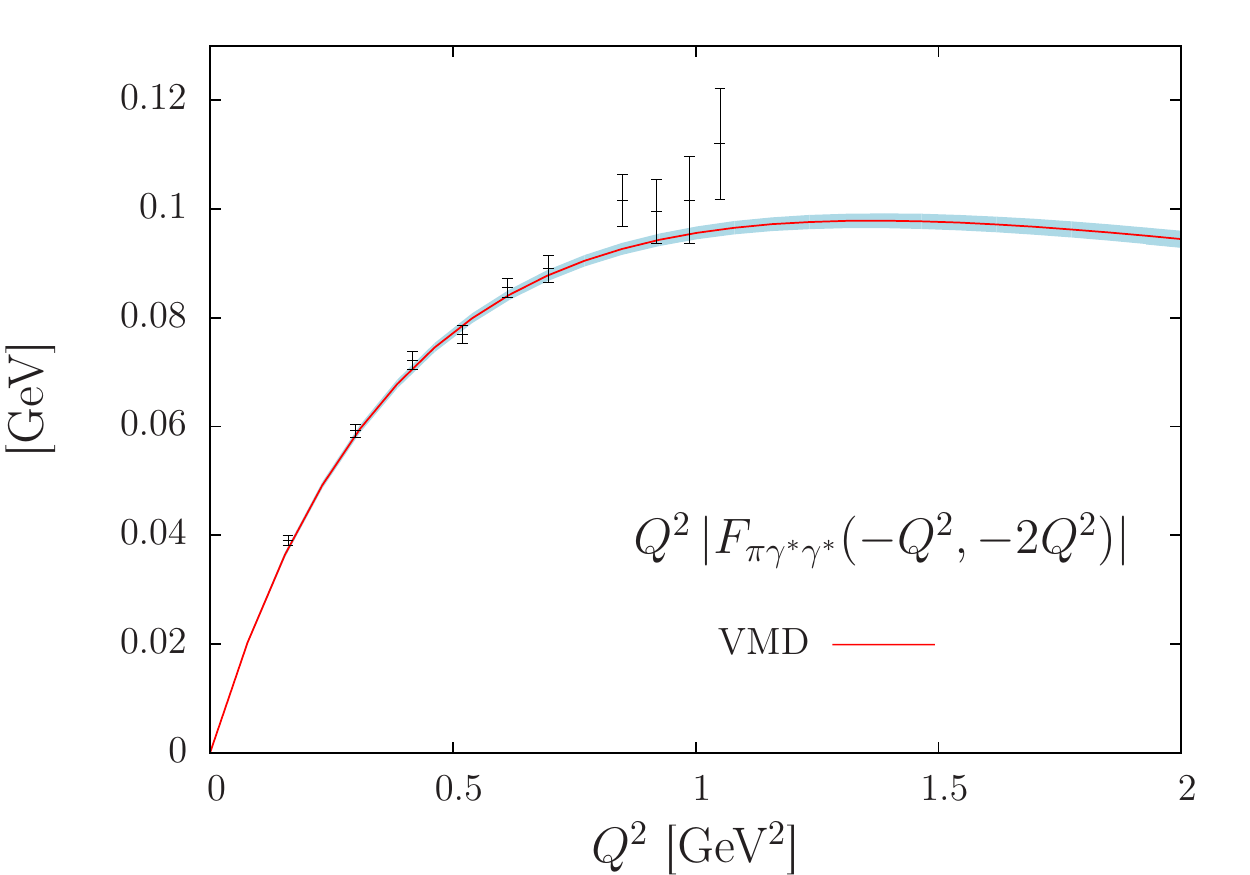}
	\end{minipage}
	\begin{minipage}[c]{0.32\linewidth}
	\centering 
	\includegraphics*[width=\linewidth]{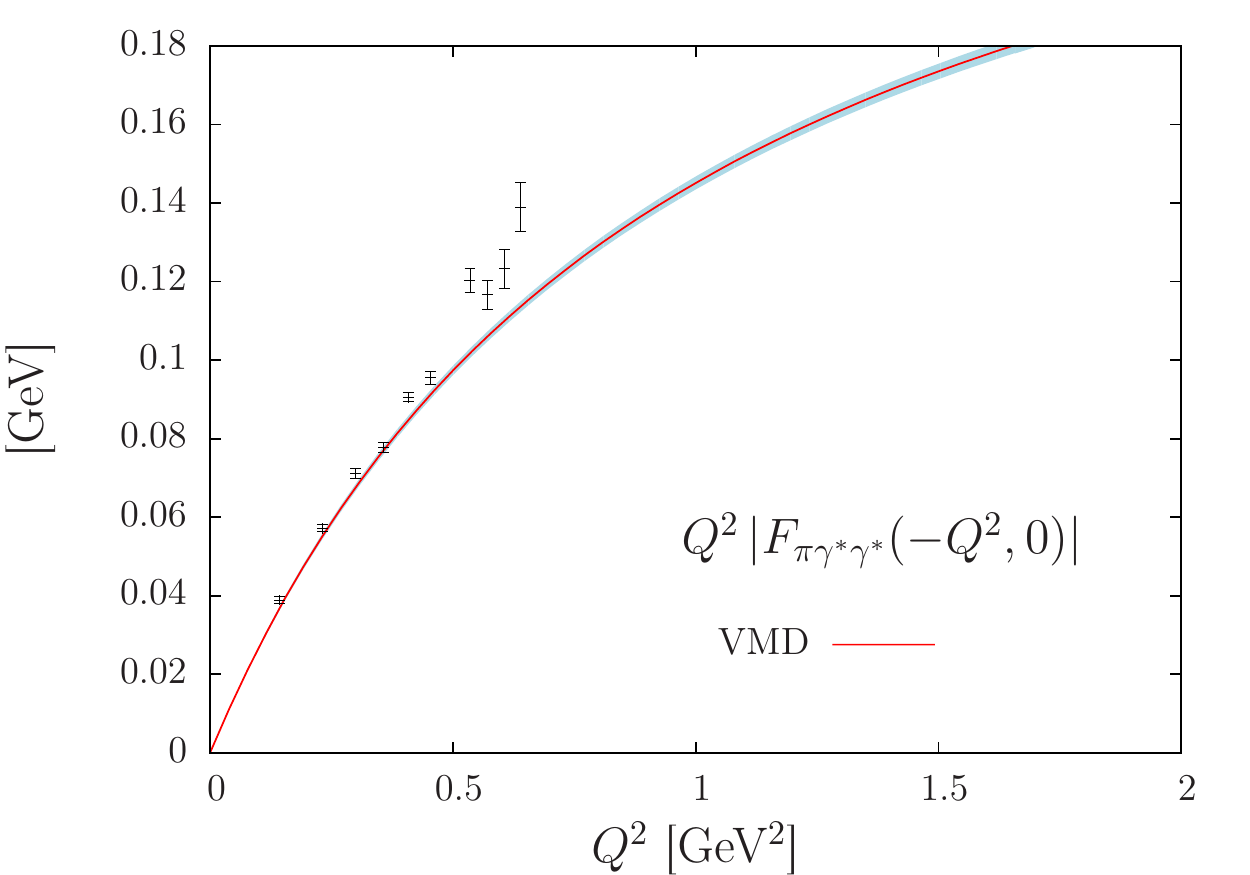}
	\end{minipage}
	
	\begin{minipage}[c]{0.32\linewidth}
	\centering 
	\includegraphics*[width=\linewidth]{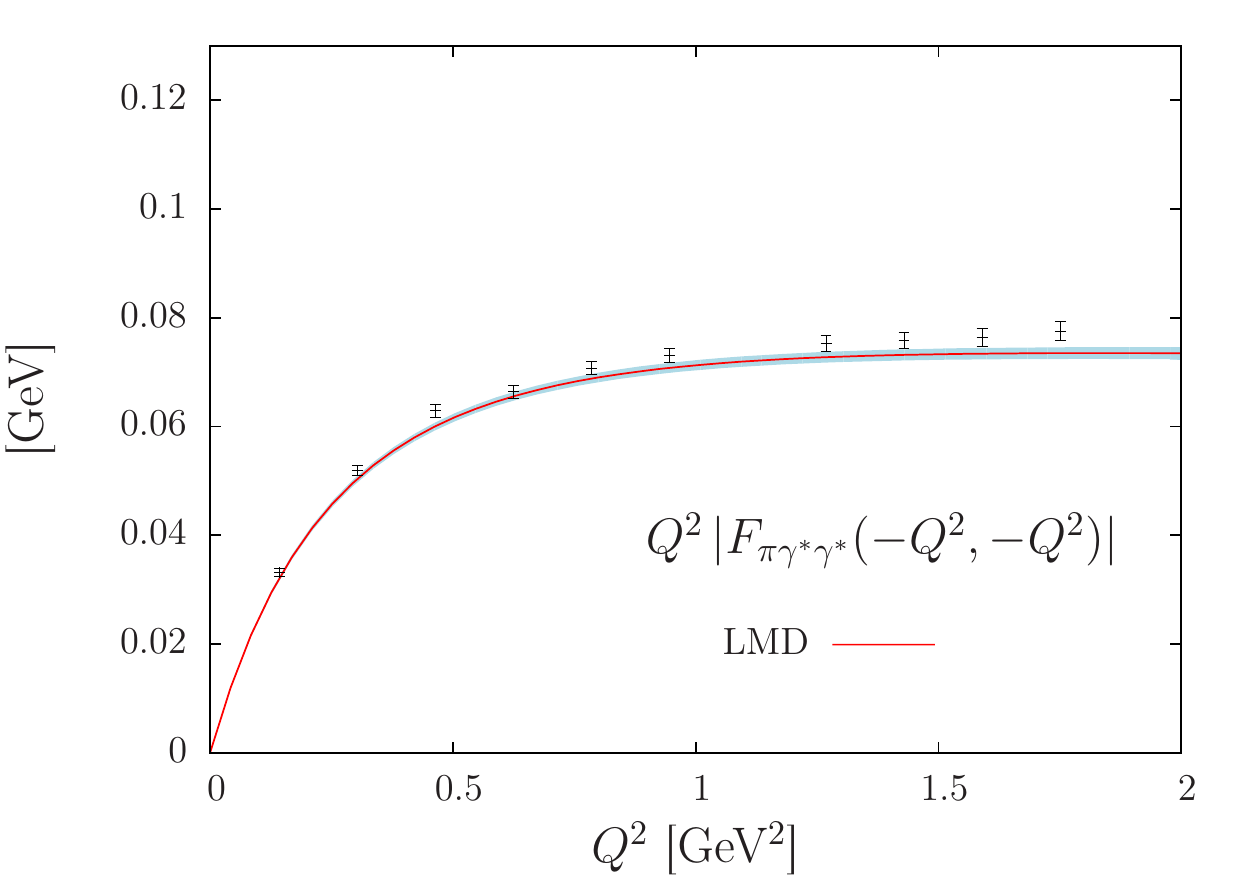}
	\end{minipage}
	\begin{minipage}[c]{0.32\linewidth}
	\centering 
	\includegraphics*[width=\linewidth]{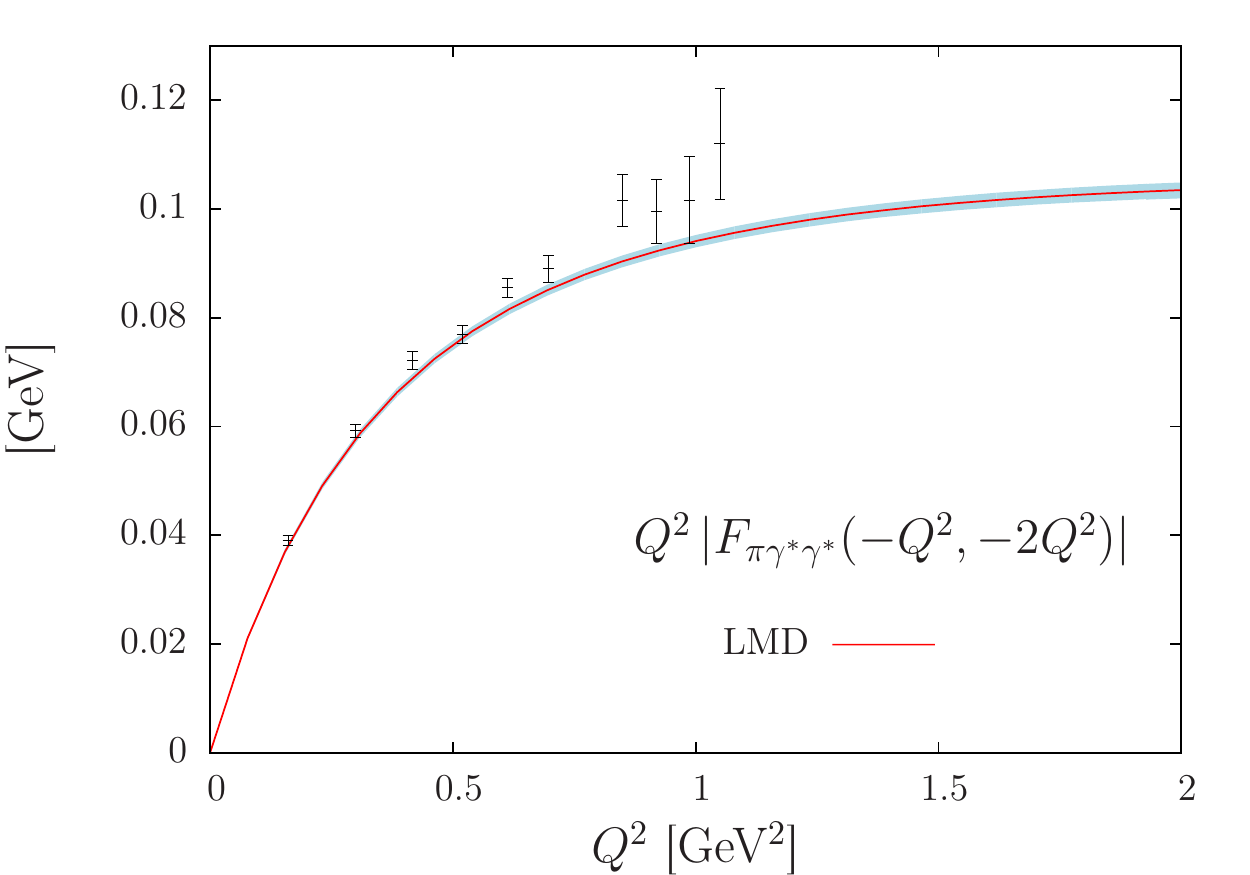}
	\end{minipage}
	\begin{minipage}[c]{0.32\linewidth}
	\centering 
	\includegraphics*[width=\linewidth]{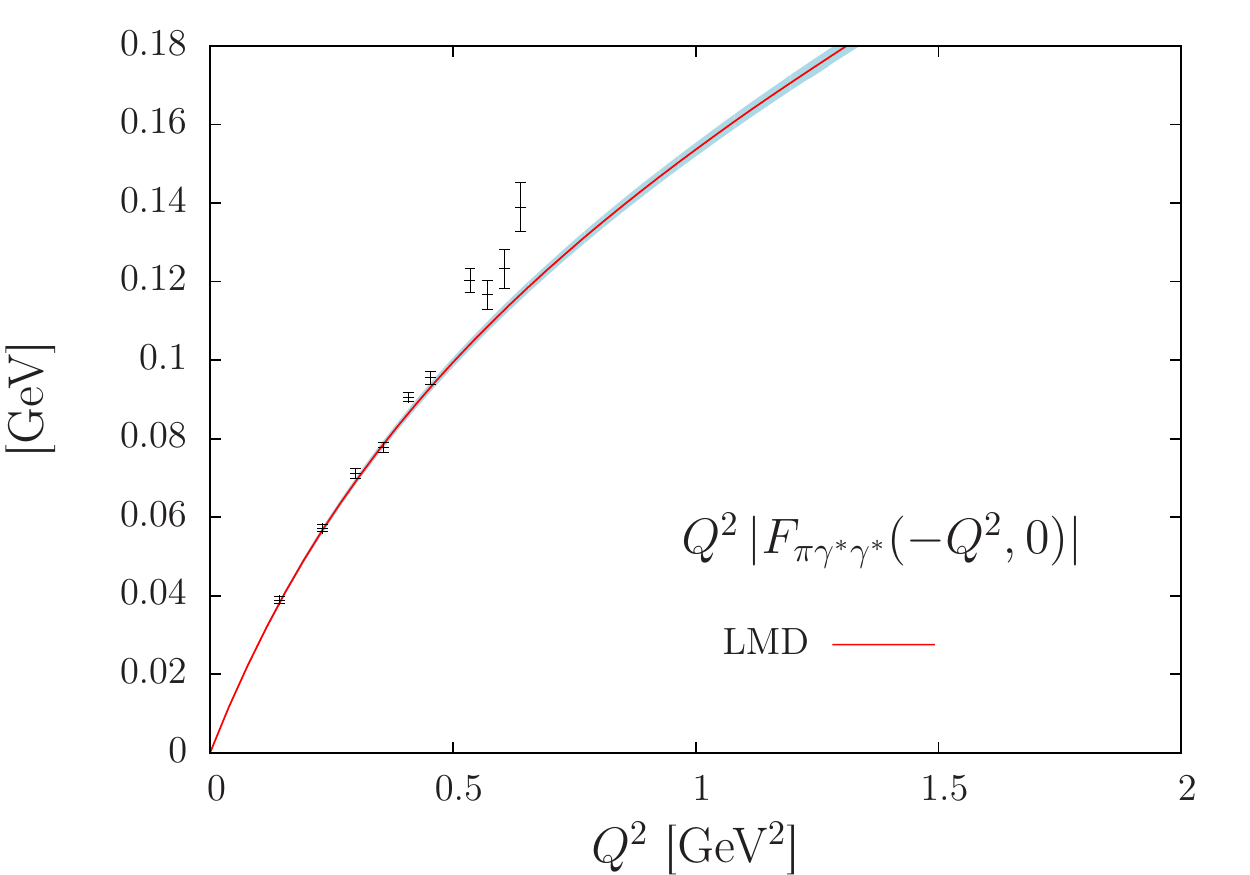}
	\end{minipage}
	
	\begin{minipage}[c]{0.32\linewidth}
	\centering 
	\includegraphics*[width=\linewidth]{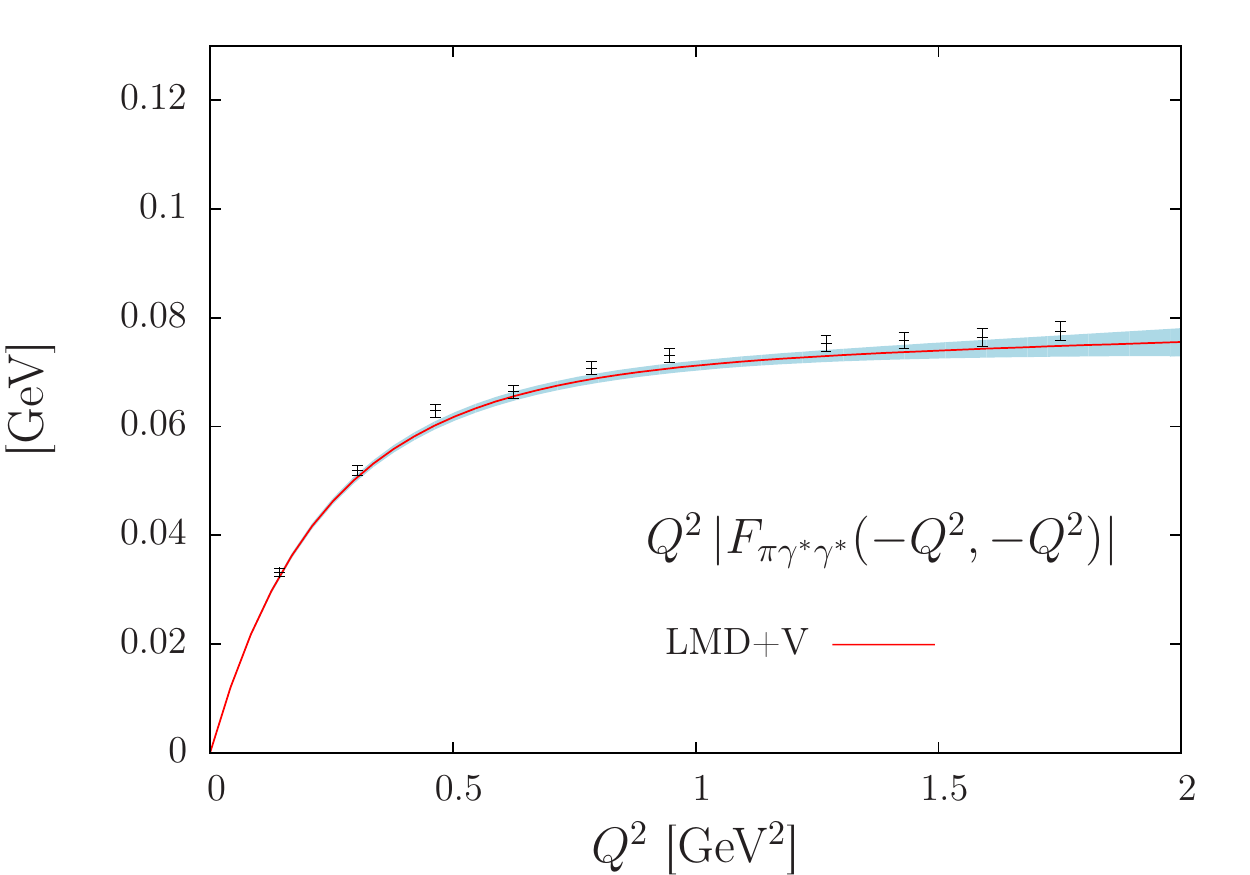}
	\end{minipage}
	\begin{minipage}[c]{0.32\linewidth}
	\centering 
	\includegraphics*[width=\linewidth]{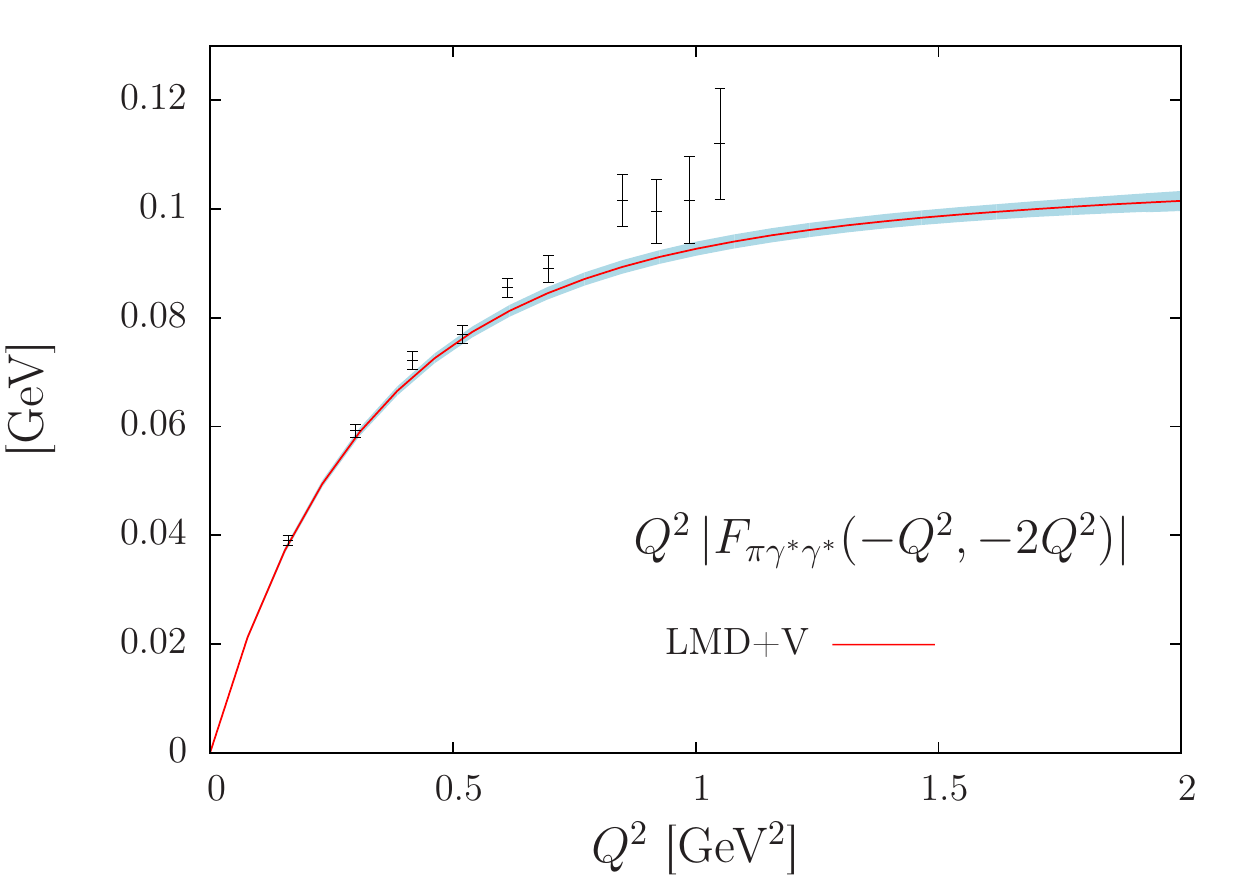}
	\end{minipage}
	\begin{minipage}[c]{0.32\linewidth}
	\centering 
	\includegraphics*[width=\linewidth]{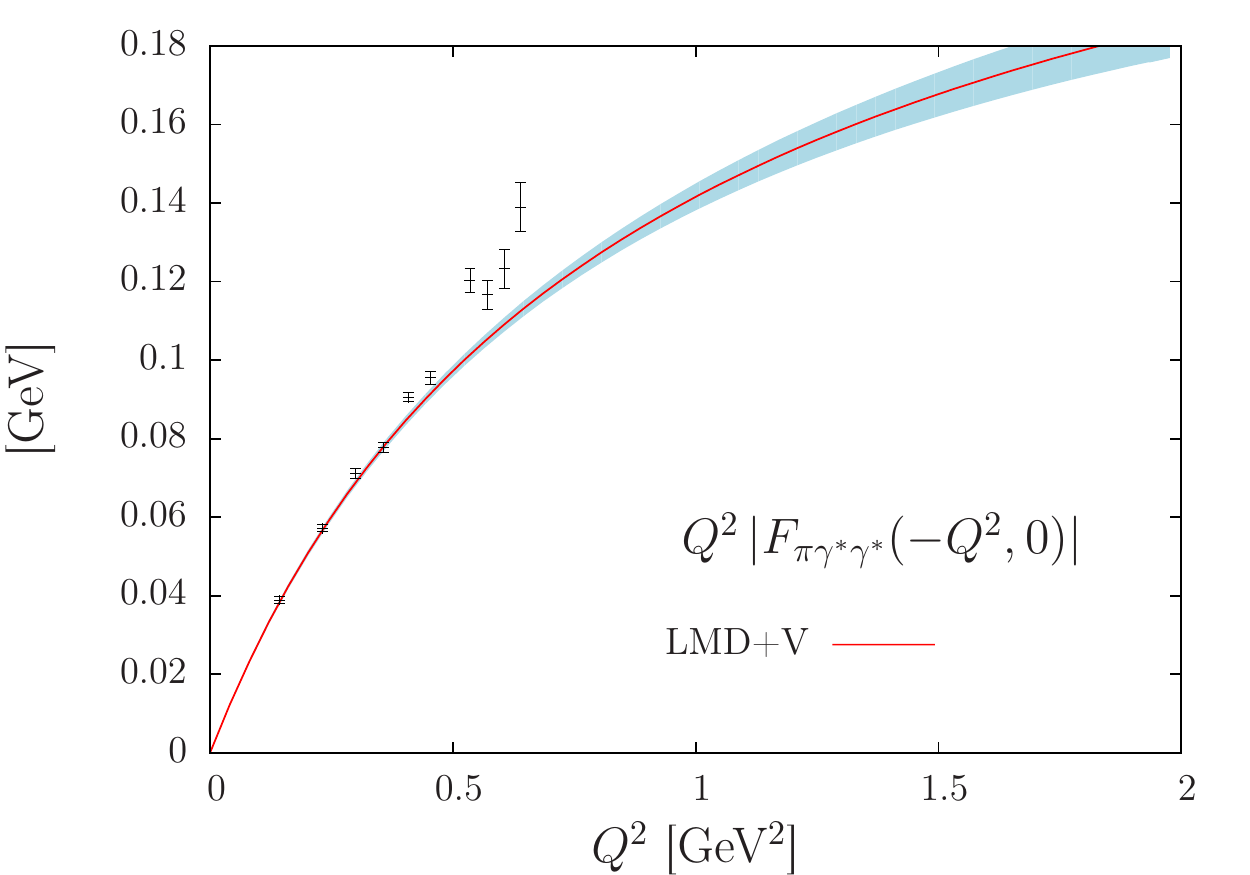}
	\end{minipage}
	
	\caption{Comparison of the VMD (top panel), LMD (mid panel) and LMD+V (bottom panel) fits for the lattice ensemble O7. The red line corresponds to the results using the global fit (method 2) corresponding to (\ref{eq:resVMD}), (\ref{eq:resLMD}), (\ref{eq:resLMDV}) where the parameters are given in Table~\ref{tab:glb_fit}. The VMD model falls off as $\FF(-Q^2, -Q^2) \sim 1/Q^4$ in the double virtual case and fails to describe the numerical data. Note that the points at different $Q^2$ are correlated.}	
	\label{fig:fit_ff}
\end{figure}

%
%
We first compare our data with the VMD model. Two fitting procedures have been used. In the first method, each lattice ensemble is fitted independently using Eq.~(\ref{eq:VMD_model}) with $\alpha$ and $M_V$ treated as free parameters. Then, in a second step, the two parameters are extrapolated to the chiral and continuum limit assuming a linear dependence in both the lattice spacing $a/a_{\beta=5.3}$ and $\widetilde{y} = m^2_{\pi} / 8 \pi^2 F^2_{\pi}$. The results are summarized in Table~\ref{tab:loc_fit} (Appendix~\ref{app:tables}). In the second fitting procedure, a global fit is performed where all lattice ensembles are fitted simultaneously assuming a linear dependence in both $a/a_{\beta=5.3}$ and $\widetilde{y}$ for each parameter of the model. In this case, we are left with only six fit parameters and the results are given in Table~\ref{tab:glb_fit} (Appendix~\ref{app:tables}). Both methods give similar results and choosing the second method, with a reduced number of fit parameters, as our preferred estimate, we obtain at the physical point
\begin{equation}
\alpha^{\VMD} = 0.243(18)~\GeV^{-1} \,, \quad M_V^{\VMD} = 0.944(34)~\GeV \,, 
\label{eq:resVMD}
\end{equation}
where the covariance matrix is (in appropriate units of GeV)
\begin{equation}
\sigma_{ij}^{\VMD}(\alpha, M_V) = \begin{pmatrix}
	+3.16 \times 10^{-4} 	&	-3.62 \times 10^{-4}	 \\
	-3.62  \times 10^{-4}	&	+1.14 \times 10^{-3}
	\end{pmatrix} \,.
\label{eq:covVMD}
\end{equation}
The covariance matrix is estimated from a jackknife procedure and used in Sec.~\ref{sec:pheno} for error propagation, but the fits are uncorrelated fits. As can be seen in Fig.~\ref{fig:fit_ff} (top panel), the VMD model leads to a poor description of our data ($\chi^2/\dof = 2.94$), especially in the double virtual case and at large Euclidean momenta. It is a direct evidence that the wrong asymptotic behavior of this model, compared to the OPE prediction in Eq.~(\ref{eq:OPE}), already matters at Euclidean momenta of order $Q^2\sim 1~\GeV^2$. In particular we do not recover the anomaly result in the chiral and continuum limit. However, fitting our data with the constraint $|Q_i^2|< 0.5~\GeV^2$ ($i=1,2$) leads to $\alpha = 0.268(21)~\GeV^{-1}$ and $M_V=0.870(45)~\GeV$ where $\alpha$ is now compatible with the theoretical prediction $\alpha_{\rm th} = 0.274~\GeV^{-1}$. Also, in the latter case we get a much better chi-squared $\chi^2 / \dof=1.29$. It confirms that the VMD model is unable to describe our data in the whole kinematical range studied here.

%
%
We have repeated the same analysis for the LMD model~(\ref{eq:LMD_model}) using $\alpha$, $\beta$ and $M_V$ as free parameters and the results are summarized in Tables~\ref{tab:loc_fit} and \ref{tab:glb_fit} (Appendix~\ref{app:tables}). The first fitting procedure suggests that lattice artifacts for the vector mass $M_V$ and chiral corrections for the parameter $\beta$ are both small. They are therefore neglected in the global fit, reducing further the number of fit parameters. In this case the global fit leads to a good description of our data, in the whole kinematical range, with $\chi^2 / \dof = 1.30$ (mid panel in Fig.~\ref{fig:fit_ff}). The results at the physical point read
\begin{equation}
\alpha^{\LMD} = 0.275(18)~\GeV^{-1} \,, \quad \beta = -0.028(4)~\GeV \,, \quad M_V^{\LMD} = 0.705(24)~\GeV \,,
\label{eq:resLMD}
\end{equation}
where the covariance matrix (in appropriate units of GeV) is 
\begin{equation}
\sigma_{ij}^{\LMD}(\alpha, \beta, M_V) = \begin{pmatrix}
	+3.33 \times 10^{-4} 	&	+2.13 \times 10^{-5}	&	-5.01 \times 10^{-5}	 \\
	+2.13 \times 10^{-5}	&	+1.77 \times 10^{-5}	&	+5.15 \times 10^{-6}	\\
	-5.01 \times 10^{-5}	&	+5.15 \times 10^{-6}	&	+5.68 \times 10^{-4}
	\end{pmatrix} \,.
\label{eq:covLMD}
\end{equation}
In particular, the anomaly constraint is recovered with a statistical error of 7\% and $\beta$ is in good agreement with the OPE asymptotic result given in Eq.~(\ref{eq:OPE}). This might be surprising as the LMD model fails to reproduce the Brodsky-Lepage behavior. However, as can be seen in Fig.~\ref{fig:kin}, all our data points in the single virtual case lie below $Q^2 \approx 0.5~\GeV^2$ and we are not probing the asymptotic behavior of the single-virtual form factor where the model is expected to fail. 

%
%
Finally, we consider the LMD+V model (\ref{eq:LMDV_model}) with $\widetilde{h}_1 = 0$ which fulfills all the theoretical constraints discussed in Sec.~\ref{sec:pion_ff}. In this case, there are too many parameters to make fits for individual ensembles with all the model parameters and including $\mathcal{O}(a)$ and chiral corrections. Therefore we perform only a global fit (Method 2, Table~\ref{tab:glb_fit}) and even there fix some of the parameters from theory or the masses from the PDG (Particle Data Group). 
In particular, we use the constraint $M_{V_1} = m_{\rho}^{\exp}$ at the physical point where $m_{\rho}^{\exp} = 0.775~\GeV$ is the experimental $\rho$ mass but still allowing for chiral corrections on each lattice ensemble. For the second vector mass $M_{V_2}$, inspired by quark models, we assume a constant shift in the spectrum and set $M_{V_2}(\widetilde{y}) = m^{\exp}_{\rho^{\prime}} + M_{V_1}(\widetilde{y}) - m^{\exp}_{\rho}$ with $m^{\exp}_{\rho^{\prime}} = 1.465~\GeV$. Finally, since we do not have data above $Q^2 \approx 1.5~\GeV^2$, we are not sensitive to the asymptotic behavior of the double-virtual form factor. We therefore impose the theoretical constraint $\widetilde{h}_0 = -F_{\pi}/3$ in the continuum and chiral limit. Using these assumptions, the LMD+V fit leads to 
\begin{equation}
\alpha^{\LMDV} = 0.273(24)~\GeV^{-1} \,, \quad \widetilde{h}_2 = 0.345(167)~\GeV^3 \,, \quad \widetilde{h}_5 = -0.195(70)~\GeV \,,
\label{eq:resLMDV}
\end{equation}
with $\chi^2 / \dof = 1.36$ and where the covariance matrix (in appropriate units of GeV) is
\begin{equation}
\sigma_{ij}^{\LMDV}(\alpha, \widetilde{h}_2, \widetilde{h}_5) = \begin{pmatrix}
	+5.59 \times 10^{-4}	&	+1.71 \times 10^{-3}	&	+8.04 \times 10^{-4} \\
	+1.71 \times 10^{-3}	&	+2.80 \times 10^{-2} 	&	+9.88 \times 10^{-3}	\\
	+8.04 \times 10^{-4}	&	+9.88 \times 10^{-3}	&	+4.87 \times 10^{-3}
	\end{pmatrix} \,.
\label{eq:covLMDV}
\end{equation}

\noindent This model also gives a good description of our data as can be seen in the bottom panel of Fig.~\ref{fig:fit_ff}. The details of the fit  are summarized in Table~\ref{tab:glb_fit} (Appendix~\ref{app:tables}) where a cross indicates that the parameter is not fitted but set to zero and where numbers quoted without error are fixed to a constant. Again, the anomaly constraint is recovered within statistical error bars and the values of $\widetilde{h}_2$ and $\widetilde{h}_5$ are compared to phenomenology in Sec.~\ref{sec:pheno}. To test the dependence of our results on our assumptions on $M_{V_1}$ and $M_{V_2}$, we have performed two more fits. In the first one, the first vector mass $M_{V_1} = 0.705~\GeV$ is set to its preferred LMD value obtained in the previous fit in Eq.~(\ref{eq:resLMD}) instead of its physical value (corresponding roughly to a shift of $10\%$). The results
\begin{equation}
\alpha^{\LMDV} = 0.277(24)~\GeV^{-1} \,, \quad \widetilde{h}_2 = 0.329(148)~\GeV^3 \,, \quad \widetilde{h}_5 = -0.222(65)~\GeV \,,
\label{eq:resLMDV1}
\end{equation}
%
%
are rather stable and differ at most by 40\% of the statistical error. Then, in the second fit, the first vector mass $M_{V_1}=m^{\exp}_{\rho}$ is set to its experimental value again but instead of a constant shift in the spectrum, we set $M_{V_2}(\widetilde{y}) = m_{\rho^{\prime}}^{\exp}$ to a constant for all lattice ensembles. Again, the results
\begin{equation}
\alpha^{\LMDV} = 0.269(24)~\GeV^{-1} \,, \quad \widetilde{h}_2 = 0.288(133)~\GeV^3 \,, \quad \widetilde{h}_5 = -0.214(65)~\GeV \,,
\label{eq:resLMDV2}
\end{equation}
%
%
do not change significantly within our statistical error bars.\\

\begin{figure}[t]
	\begin{minipage}[c]{0.49\linewidth}
	\centering 
	\includegraphics*[width=0.95\linewidth]{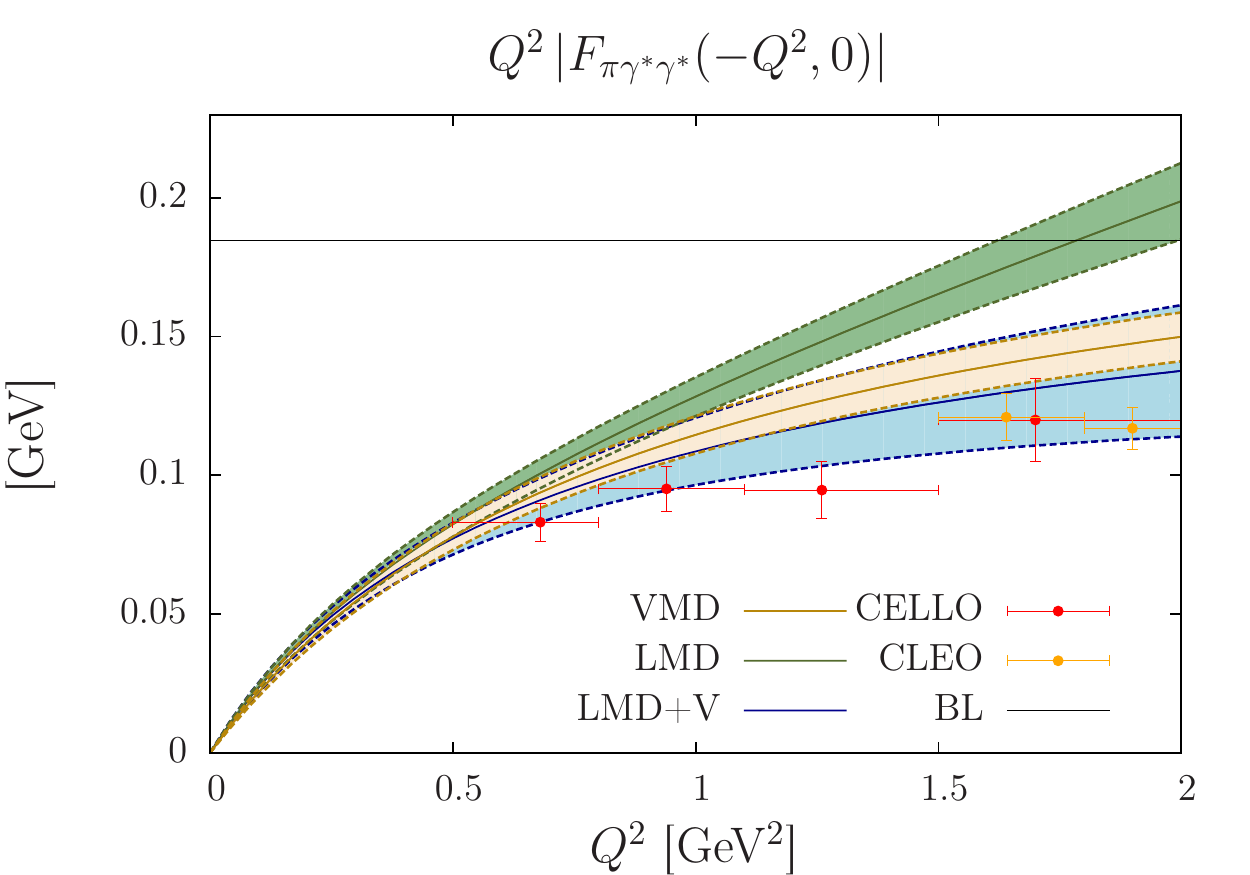}
	\end{minipage}
	\begin{minipage}[c]{0.49\linewidth}
	\centering 
	\includegraphics*[width=0.95\linewidth]{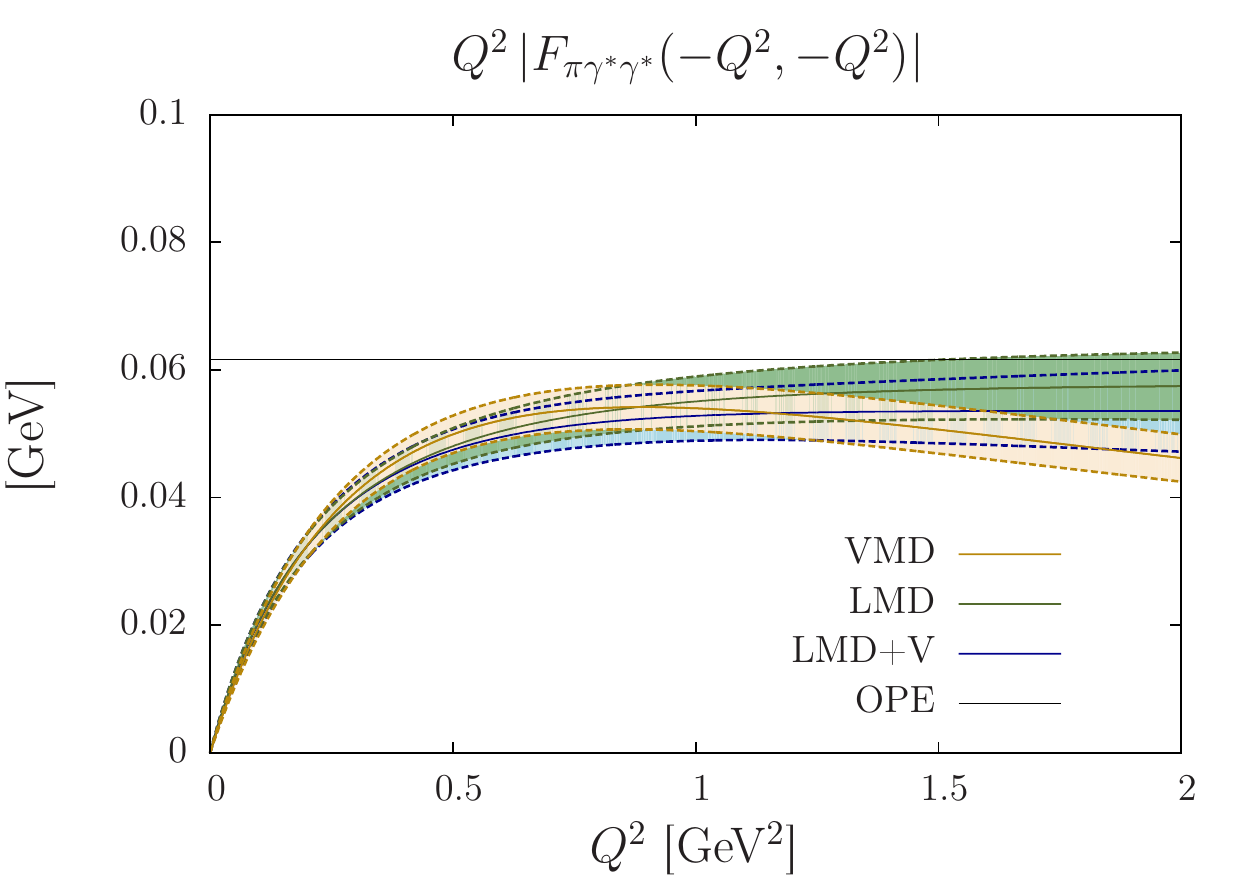}
	\end{minipage}
	
	\caption{Lattice extrapolations for the VMD, LMD and LMD+V models corresponding to the fit parameters in Table~\ref{tab:glb_fit}. Left: Single-virtual form factor. The horizontal black line corresponds to the prediction from Brodsky-Lepage in Eq.~(\ref{eq:BL}). Experimental results from CELLO and CLEO are also depicted. Right: Double-virtual form factor at $Q_1^2=Q_2^2$, the horizontal black line corresponds to the OPE prediction given by Eq.~(\ref{eq:OPE}).}
	\label{fig:FF_cmp}
\end{figure}

Finally, the form factor extrapolated at the physical point is shown in Fig.~\ref{fig:FF_cmp} for the three models considered here. We also show the theoretical predictions (horizontal lines) for the asymptotic behaviors of the form factor and the experimental results available in the single-virtual case. In the single-virtual case, the VMD and LMD+V models agree with each other and are in good agreement with experimental data. The LMD model, which has the wrong asymptotic behavior, starts to deviate from the LMD+V result at $Q^2 > 1~\GeV^2$. In the double-virtual case with $Q_1^2=Q_2^2$, the form factor for the LMD and LMD+V model is already close to its asymptotic behavior at $Q^2 \sim 1.5~\GeV^2$ where we have lattice data.

\subsubsection{Fit in the time-momentum representation \label{subsec:fit_res2}}

In the previous section, the form factor was first computed using Eqs.~(\ref{eq:form_factor}) and (\ref{eq:lat_M}) and then compared to some phenomenological models. However, to test the validity of a particular model, one can directly fit the function $\widetilde{A}(\tau)$ given in Eq.~(\ref{eq:Amunu}) in the time-momentum representation. One advantage of this method is that it becomes unnecessary to model the tail of the function $\widetilde{A}(\tau)$ to perform the integration up to $\pm \infty$ in Eq.~(\ref{eq:lat_M}) where we have no lattice data. Moreover, this method could benefit from $\mathcal{O}(a)$-improvement if it would be fully implemented. However, it is then more difficult to compare lattice data with phenomenology where one is eventually interested in the form factor.
In the case of the LMD model, the expression of $\widetilde{A}^{\LMD}(\tau)$ is given in Eq.~(\ref{eq:Apos}) of Appendix~\ref{app:A}. As for the four-momentum analysis in the previous section, we have performed both a local (method 1) and a global (method 2) fit and the results are summarized in Tables~\ref{tab:loc_fit_TMR} and \ref{tab:glb_fit_TMR} of Appendix~\ref{app:tables2}. One expects the small $\tau$ region to be more affected by lattice artifacts. Therefore, we tried two fits by excluding data points with $\tau/a<\tau_{\rm min}/a=2,3$. The results at $\tau/a=3$ are $\alpha^{\LMD} = 0.297(17)~\GeV^{-1}$, $\beta = 0.025(4)~\GeV$ and $M_V^{\LMD} = 0.682(20)~\GeV$ and are compatible with fits in the four-momentum representation within statistical error bars. This is a hint that the part of the tail of $A(\tau)$ for $\tau > \tau_c$, which is estimated using a VMD fit, is not relevant in our calculation. We will come back to this issue in Sec.~\ref{subsubsec:FTE}.

\subsection{Systematic errors \label{sec:syst_err}}

\subsubsection{Sampling \label{subsubsec:Sampling}}

We have performed a second analysis using a different sampling of our data. Instead of using data points regularly distributed along each curve in the ($q_1^2,q_2^2)$ plane, we select points using a constant step in $\omega_1$ in Eq.~(\ref{eq:kin}).  More details and fit results are given in Appendix~\ref{app:sampling} and Table~\ref{tab:glb_fit_s2}. An illustration of the two samplings is given in Fig.~\ref{fig:sampling} (Appendix \ref{app:tables}). For the VMD model, with the worst $\chi^2$, the results differ by $4\%$ for the anomaly and $25~\MeV$ for the vector mass $M_V$. For the LMD model, the result for the anomaly is stable and differs by less than $1\%$ at the physical point while the vector mass $M_V$ varies by about $10~\MeV$. Finally, for the LMD+V model, the anomaly is again stable ($2\%$) and we observe a difference of $0.050~\GeV^3$ for $\widetilde{h}_2$ and $0.027~\GeV$ for $\widetilde{h}_5$.

\subsubsection{Finite-time extent \label{subsubsec:FTE}}

To perform the integration in Eq.~(\ref{eq:lat_M}), a cutoff $\tau_c$ has been introduced and for $\tau > \tau_c$ the integrand $A(\tau)$ is obtained from a fit to the data using a VMD \textit{Ansatz} as explained in Sec.~\ref{subsec:fte}. In particular, due to the exponential factor in Eq.~(\ref{eq:lat_M}), the large $\tau$ region contributes more in the single-virtual case. However, we have checked that even in the less favorable case the contribution from the fitted tail is less than 20\% of the total contribution. To further investigate this issue we have fitted the tail using the LMD \textit{Ansatz} rather than the VMD \textit{Ansatz}. The fit parameters are collected in Table~\ref{tab:glb_fit_fte} (Appendix~\ref{app:tables}) and the results at the physical point do not change within statistical error bars. The results differ by less than $1\%$ for the anomaly at the physical point in all cases. In the LMD case the vector mass differs by about $20~\MeV$ and for the LMD+V model, the parameters $\widetilde{h}_{2,5}$ are rather stable within the large error bars (we observe a deviation of $0.066~\GeV^3$ and $0.021~\GeV$ respectively). 

\subsubsection{Excited pseudoscalar states}

\renewcommand{\arraystretch}{1.1}
\begin{table}[t]
\caption{Study of the excited state contamination. We collect our fit parameters for the ensembles E5 and F6 for different values of $t=t_f-t_0$. The data are fitted using a LMD fit (method 1).}
\begin{center}
	\begin{tabular}{l@{\hskip 0.1in}S[table-format=0.5] cc @{\hskip 0.2in} S[table-format=0.5] cc}
	\hline 
		&	\multicolumn{3}{c}{E5} & 	\multicolumn{3}{c}{F6}			 \\
	\cline{2-4} \cline{5-7}
		$t$	&	$\alpha~[\GeV^{-1}]$ & $\beta~[\MeV]$	& 	$M_V~[\MeV]$	&	$\alpha~[\GeV^{-1}]$ & $\beta~[\MeV]$	& 	$M_V~[\MeV]$		\\ 
	\hline 
	15		&	0.282(6) 	&	-34(2)	& 	945(18)	&	$\times$	&	 $\times$	&	$\times$	 \\
	17		&	0.290(5) 	&	-36(1)	& 	925(17)	&	$\times$	&	 $\times$	&	$\times$	 \\
	19 		& 	0.291(6)	&	-36(2)	& 	925(23)	&	0.326(10)	& 	-34(3)	& 	801(24)		\\
	21		&	0.290(9)	&	-35(2)	&	920(29) 	&	0.337(9)	& 	-35(3)	&	795(21) 	\\ 
	23		&	0.288(9)	&	-35(3)	&	926(30) 	&	0.329(10)	& 	-36(3)	&	787(27)	\\ 
	25		&	0.277(14)	&	-33(4)	&	959(49) 	&	0.331(10)	& 	-36(3)	&	783(28)	\\ 
	\hline 
	
	\end{tabular}
\label{tab:excited_states}
\end{center}
\end{table}

For the lattice ensembles E5 and F6, the form factor has been computed for different values of $t=t_f-t_0$ in the range $[1.0-1.65]~\fm$. The values of the LMD fit parameters (using the fit method 1, as explained in Sec.~\ref{subsec:fit_res}) are summarized in Table~\ref{tab:excited_states}. The results do not depend on $t \geq 17$ within our statistical error bars which make us confident that the excited states contribution in the pseudoscalar channel can be neglected at our level of precision.

\subsubsection{Disconnected contribution \label{subsub:disc}}

\begin{figure}[t!]

	\begin{minipage}[c]{0.33\linewidth}
	\centering 
	\includegraphics*[width=0.99\linewidth]{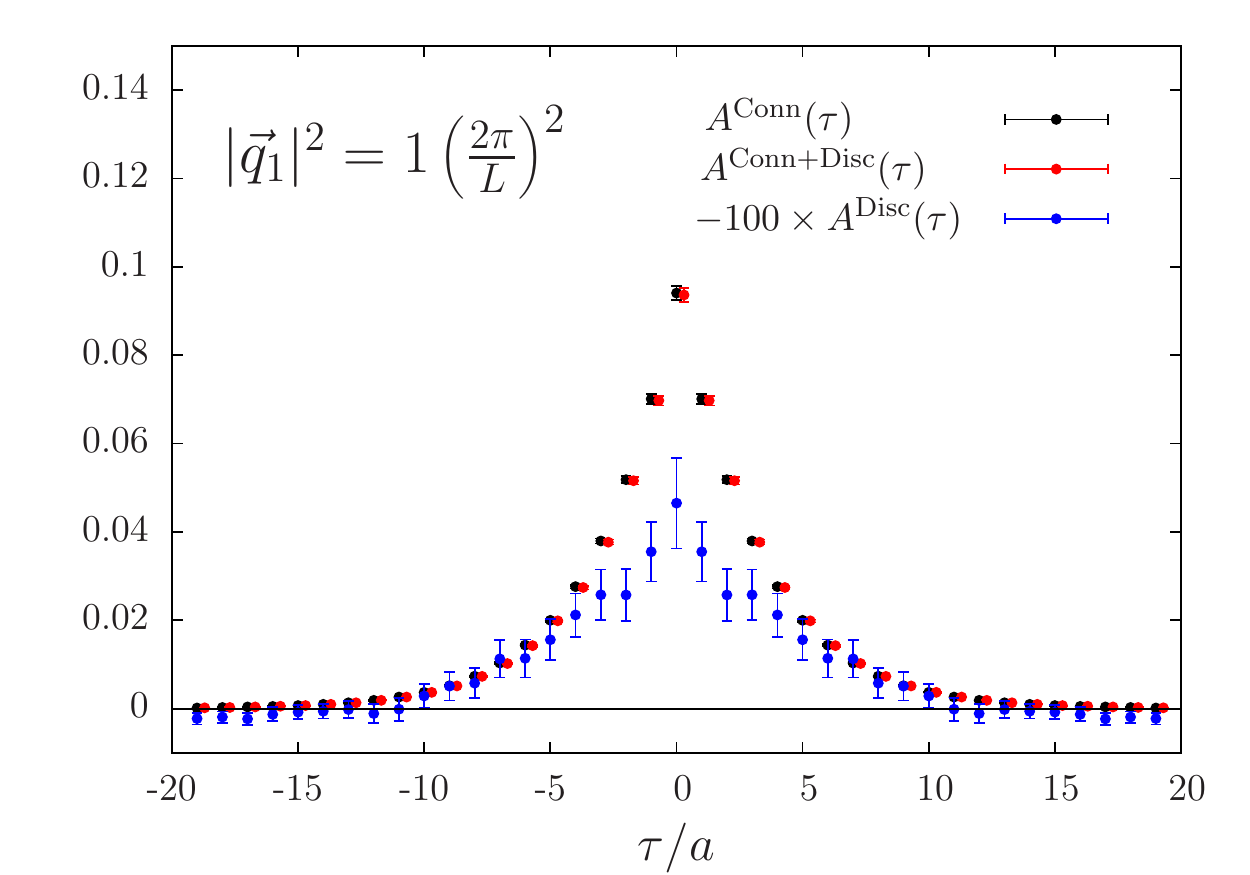}
	\end{minipage}
	\begin{minipage}[c]{0.32\linewidth}
	\centering 
	\includegraphics*[width=0.99\linewidth]{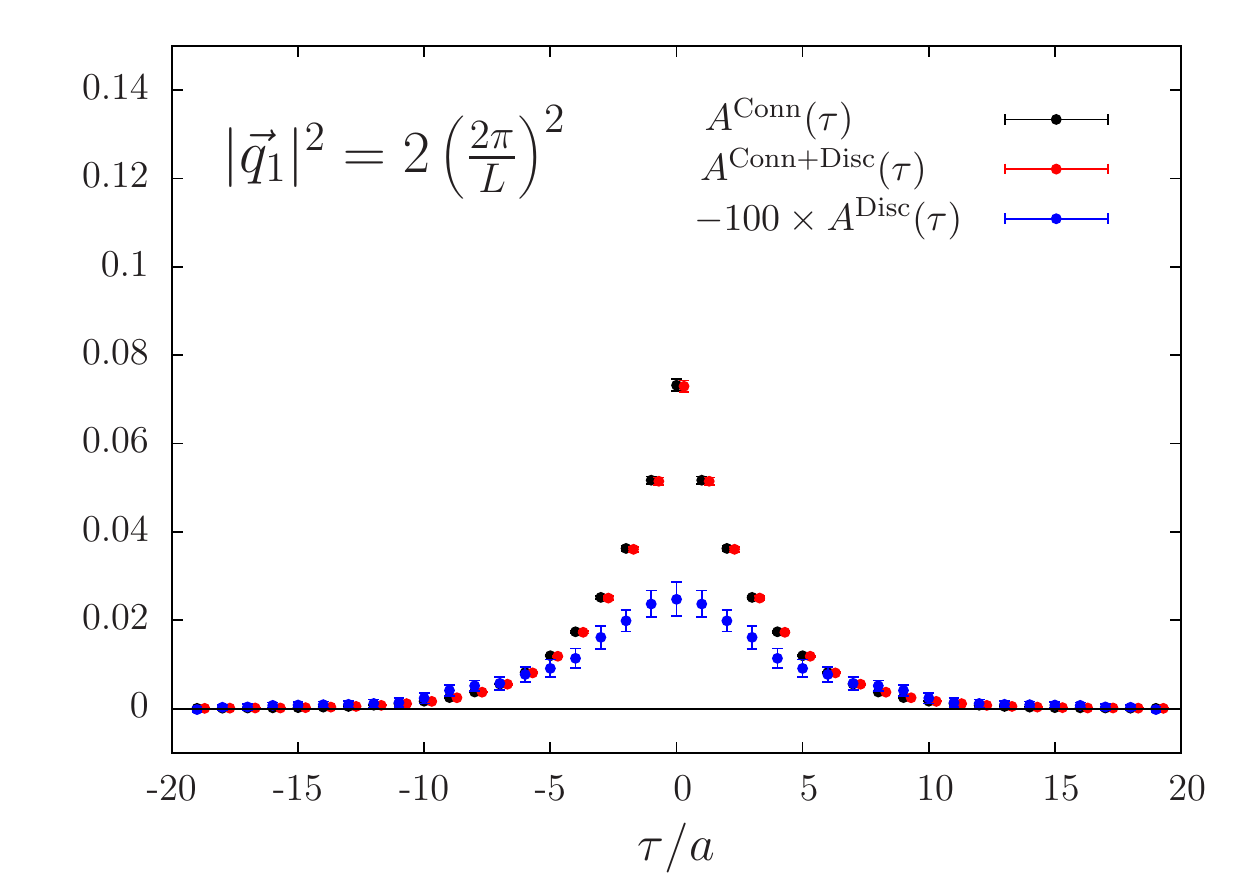}
	\end{minipage}
	\begin{minipage}[c]{0.33\linewidth}
	\centering 
	\includegraphics*[width=0.99\linewidth]{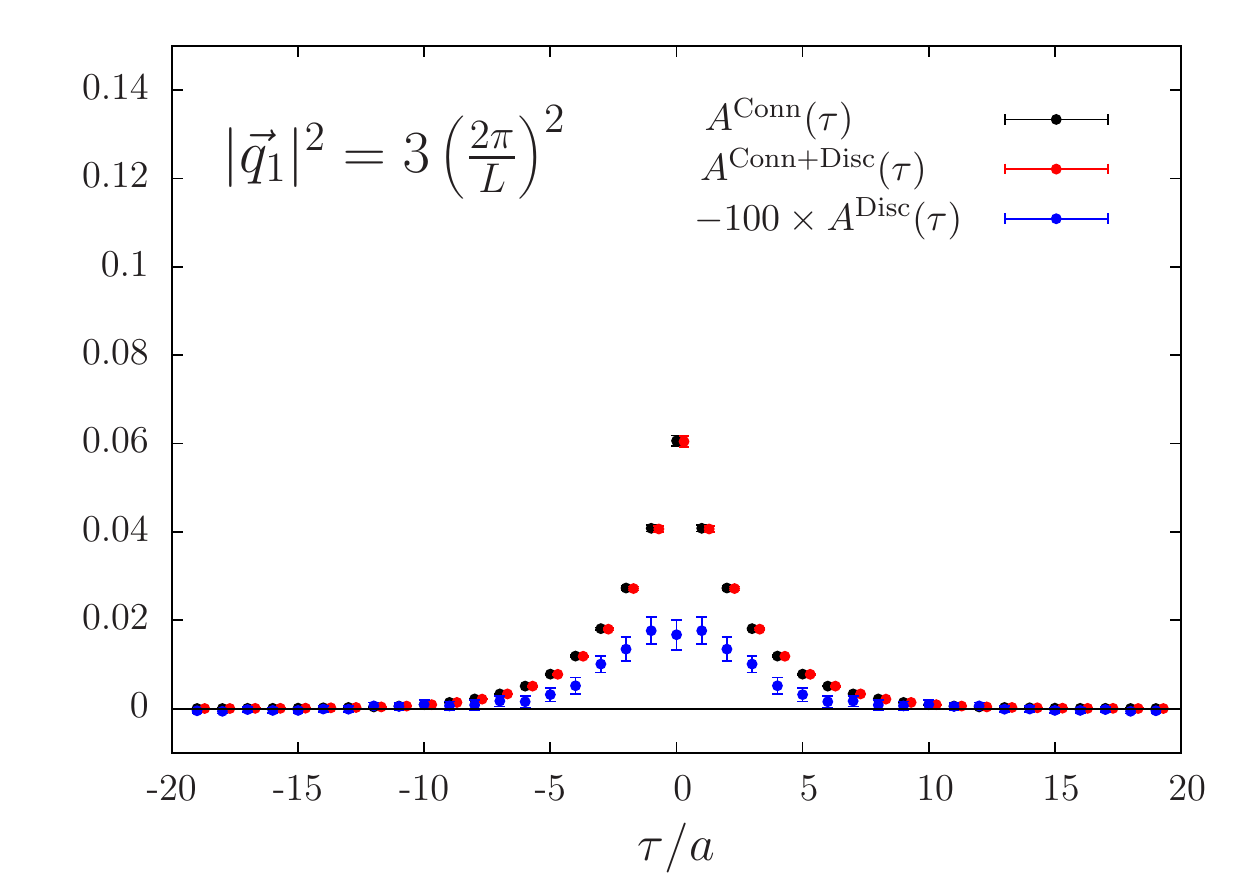}
	\end{minipage}
	
	\caption{Disconnected contribution to the function $A(\tau)$ for the lattice ensemble E5 and the first three values of the discretized spatial lattice momenta. Black points correspond to the connected contribution, red points to the total contribution including disconnected diagrams and blue points correspond to the disconnected contribution only multiplied by a factor -100.}	
	\label{fig:disc_contrib}
\end{figure}

In the previous results, only the connected contribution given in Eq.~(\ref{eq:connected_contrib}) has been considered and the disconnected contributions, given in Eqs.~(\ref{eq:C3disc1}) and (\ref{eq:C3disc2}), were neglected. Those contributions are much more difficult to evaluate numerically because of their poor signal-to-noise ratio. In this study, the calculations of the disconnected contribution have been performed on one lattice ensemble E5 and only for the first three values of the spatial momentum $|\vec{q}_1|^2 = \vec{n}^2 (2\pi/L)^2$, $\vec{n}^2=1,2,3$. 
The loops (Fig.~\ref{fig:disconnected}) were computed using 75 stochastic sources with full-time dilution ($75\times T$ inversions of the Dirac operator) and a generalized Hopping Parameter Expansion to sixth order~\cite{Gulpers:2013uca,Francis:2014hoa}. For the two-point correlation functions, we used seven stochastic sources with full-time dilution and stored the results for all possible values of the time source and time sink locations. Also, in this case, we used a larger set of gauge configurations compared to the connected part ($\#1000$).
The results are depicted in Fig~\ref{fig:disc_contrib} where we compare the disconnected to the connected contribution. The disconnected contribution is below $1\%$ of the total contribution and does not show any clear dependence on the value of the spatial momentum $|\vec{q}_1^{\ 2}|$. We conclude that the disconnected contribution is negligible at our level of accuracy, even though its size could be quite strongly pion mass dependant.

\subsubsection{Finite-size effects}

A potentially significant source of systematic error are the
finite-size effects.  Indeed, in the correlator
$C^{(3)}_{ij}(\tau,t_\pi)$, the states dominating at large separation
$\tau$ between the two vector currents are not one-particle states,
but rather $\pi^+ \pi^-$ states. Since their spectrum is discrete on
the torus used in our simulations, this long-distance contribution is
distorted relative to the infinite-volume correlator. For increasing
$\omega_1$, the long-distance contribution is enhanced.  

An empirical look at our data sets does not seem to indicate a major
issue in the determination, for instance of the LMD model parameters.
Unfortunately, we do not have a dedicated finite-volume study, where
the volume is the only parameter varying. For now, we may compare
ensembles at different lattice spacings. First, comparing the results
of ensembles A5 and N6, which have the same pion mass and volume, we
observe a $14\%$ discretization error on the parameter $\alpha$, and
no significant effect on $M_V$ and $\beta$.  If we then compare
ensembles B6 and O7, which have pion masses 283~MeV and 269~MeV
respectively, but different volumes (respectively $m_\pi L=5.2$ and
4.2), we observe compatible values of the parameters $M_V$ and
$\beta$, while the $\alpha$ parameters differ by the same factor as A5
and N6, which we interpret as a discretization error. Thus no major
finite-size effect on the LMD parameters is observed.

From a more theoretical perspective, we may try to predict the
magnitude of the finite-size effects.  The situation is similar to the
calculation of the hadronic vacuum polarization at low momentum
transfer via the correlator $G(x_0)=\frac{-1}{3}\int d^3x
\,\<V_k(x)V_k(0)\>$.  The finite-size effects on the latter were
analyzed in~\cite{Francis:2013qna} by using a spectral representation
of the correlator both in finite and in infinite volume, and by using
the L\"uscher formalism to relate the discrete finite-volume spectrum
and the corresponding matrix elements to the $I=\ell=1$ $\pi\pi$ phase
shift and the timelike pion form
factor~\cite{Luscher:1991cf,Meyer:2011um,Feng:2014gba}. A similar approach is
possible here, where the three-point function can be written in a
dispersive way in terms of the same timelike pion form factor and the
amplitude for the process $\pi^0\gamma^*\to \pi^+\pi^-$. The procedure
is illustrated in Appendix~\ref{app:FSE}, however we leave the
quantitative study of finite-volume effects for the future.  We note
that a similar dispersion relation was presented
in~\cite{Hoferichter:2012pm} for the transition form factor $\FF(q_1^2,q_2^2)$, and that the amplitude $\pi^0\gamma^*\to
\pi^+\pi^-$ has recently been investigated in lattice QCD for the
first time~\cite{Briceno:2016kkp}.

\subsection{Final results \label{subsec:final_res}}

The VMD model fails to describe our data in the whole kinematical range. On the contrary, the LMD model gives a good description of the lattice data in the kinematical region considered here despite its wrong asymptotic behavior in the single-virtual case. Finally, the LMD+V model has a larger number of free parameters but fulfills all the theoretical constraints, it leads to larger error bars but gives a good description to the lattice data. Therefore, we quote as our final results
\begin{equation}
\alpha^{\LMD} = 0.275(18)(3)~\GeV^{-1} \,, \quad \beta = -0.028(4)(1)~\GeV \,, \quad M_V^{\LMD} = 0.705(24)(21)~\GeV \,,
\label{eq:resLMD_final}
\end{equation}
for the LMD model and
\begin{gather}
\alpha^{\LMDV} = 0.273(24)(7)~\GeV^{-1} \,, \quad \widetilde{h}_2 = 0.345(167)(83)~\GeV^3 \,, \quad \widetilde{h}_5 = -0.195(70)(34)~\GeV \,, \\
\nonumber [ \bar h_2 = -11.2(5.4)(2.7)~\GeV^2 \,, \quad \bar h_5 = 8.2(2.9)(1.4)~\GeV^4  ]
\label{eq:resLMDV1_final}
\end{gather}
for the LMD+V model where $\widetilde{h}_0 = -F_{\pi}/3 = -0.0308~\GeV$, $\widetilde{h}_1=0$, $M_{V_1} = 0.775~\GeV$ and $M_{V_2} = 1.465~\GeV$ are fixed parameters at the physical point. The first error is statistical and the second error includes the systematics discussed in Sec.~\ref{sec:syst_err}. The systematic errors are estimated in the previous subsections and added quadratically.

\section{Phenomenology}
\label{sec:pheno}

\subsection{Comparison with experimental data \label{subsec:experimental_data}}

The normalization of the form factor at zero momentum $\FF(0,0)$ is
related to the decay width $\Gamma(\pi^0 \to \gamma\gamma)$ as shown
in Eq.~(\ref{eq:pion_decay}) and the experimental result is well
reproduced with the value from the chiral anomaly from
Eq.~(\ref{eq:ABJ}). The VMD fit from Eq.~(\ref{eq:resVMD}) does not
reproduce the anomaly at the $1\sigma$ level, while the LMD fit from
Eq.~(\ref{eq:resLMD}) and the LMD+V fit from Eq.~(\ref{eq:resLMDV}) do
so. However, the statistical precision of 9\% for LMD+V cannot compete
with the experimental precision from the PrimEx
experiment~\cite{Larin:2010kq} which translates into a 1.4\%
determination of the normalization of the form factor.

\begin{figure}[t!]
	\centering 
	\includegraphics*[width=0.65\linewidth]{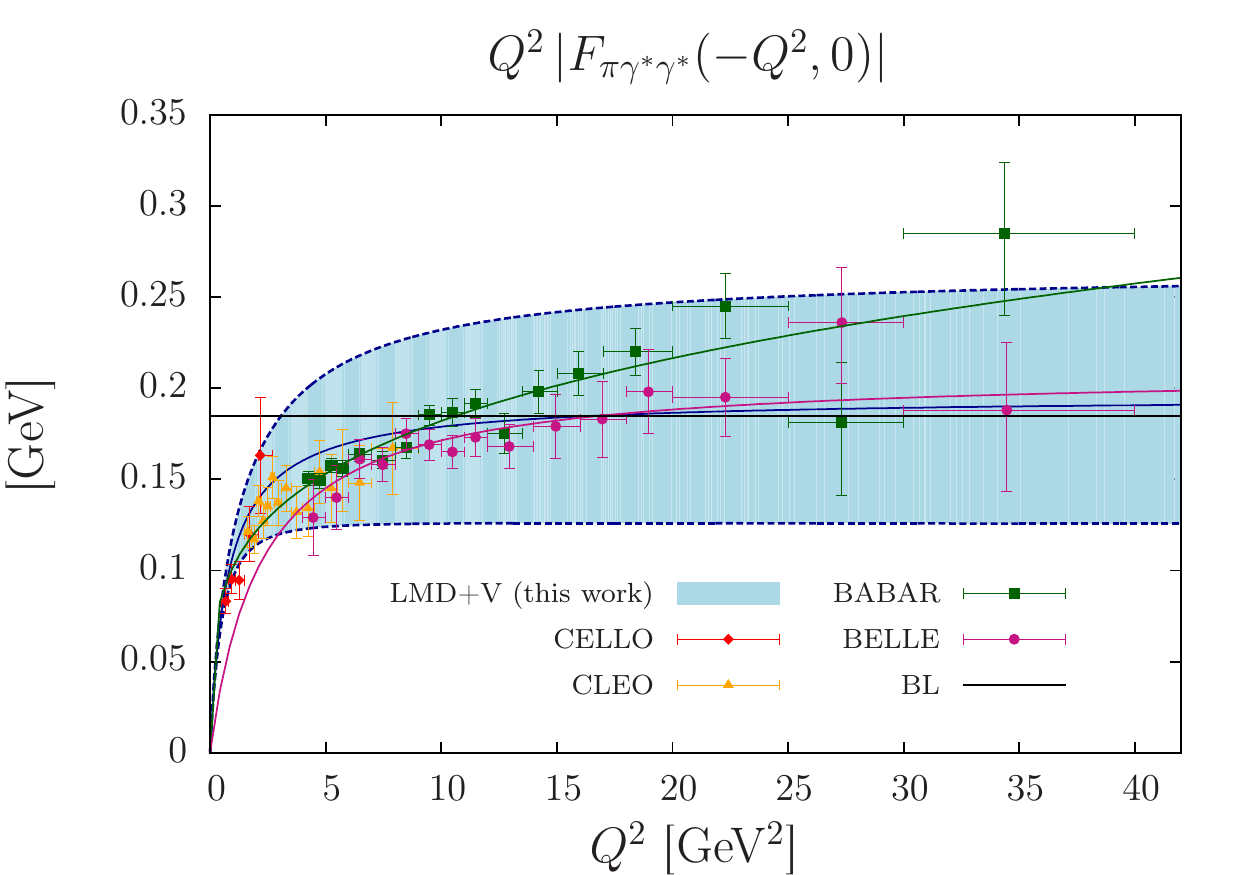}
	
	\caption{Comparison of the experimental data for the
          single-virtual form factor with the fit to the lattice data
          for the LMD+V model extrapolated to the continuum and at the
          physical pion mass. The curves for \babar{} and \belle{} show 
          the fits given in the experimental papers. The
          horizontal black line corresponds to the prediction from
          Brodsky-Lepage (BL).}
	
	\label{fig:FF_cmp_exp}
\end{figure}

For the single-virtual form factor $\FF(-Q^2,0)$ there are
experimental data available in the spacelike region from several
experiments: CELLO~\cite{Behrend:1990sr}, CLEO~\cite{Gronberg:1997fj},
\babar~\cite{Aubert:2009mc} and \belle~\cite{Uehara:2012ag}. The
experimental data in the region $[0-2]~\GeV^2$ have been plotted
already in Fig.~\ref{fig:FF_cmp} together with the fit results for
VMD, LMD and LMD+V and the full region $[0-40]~\GeV^2$, where there are
currently data available, is shown in Fig.~\ref{fig:FF_cmp_exp}
together with the LMD+V fit representing our lattice data.

An important experimental information is the slope of the form factor
at the origin. Following Ref.~\cite{Landsberg}, one
defines
\be \label{TFF_slope} 
b_{\pi^0} = \left. \frac{1}{\FF(0,0)} \frac{d \FF(q^2,0)}{dq^2}
\right|_{q^2 = 0}. 
\ee
For our three form factor models, one obtains from
Eqs.~(\ref{eq:VMD_model}), (\ref{eq:LMD_model}) and
(\ref{eq:LMDV_model}) the  
expressions  
\bea
b_{\pi^0}^{\mathrm{VMD}} & = & \frac{1}{M_V^2}, \label{eq:slope_VMD}
\\ 
b_{\pi^0}^{\mathrm{LMD}} & = & \frac{1}{M_V^2} + \frac{\beta}{\alpha
  M_V^4}, \\ 
b_{\pi^0}^{\mathrm{LMD+V}} & = & \frac{1}{M_{V_1}^2} +
\frac{1}{M_{V_2}^2} + \frac{\widetilde{h}_5}{\alpha M_{V_1}^2
  M_{V_2}^2}.  \label{slope_LMDV}
\eea

Although the normalization $\alpha$ drops out in the slope for the VMD
form factor in Eq.~(\ref{eq:slope_VMD}), the fit results from
Eq.~(\ref{eq:resVMD}) have a bad $\chi^2 / \dof$ and lead to a
suppression of $\alpha^{\mathrm{VMD}}$ and an enhanced value for
$M_V^{\mathrm{VMD}}$ compared to the fits with LMD in
Eq.~(\ref{eq:resLMD}) and LMD+V in Eq.~(\ref{eq:resLMDV}). This leads
to a distortion of the function near the origin and we will therefore
not evaluate the slope of the form factor with the VMD model. Since
the fits with the LMD and LMD+V models work fine, we calculate those
slopes from the fitted model parameters, taking into account the
correlations from Eqs.~(\ref{eq:covLMD}) and (\ref{eq:covLMDV}), to
obtain the following estimates with their statistical uncertainty
\bea 
b_{\pi^0}^{\mathrm{LMD, fit}} & = & (1.60 \pm 0.11)~\GeV^{-2} \quad
(\pm 6.8\%), \\  
b_{\pi^0}^{\mathrm{LMD+V, fit}} & = & (1.58 \pm 0.23)~\GeV^{-2} \quad
(\pm 14.3\%),  
\eea
which agree very well with each other. This is an indication that the
lattice data at low momenta are well represented by these two
fits. Note that the error for LMD+V does not include variations of the
vector meson masses which enter in the expression~(\ref{slope_LMDV}),
which we fixed to $M_{V_1} = M_\rho$ and $M_{V_2} = M_{\rho^\prime}$.

For comparison, the PDG~\cite{Agashe:2014kda} uses the determination
of the slope of the form factor by the CELLO Collaboration as their
average $b_{\pi^0}^{\mathrm{PDG}} = (1.76 \pm 0.22)~\GeV^{-2}$, with a
$12.5\%$ precision. Within the large uncertainties our numbers agree
with the PDG. The PDG error is based on the assumption that the
systematic error is of the same size as the statistical error as
stated by the CELLO Collaboration. This systematic error does not,
however, take into account a potentially large bias from the
extrapolation (modeling) of the experimental data from $Q^2 \geq
0.5~\GeV^2$ to zero~\cite{Knecht:2001xc,Masjuan_12}. The CELLO
Collaboration simply uses a VMD fit to their data and from this they
calculate the slope of the form factor at the origin.

Recently, a phenomenological determination of the slope has been
obtained in Ref.~\cite{Masjuan_12} from a sequence of Pad\'e
approximants to form factor data from CELLO, CLEO, \babar{} and
\belle{} and the normalization from PrimEx, with the result
$b_{\pi^0}^{\textrm{Pad\'e}} = (1.78 \pm 0.12)~\GeV^{-2}$ with $6.9\%$
precision. Furthermore the dispersion relation for the form
factor~\cite{DR_pion_TFF} predicts the slope with $2\%$ precision:
$b_{\pi^0}^{\textrm{DR}} = (1.69 \pm 0.03)~\GeV^{-2}$. Although we
cannot compete with the precision from the dispersive approach, our
values are fully compatible with the latter result.

The low-to-intermediate momentum region in Fig.~\ref{fig:FF_cmp} shows
that the VMD model is a bit higher than the data points but the error
band still touches most points. The LMD model clearly fails to
describe the data which only start at $0.5~\GeV^2$. The LMD+V model is
again a bit higher than the data, but the relatively large error band at
least touches the central values of the data points, except the third
lowest point from CELLO. In the full momentum region in
Fig.~\ref{fig:FF_cmp_exp} the large error band for LMD+V covers
essentially all data points above $2~\GeV^2$. Even the highest data
point of \babar{}, where the data do not show a $1/Q^2$ falloff for
the single-virtual form factor $\FF(-Q^2,0)$ (see also the fit to the
\babar{} data from the experimental paper) is within $1\sigma$ from
the LMD+V band.

Surprisingly, the central curve of the LMD+V fit has an asymptotic
value at large momenta which is rather close to the prediction from
Brodsky-Lepage in Eq.~(\ref{eq:BL}), although only lattice data below
$1.5~\GeV^2$ are fitted (even below $0.5~\GeV^2$ for the
single-virtual form factor). Of course, the large uncertainty in the
error band does not allow any firm conclusions about the asymptotic
value. Finally, the LMD+V fit result for $\widetilde{h}_5$ from
Eq.~(\ref{eq:resLMDV}) translates to ${\bar h}_5 = (8.2 \pm
2.9)~\mbox{GeV}^4$ which is consistent with the phenomenological value
from Eq.~(\ref{h5}). The latter value was obtained in
Ref.~\cite{Knecht:2001xc} by fitting the LMD+V model to the CLEO data.

Unfortunately, there are currently no experimental data available for
the double-virtual form factor $\FF(-Q_1^2, -Q_2^2)$ in the spacelike
region (nor in the timelike region, e.g.\ from double-Dalitz
decays of the pion $\pi^0 \to \gamma^* \gamma^* \to e^+ e^- e^+
e^-$). As shown in Fig.~\ref{fig:FF_cmp}, for $Q_1^2 = Q_2^2 = Q^2$,
the LMD and LMD+V model roughly agree, within their large error bands,
for momenta below $2~\GeV^2$, while the VMD model clearly falls off
too fast above $1.5~\GeV^2$. This is reflected in the bad quality of
the VMD fit. It will be interesting to compare our LMD+V fit with planned
measurements of the double-virtual form factor at BESIII in the range
$Q_{1,2}^2 \in [0.3,3]~\GeV^2$~\cite{BESIII_double_virtual} and with
results using the dispersion relation from Ref.~\cite{DR_pion_TFF},
once it has been evaluated for the double-virtual case, which should
be particularly precise at very low momenta $Q_{1,2}^2 \leq
0.5~\GeV^2$.

It is, however, reassuring that the LMD+V fit yields a value for the
parameter $\widetilde{h}_2$ in Eq.~(\ref{eq:resLMDV}) which
corresponds to ${\bar h}_2 = (-11.2 \pm 5.4)~\mbox{GeV}^2$ which is
again in agreement with the theoretically preferred value from
Eq.~(\ref{h2}). That prediction is obtained from higher-twist
corrections in the OPE; see Eq.~(\ref{eq:OPE}). As stressed in
Ref.~\cite{HLbL_PS_vs_lepton_pair_decay_bivariate}, such a negative
value for ${\bar h}_2$ leads, however, to tensions when one tries to
simultaneously explain the radiative decay $\pi^0 \to e^+ e^-$ with
the LMD+V model or some generalization of it using bivariate
approximants. But there are also issues with radiative corrections to
extract the decay rate from the measured data; see
Ref.~\cite{pi0_lepton_pair_rad_corr}.  The connection between the
pseudoscalar decay into a lepton pair and the pseudoscalar-pole
contribution to HLbL was already pointed out in
Refs.~\cite{KN_02,Knecht_et_al_PRL_02,HLbL_PS_vs_lepton_pair_decay}.

\subsection{Lattice estimate of the pion-pole contribution $\amu$ \label{subsec:amu}} 

In this section, we use the results from Secs.~\ref{subsec:fit_res} and
\ref{subsec:experimental_data} of the fits to the lattice data in the
different models to estimate the pion-pole contribution $\amu$ to
hadronic light-by-light scattering in the muon $g-2$, thought to be
numerically dominant. As shown in Ref.~\cite{Jegerlehner:2009ry},
starting from the two-loop integrals in Fig.~\ref{fig:hlbl}, one can
perform, after a Wick rotation to Euclidean momenta, all angular
integrals except one for general pion transition form factors. The
pion-pole contribution is then given by
\be 
\amu = \left( \frac{\alpha_{e}}{\pi} \right)^3 \left(
  a_\mu^{\mathrm{HLbL}; \pi^0 (1)} + a_\mu^{\mathrm{HLbL}; \pi^0 (2)}
\right) \,, 
\ee 
where 
\bea 
a_\mu^{\mathrm{HLbL}; \pi^0 (1)} & = & \int_0^\infty \!\!\!dQ_1
\!\!\int_0^\infty \!\!\!dQ_2 \!\!\int_{-1}^{1} \!\!d\tau \, 
w_1(Q_1,Q_2,\tau) \, \FF(-Q_1^2, -(Q_1 + Q_2)^2) \, \FF(-Q_2^2,0) \,,
\label{contribution_1} \\ 
a_\mu^{\mathrm{HLbL}; \pi^0 (2)} & = & \int_0^\infty \!\!\!dQ_1
\!\!\int_0^\infty \!\!\!dQ_2 \!\!\int_{-1}^{1} \!\!d\tau \, 
w_2(Q_1,Q_2,\tau) \, \FF(-Q_1^2, -Q_2^2) \, \FF(-(Q_1+Q_2)^2,0)
\,. 
\label{contribution_2}
\eea 
The integrals run over the lengths $Q_i = |(Q_{i})_\mu|, i=1,2$ of the
two Euclidean four-momentum vectors and the angle $\theta$ between
them $Q_1 \cdot Q_2 = Q_1 Q_2 \cos\theta$ and we defined $\tau =
\cos\theta$. The analytical expressions for the model-independent
weight functions $w_i(Q_1, Q_2,\tau), i=1,2$ can be found in
Ref.~\cite{Jegerlehner:2009ry}. Their properties have been analyzed in
detail recently in Ref.~\cite{Nyffeler:2016gnb}. These functions
vanish for $Q_i \to 0, Q_i \to \infty~(i = 1,2)$ and $\tau \to \pm 1$
and for the pion they are concentrated at small momenta below
$1~\GeV$. This explains that the main contribution to $\amu$ arises
from the low-energy region of the double-virtual pion transition form
factor, which has been studied in this paper. But $w_1$ also has a slow falloff (ridge) in one direction of the $(Q_1, Q_2)$ plane, which has to be
dampened by the form factors. Therefore there is some dependence of
the final result on the behavior of the single- and double-virtual
form factors according to Brodsky-Lepage in Eq.~(\ref{eq:BL}) and the
OPE in Eq.~(\ref{eq:OPE}), which explains the different results
(central values) for VMD, LMD and LMD+V given below.

Using the results of the fitted model parameters for VMD, LMD and
LMD+V from Eqs.~(\ref{eq:resVMD}), (\ref{eq:resLMD}) and
(\ref{eq:resLMDV}) and integrating Eqs.~(\ref{contribution_1}) and
(\ref{contribution_2}) numerically, we obtain the results collected in
Table~\ref{Tab:amu_results} where the correlations from
Eqs.~(\ref{eq:covVMD}), (\ref{eq:covLMD}) and (\ref{eq:covLMDV}) have
been taken into account to estimate the statistical error.\footnote{We
  use $m_\mu = 105.6583715~\MeV$, $m_{\pi^0} = 134.9766~\MeV$,
  $\alpha_{e} = 1/137.035999$ and for the theory calculations $M_\rho
  = 775~\MeV$ and $M_{\rho^\prime} = 1465~\MeV$.}
\begin{table}[t]
\begin{center}
\caption{Results for $\amu$ with statistical errors (including
  correlations) using the model parameters from the fits to the
  lattice data and comparison with theory predictions with the
  parameters chosen as discussed in the text.  
  \label{Tab:amu_results}}
    \vskip 0.1in
\renewcommand{\arraystretch}{1.3}
\begin{tabular}{lc} 
\hline 
Model & $\amu \times 10^{11}$ \\  
\hline 
VMD                            & 56.7(7.1) \\ 
LMD                            & 68.2(7.4) \\
LMD+V                          & 65.0(8.3) \\ 
\hline 
VMD (theory)          	       & 57.0 \\ 
LMD (theory)          	       & 73.7 \\ 
LMD+V (theory + phenomenology) & 62.9 \\ 
\hline 
\end{tabular} 
\end{center}
\end{table}

Note that the VMD model yields a bad fit to the lattice data. The
corresponding estimate is therefore only given for illustration. In
Table~\ref{Tab:amu_results} we also compare our results with those
obtained with the theoretically preferred model parameters $\alpha =
\alpha_{\mathrm{th}} = 1 / (4 \pi^2 F_\pi)$, $\beta =
\beta^{\mathrm{OPE}} = -F_\pi / 3$, $\widetilde{h}_0 =
\widetilde{h}_0^{\mathrm{OPE}} = -F_\pi / 3$, $\widetilde{h}_1 = 0$
and the phenomenologically determined parameters $\widetilde{h}_2$ and
$\widetilde{h}_5$ from Eqs.~(\ref{h2}) and (\ref{h5}) discussed in
Sec.~\ref{sec:pion_ff} (we only quote the central value for the
theoretical estimates). The fit results with their relatively large
statistical errors of about 13\% agree well with the corresponding
theoretical estimates.

The agreement of the results for VMD with the fitted and the
theoretically preferred model parameters is a pure coincidence, since
the fitted parameters $\alpha^{\VMD}, M_V^{\VMD}$ in
Eq.~(\ref{eq:resVMD}) differ significantly from $\alpha_{\mathrm{th}}$
and $M_\rho$. The form factor with the fitted parameters is smaller at
small momenta compared to the form factor with the theoretical
parameters ($\FF^{\VMD, \mathrm{fit}}(0,0) = \alpha^{\mathrm{VMD}} =
0.243~\GeV^{-1}$ vs $\FF^{\VMD, \mathrm{theory}}(0,0) =
\alpha_{\mathrm{th}} = 0.274~\GeV^{-1}$), but then falls off slower
beyond about $0.5~\GeV$.

For illustration, we show in Table~\ref{Tab:cutoffdependence} how the
value for $\amu$ changes in the different models obtained from the
fits, if we use a momentum cutoff $\Lambda$ in the integrals in
Eqs.~(\ref{contribution_1}) and (\ref{contribution_2}).  As already
observed in Refs.~\cite{Nyffeler_13,Nyffeler:2016gnb}, the bulk of the
pion-pole contribution to HLbL comes from the region below $1~\GeV$,
around $85-90\%$, depending on the model. The absolute values for the LMD
and LMD+V models start to differ more and more above $\Lambda =
0.75~\GeV$, since the LMD model has the wrong asymptotics for the
single-virtual case, while the relative contributions only differ by
about $1-2$ percentage points. The absolute values of VMD are always
smaller than for LMD and LMD+V, because of the smaller normalization at
vanishing momenta. This latter behavior differs from the observations
made in Ref.~\cite{Nyffeler:2016gnb} where the normalization with the
chiral anomaly was used for VMD and LMD+V, so that the form factors
themselves only differed little for momenta below $0.75~\GeV$ and thus
also the contributions to $\amu$ were very similar for values of the
cutoff $\Lambda \leq 0.75~\GeV$. If the cutoff is higher, then the
wrong high-momentum behavior of the VMD form factor with a $1/Q^4$
falloff in the double-virtual case leads to a further suppression of
the contribution. 
\begin{table}[t] 
\begin{center} 
\caption{Pion-pole contribution $\amu \times 10^{11}$ for different form
    factor models fitted to the lattice data as function of a momentum
    cutoff~$\Lambda$. In brackets, relative contribution of the total
    obtained with $\Lambda = 20~\GeV$. 
    \label{Tab:cutoffdependence}}
    \vskip 0.1in
\renewcommand{\arraystretch}{1.3}
\begin{tabular}{S[table-format=2.2]@{\hskip 0.2in}cc@{\hskip 0.2in}cc@{\hskip 0.2in}cc}
\hline 
\mc{$\Lambda$ [GeV]} & 	\multicolumn{2}{c}{VMD}           & \multicolumn{2}{c}{LMD}           & \multicolumn{2}{c}{LMD+V} \\ 
\hline 
 0.25           & 11.8 &(20.8\%) & 14.6&(21.4\%) & 14.4&(22.1\%) \\  
 0.5            & 32.1 &(56.7\%) & 37.9&(55.5\%) & 37.2&(57.2\%) \\  
 0.75           & 44.1&(77.8\%) & 50.7&(74.4\%) & 49.5&(76.1\%) \\  
 1.0            & 50.1&(88.4\%) & 57.3&(84.0\%) & 55.5&(85.4\%) \\  
 1.5            & 54.6&(96.3\%) & 62.9&(92.3\%) & 60.6&(93.1\%) \\  
 2.0            & 55.9&(98.6\%) & 65.1&(95.5\%) & 62.5&(96.1\%) \\  
 5.0            & 56.7&(100\%)  & 67.7&(99.2\%) & 64.6&(99.4\%) \\  
20.0            & 56.7&(100\%)  & 68.2&(100\%)  & 65.0&(100\%)  \\  
\hline 
\end{tabular} 
\end{center}
\end{table}

Our preferred estimate for $\amu$ is obtained with the fitted LMD+V
model,  
\be \label{amupi0LMDV_lattice}
a_{\mu; \LMDV}^{\mathrm{HLbL}; \pi^0} = (65.0 \pm 8.3) \times
10^{-11} \,.  
\ee 
Although this model yields a good fit to the lattice data, not all model
parameters can be fitted simultaneously: some parameters are fixed to
constraints from theory. On the other hand, the LMD model yields an
even slightly better fit to the data in the limited kinematical range
where there are lattice data, up to $1.5~\GeV^2$ in the double-virtual
case and only up to $0.5~\GeV^2$ in the single-virtual case. But for
large momenta the single-virtual LMD form factor does not fall off
like $1/Q^2$ according to the Brodsky-Lepage condition in
Eq.~(\ref{eq:BL}). It seems doubtful to then simply perform the
integration in Eqs.~(\ref{contribution_1}) and (\ref{contribution_2})
up to infinite momenta for the LMD form factor. This partly explains
the larger result for LMD compared to VMD and LMD+V which both fulfill
the Brodsky-Lepage prediction.

Before drawing any further conclusions, an estimate of the systematic
error should be obtained. A sophisticated error analysis, including
effects from discretization, finite volume and the used fit \textit{Ans\"atze}
(different form factor models) is beyond the scope of this paper. If
we use the results for the LMD+V model parameters from the additional
fits in Eqs.~(\ref{eq:resLMDV1}) and (\ref{eq:resLMDV2}), we obtain
results for $a_{\mu; \LMDV}^{\mathrm{HLbL}; \pi^0}$ which differ by
about $\pm 1.2 \times 10^{-11}$ from the result given in
Eq.~(\ref{amupi0LMDV_lattice}), whereas the statistical uncertainty
stays about the same, if one uses the covariance matrices for these
fits. This variation does cover different ways to vary the vector
meson masses $M_{V_1}$ and $M_{V_2}$, but it does not take into
account that not all LMD+V model parameters have been fitted. On the
other hand, since VMD and LMD do not obey important short-distance
constraints from QCD, in contrast to LMD+V, one should not take the
difference of these results from LMD+V as indication of an additional
systematic error.

For comparison, we note that most model calculations yield results for
the pion-pole contribution in the range $\amu = (50-80) \times
10^{-11}$ (central values) with rather arbitrary, model-dependent
error estimates, see
Refs.~\cite{Jegerlehner:2009ry,Bijnens:2015jqa,Nyffeler:2016gnb} and
references therein.

%
%
\section{Conclusion}

We have performed a calculation of the double-virtual pion transition
form factor ${\cal F}_{\pi^0\gamma^*\gamma^*}(q_1^2, q_2^2)$ in
lattice QCD with two flavors of quarks. We find that we are able to
describe the lattice data by performing a three-parameter fit, either
using the LMD model or the LMD+V model defined in Eqs.\
(\ref{eq:LMD_model}) and (\ref{eq:LMDV_model}). In both cases, the
overall normalization of the form factor, ${\cal
  F}_{\pi^0\gamma^*\gamma^*}(0,0)$, comes out consistent with the
prediction of the chiral anomaly, with a statistical accuracy of
$8-9\%$. In the case of LMD+V, the functional form
contains a sufficient number of parameters to be consistent with the
theoretically predicted leading behavior at large $Q^2$, both in the
single-virtual and the double-virtual case. Being unable to fit all
the parameters, we have set some of these parameters to their
phenomenological or to their ``preferred'' theory values. In particular,
the parameter determining the ${\cal
  F}_{\pi^0\gamma^*\gamma^*}(-Q^2,-Q^2)$ behavior at large $Q^2$ has
been set to the OPE prediction, Eq.~(\ref{eq:OPE}).  However, the
parameter determining the large $Q^2$ behavior in the single-virtual
case comes out consistent, albeit with large uncertainties, with the
Brodsky-Lepage expectation as given by Eq.~(\ref{eq:BL}) and the value
for ${\bar h}_5$ in the LMD+V fit is consistent with a fit to the CLEO
data in Eq.~(\ref{h5}). Furthermore, the parameter ${\bar h}_2$ which
only enters the double-virtual and not the single-virtual form factor,
comes out as expected from theoretical expectations from higher-twist
corrections in the OPE, see Eq.~(\ref{h2}), although with rather large
uncertainty. 

On the other hand, the popular VMD form factor model yields a bad fit
to the lattice data. The extracted normalization is not consistent
with the chiral anomaly and the VMD form factor, which factorizes as
function of the two momenta $Q_1^2$ and $Q_2^2$, fails to reproduce the
double-virtual lattice data for increasing spacelike momenta.

We have presented a new value for the pion-pole contribution to
hadronic light-by-light scattering in the $g-2$ of the muon,
$a_\mu^{{\rm HLbL};\pi^0}$, using the LMD+V fit \textit{Ansatz}. The result,
given in Eq.~(\ref{amupi0LMDV_lattice}), is based for the first time
on direct nonperturbative information on the double-virtual form
factor. It is well in line with other phenomenological estimates; see
Table \ref{Tab:amu_results} and
Refs.~\cite{Prades:2009tw,Jegerlehner:2009ry,Bijnens:2015jqa}.

As for the technical aspects of the lattice calculation, we have
demonstrated that fairly accurate results can be obtained for the
transition form factor, particularly in the doubly spacelike regime.
In this respect, our calculation is complementary both to existing
experimental data and to~\cite{Feng:2012ck}, which focused mainly on
the chiral anomaly prediction for ${\cal
  F}_{\pi^0\gamma^*\gamma^*}(0,0)$ and neutral pion decay $\pi^0 \to
\gamma\gamma$. We have shown
that the cusp in the matrix element of a short-distance product of two
vector currents is directly related to the coefficient of the $1/Q^2$
falloff of ${\cal F}_{\pi^0\gamma^*\gamma^*}(-Q^2,-Q^2)$.
Finite-size effects can be a significant issue, especially in the
single-virtual kinematics, because the tail of the correlator when the
vector currents are far apart is strongly affected by finite-size
corrections. Although we have described a way to potentially correct
for these effects, a quantitative study is left for the future.  We
have also found that the disconnected contributions are at the
subpercent level on an ensemble with $m_\pi=440~\MeV$ and tend to
reduce the form factor.  Although further calculations at smaller pion
masses are needed, these contributions appear to be negligible at our
current level of precision. In the future, the time-momentum representation
will probably be our preferred method, especially if $\mathcal{O}(a)$-improvement
is fully implemented. Finally, the calculation should be
repeated with a dynamical strange quark, even though the strange quark
contributes to ${\cal F}_{\pi^0\gamma^*\gamma^*}$ only via diagrams
with disconnected quark lines.


\begin{acknowledgments}
We are grateful to Vera G\"ulpers, Georg von Hippel 
and Hartmut Wittig for providing the 
disconnected vector quark loops used in this study and helpful
discussions.  We are also grateful for the access to the ensembles
used here, made available to us through CLS.  We acknowledge the use
of computing time for the generation of the gauge configurations on
the JUGENE and JUQUEEN computers of the Gauss Centre for
Supercomputing located at Forschungszentrum J\"ulich, Germany. All
correlation functions were computed on the dedicated QCD platforms
``Wilson'' at the Institute for Nuclear Physics, University of Mainz,
and ``Clover'' at the Helmholtz-Institut Mainz. This work is partly
supported by the DFG through CRC 1044.
\end{acknowledgments}


\appendix

\section{Analytic expression of $\widetilde{A}_{\mu\nu}^{\rm VMD}(\tau)$ and $\widetilde{A}_{\mu\nu}^{\rm LMD}(\tau)$ \label{app:A} } 

In this appendix, we calculate the explicit expression for $\widetilde{A}_{\mu\nu}^{\rm LMD}(\tau)$ introduced in Eq.~(\ref{eq:lat_M}) corresponding to the LMD form factor. The same expression for the VMD form factor is easily obtained by setting $\beta=0$. Starting from 
\begin{align}
M_{\mu\nu}^{\rm E} 
=  \frac{2 E_{\pi}}{ Z_{\pi} }  \int_{-\infty}^{\infty} \, \mathrm{d}\tau \, e^{\omega_1 \tau} \, \widetilde{A}_{\mu\nu}(\tau) \,,
\end{align}
which holds for all real values of $\omega_1$, 
we consider the analytic continuation for all complex values of $\omega_1 = i \widetilde{\omega}$. Then
\begin{align}
\frac{1}{\sqrt{2\pi}} \int_{-\infty}^{\infty} \, \mathrm{d}\tau \, e^{i \widetilde{\omega} \tau} \, \widetilde{A}_{\mu\nu}(\tau)  = \frac{ Z_{\pi}  }{ 2 \sqrt{2\pi}\, E_{\pi} }  M_{\mu\nu}^{\rm E},
\end{align}
and
\begin{align}
\widetilde{A}_{\mu\nu}(\tau)  = \frac{ Z_{\pi}  }{ 4 \pi E_{\pi} } \int_{-\infty}^{\infty} \, \mathrm{d} \widetilde{\omega} \, 
M_{\mu\nu}^{\rm E}
 \, e^{- i \widetilde{\omega} \tau} \,.
\label{eq:A_tilde1}
\end{align}
Since $M_{\mu\nu}^{\rm E}$ is directly proportional to the pion transition form factor 
(see Eqs.\ (\ref{eq:MFF}) and (\ref{eq:Mminkowski})), this equation shows that 
$\widetilde{A}_{\mu\nu}(\tau)$ is in essence the Fourier transform of the form factor. More precisely, 
consider the case where the pion is at rest, and $\mu=k$, $\nu=l$ spatial indices, so that $M_{kl}^{\rm E} = M_{kl}$.
We have $q_1 = (\omega_1, \vec{q}_1)$ and $q_2 = (m_{\pi}-\omega_1, - \vec{q}_1)$. Then, using the definition (\ref{eq:Ascalar})
and the fact that $q_{kl}$ is independent of $\widetilde\omega$, Eq.\ (\ref{eq:A_tilde1}) becomes 
\be
\widetilde{A}(\tau)  = \frac{i Z_{\pi} }{ 4 \pi m_{\pi} } \int_{-\infty}^{\infty} \, \mathrm{d} \widetilde{\omega} \, 
 \FF(q_1^2, q_2^2)  \, e^{- i \widetilde{\omega} \tau} \,.
\ee
Using the LMD form factor given in Eq.~(\ref{eq:LMD_model}), we obtain
\begin{align}
\nonumber \widetilde{A}^{\LMD}(\tau) &= \frac{ i Z_{\pi} \, }{ 4 \pi m_{\pi}} \int_{-\infty}^{\infty} \, \mathrm{d} \widetilde{\omega}  \, \frac{ \alpha \, M_V^4 + \beta \, (q_1^2 + q_2^2) }{\left( M_V^2-q_1^2 \right) \left( M_V^2-q_2^2 \right)} \, e^{-i\widetilde{\omega} \tau}\\
&= \frac{i Z_{\pi} }{ 4 \pi m_{\pi}} \int_{-\infty}^{\infty} \, \mathrm{d} \widetilde{\omega}  \, \frac{ \alpha \, M_V^4 + \beta \, (q_1^2 + q_2^2) }{\left( M_V^2+|\vec{q}_1|^2+\widetilde{\omega}^2 \right) \left( M_V^2+|\vec{q}_1|^2-(m_{\pi}-i\widetilde{\omega})^2 \right)} \, e^{- i \widetilde{\omega} \tau} \,.
\label{eq:A_tilde2}
\end{align}
The integrand has four distinct simple poles 
\begin{align}
\widetilde{\omega}_1^{(\pm)} = \pm i\sqrt{M_V^2+|\vec{q}_1|^2} \quad , \quad \widetilde{\omega}_2^{(\pm)} = - i\left( m_{\pi}  \mp \sqrt{M_V^2+|\vec{q}_1|^2}  \right)  \,,
\end{align}
such that
\begin{align}
\widetilde{A}^{\LMD}(\tau) = \frac{i Z_{\pi} }{ 4 \pi m_{\pi}} \int_{-\infty}^{\infty} \, \mathrm{d} \widetilde{\omega}  \, \frac{ \alpha \, M_V^4 + \beta \, (q_1^2 + q_2^2) }{ \left( \widetilde{\omega} - \widetilde{\omega}_1^{(+)} \right) \left(\widetilde{\omega} - \widetilde{\omega}_1^{(-)} \right)   \left( \widetilde{\omega} - \widetilde{\omega}_2^{(+)} \right) \left(\widetilde{\omega} - \widetilde{\omega}_2^{(-)} \right) } \, e^{- i\widetilde{\omega} \tau} \,.
\end{align}
%
\underline{Case $\tau > 0$ :} 
%
\begin{align}
\nonumber \widetilde{A}^{\LMD}(\tau > 0) =  \frac{i Z_{\pi} }{ 4 m_{\pi}} & \left[  \frac{\alpha\, M_V^4 + \beta \left( 2 M_V^2+m_{\pi}^2 - 2 m_{\pi}\sqrt{M_V^2 + |\vec{q}_1|^2} \, \right)}{ m_{\pi} \sqrt{M_V^2+|\vec{q}_1|^2} \left( 2 \sqrt{M_V^2+|\vec{q}_1|^2 } - m_{\pi} \right)  } \, e^{- \sqrt{M_V^2+|\vec{q}_1|^2} \tau}   \right. \\
& \left. - \frac{\alpha\, M_V^4 + \beta \left( 2M_V^2 + m_{\pi}^2 + 2 m_{\pi}\sqrt{M_V^2 + |\vec{q}_1|^2} \,  \right)}{  m_{\pi} \sqrt{M_V^2+|\vec{q}_1|^2}  \left(2 \sqrt{M_V^2+|\vec{q}_1|^2} + m_{\pi} \right) } \, e^{-  \left( m_{\pi}  + \sqrt{M_V^2+|\vec{q}_1|^2}  \right)    \tau} \right] \,.
\label{eq:Apos}
\end{align}

\underline{Case $\tau < 0$ :} 
%
\begin{align}
\nonumber \widetilde{A}^{\LMD}(\tau < 0) =  \frac{ i Z_{\pi} }{ 4 m_{\pi }} & \left[  - \frac{\alpha\, M_V^4 + \beta \left( 2 M_V^2+m_{\pi}^2 + 2 m_{\pi}\sqrt{M_V^2 + |\vec{q}_1|^2} \, \right)   }{ m_{\pi} \sqrt{M_V^2+|\vec{q}_1|^2}  \left( 2 \sqrt{M_V^2+|\vec{q}_1|^2 } + m_{\pi} \right) } \, e^{ \sqrt{M_V^2+|\vec{q}_1|^2} \tau}   \right. \\
& \left. + \frac{\alpha\, M_V^4  + \beta \left(2M_V^2 + m_{\pi}^2 - 2 m_{\pi}\sqrt{M_V^2 + |q_1|^2} \,  \right) }{  m_{\pi} \sqrt{M_V^2+|\vec{q}_1|^2} \left( 2 \sqrt{M_V^2+|\vec{q}_1|^2} - m_{\pi} \right)   } \, e^{-     \left( m_{\pi}  - \sqrt{M_V^2+|\vec{q}_1|^2}  \right)    \tau} \right] \,.
\label{eq:Aneg}
\end{align}
In particular, comparing Eqs.~(\ref{eq:Apos}) and (\ref{eq:Aneg}), the symmetry under $\tau \to - \tau$ of $A(\tau)$ defined in Eqs.~(\ref{eq:Amunu}) and (\ref{eq:Ascalar}) is now explicit in this particular model. The first exponential in Eq.~(\ref{eq:Apos}) describes a vector meson with spatial momentum $\vec{q}_1$ whereas the second exponential describes a vector meson plus a pion.

We note that $\widetilde{A}_{kl}$ in general admits a cusp at $\tau=0$. The discontinuity in the derivative
is given by 
\be
\frac{ \mathrm{d}\widetilde{A}^{\LMD}_{kl} }{ \mathrm{d}\tau }\Big|_{\tau=0^{-}} - \frac{ \mathrm{d}\widetilde{A}^{\LMD}_{kl} }{ \mathrm{d}\tau }\Big|_{\tau=0^{+}} =  - Z_{\pi} \, {\beta }\,\epsilon_{kli} \, q_1^i \, .
\label{eq:discLMD}
\ee
In particular, for the VMD model where $\beta=0$, the cusp vanishes. This comes from the fact that we have ${\cal F}_{\pi^0\gamma^*\gamma^*}^{\rm VMD}(-Q^2,-Q^2)\stackrel{Q^2\to+\infty}{\sim} Q^{-4}$,
while the cusp is proportional to the coefficient of the $1/Q^2$ term.
As we show in Appendix \ref{app:OPE}, the cusp is directly calculable using the operator-product expansion
in the Euclidean theory.
 
\section{Results of the fits in four-momentum space \label{app:tables}} 

In this appendix, we collect our fit results for the form factor in four-momentum space, presented in Sec~\ref{subsec:fit_res}. As explained in the text, two fitting procedures have been used. In the first method (Table~\ref{tab:loc_fit}), the form factor on each lattice ensemble is fitted independently using either the VMD or the LMD model (Eqs.~(\ref{eq:VMD_model}) and (\ref{eq:LMD_model})). Then, in a second step, the chiral and continuum limit of each parameter is taken using the \textit{Ansatz}
\begin{equation}
\vec{p}(\widetilde{y},a) = \vec{p}(0,0) + \vec{C}_{\widetilde{y}} \, \widetilde{y} + \vec{C}_{a} \, \left(\frac{a}{a_{\beta=5.3}}\right) \,,
\label{eq:fit_param}
\end{equation}
where $\widetilde{y}=m^2_{\pi} / 8 \pi^2 F^2_{\pi}$, 
$\vec{p} = (M_V, \alpha)$ for the VMD model and $\vec{p}=(M_V, \beta, \alpha)$ for the LMD model. 
In the second method (Table~\ref{tab:glb_fit}), all lattice ensembles are fitted simultaneously using Eqs.~(\ref{eq:VMD_model}), (\ref{eq:LMD_model}) or (\ref{eq:LMDV_model}) and assuming a linear dependence in both the lattice spacing $a/a_{\beta=5.3}$ and $\widetilde{y}$ for each parameter, similar to Eq.~(\ref{eq:fit_param}). In the tables, a cross indicates that the parameter is not fitted and explicitly set to zero and a number quoted without error indicates that the parameter is not fitted but set to a constant value.

\newpage

\renewcommand{\arraystretch}{1.0}
\begin{table}[h]

\caption{Fit results using the first method. We collect the values of the fit parameters for both the VMD model (Eqs.~(\ref{eq:VMD_model})) and LMD model (Eq.~\ref{eq:LMD_model})) for each lattice ensemble and the associated $\chi_1^2/\dof$. Then the continuum and chiral extrapolation of each parameter is given in both the chiral limit ($\widetilde{y}=0$) and at the physical point ($\widetilde{y}=\widetilde{y}_{\exp}$). The last lines correspond to the slope for discretization effects and chiral corrections and the $\chi_2^2/\dof$ of the corresponding fit (see Eq.~(\ref{eq:fit_param})). }
\label{tab:loc_fit}

\begin{center}
	\begin{tabular}{l S[table-format=3.2] S[table-format=0.5] @{\quad}c}
	\hline 
				&	\multicolumn{3}{c}{VMD} 	 \\
	\cline{2-4}
				& 	$M_V~[\MeV]$	&	$\alpha~[\GeV^{-1}]$	 &	$\chi_1^2/{\dof}$ \\ 
	\hline  
	A5    			& 	1020(15)		&	0.313(8) 	&	2.09	\\
	B6    			& 	980(18)		& 	0.343(13)	& 	1.16	\\
	E5			&	1111(14) 		&	0.276(9) 	&	3.37	\\ 
	F6			&	945(11)		&	0.329(9)	&	1.53	\\
	F7    			& 	940(12)		&	0.335(8) 	& 	2.78 \\
	G8    		& 	927(17)		&	0.324(11) 	&	0.92	\\
	N6    		& 	1048(13)		&	0.266(7)	&	4.23	\\
	O7    		&	999(16)		& 	0.299(8)	& 	2.36	\\
	\hline
	\hline
	Extrap. ($\widetilde{y}=0$)  				& 	930(37)		&	0.254(20)	&	\\
	Extrap. ($\widetilde{y}=\widetilde{y}_{\exp}$)	& 	967(25)		&	0.240(19)	&	\\
	\hline
	$C_{\widetilde{y}}$					&	\mc{2.71(38)}		&	\mc{-1.03(15)}	&	\\
	$C_{a}$							&	\mc{-0.09(3)} 		&	\mc{0.12(2)}	&	\\	
	\hline 
	$\chi_2^2/{\dof}$    						& 	\mc{4.77}			&	\mc{2.02}		& 	\\
	\hline 
	\end{tabular}
\end{center}

\vspace{1cm}

\begin{center}
	\begin{tabular}{l S[table-format=3.2] S[table-format=3.2] S[table-format=0.6] @{\quad}c}
	\hline 
				&	\multicolumn{4}{c}{LMD}	 \\
	\cline{2-5} 
				&	$M_V~[\MeV]$	& $\beta~[\MeV]$	&	$\alpha~[\GeV^{-1}]$		&	$\chi_1^2/{\dof}$ \\ 
	\hline 
	A5    			&	844(25)	&	 -36(3)	&	0.328(10) 		&	0.62\\
	B6    			&	828(22)	& 	-33(3)	&	0.360(14)		& 	0.29\\
	E5			&	925(23)	& 	 -36(2)	&	0.291(9)			& 	0.20\\ 
	F6			&	795(21)	&	-35(3)	&	0.337(9) 			&	0.22\\
	F7    			&	770(18)	&	 -33(2)	&	0.355(9) 			&	0.63\\
	G8    		&	755(44)	&	-32(5)	&	0.344(14)		&	0.42\\
	N6    		&	857(18)	&	 -30(1)	&	0.287(8)			&	0.47\\
	O7    		&	805(25)	& 	-34(2)	& 	0.319(10)		&	0.58\\
	\hline 
	\hline 
	Extrap. ($\widetilde{y}=0$)  				& 	674(62) 	&	 -26(5)	&	0.291(23) 		&	\\
	Extrap. ($\widetilde{y}=\widetilde{y}_{\exp}$)	& 	712(56)	&	 -26(5)	&	0.275(22)		&	\\
	\hline 
	$C_{\widetilde{y}}$						&	\mc{2.77(65)}  	&	\mc{0.006(49)}	&	 \mc{-1.13(18)}		&	\\
	$C_{a}$								&	\mc{-0.009(46)} 	&	\mc{-0.008(4)}	&  \mc{0.10(2)}	&\\	
	\hline 
	$\chi_2^2/{\dof}$    	& \mc{0.88} 		&	\mc{1.08}		&	\mc{0.68}			&	\\
	\hline  
	
	\end{tabular}
\end{center}

\end{table}

\newpage 


\renewcommand{\arraystretch}{1.3}
\begin{table}[h!]

\caption{Fit results using the second method. The continuum and chiral extrapolation of each parameter is given both in the chiral limit ($\widetilde{y}=0$) and at the physical point ($\widetilde{y}=\widetilde{y}_{\exp}$). The last lines correspond to the slope for discretization effects and chiral corrections and the $\chi_2/\dof$ of the fit (see Eq.~(\ref{eq:fit_param})). The VMD, LMD and LMD+V models are defined by Eqs.~(\ref{eq:VMD_model}), (\ref{eq:LMD_model}) and (\ref{eq:LMDV_model}) respectively. In the case of the LMD+V model, as explained in the main text, we assume a constant shift in the spectrum and set $M_{V_2}(\widetilde{y}) = m^{\exp}_{\rho^{\prime}} + M_{V_1}(\widetilde{y}) - m^{\exp}_{\rho}$ with $m^{\exp}_{\rho^{\prime}} = 1.465~\GeV$. }
\label{tab:glb_fit}
    \vskip 0.1in
\begin{minipage}{0.49\textwidth}

\begin{center}
	\begin{tabular}{lcc}
	\hline 
				&	\multicolumn{2}{c}{VMD} 	 \\
	\cline{2-3} 
				& 	$M_V~[\MeV]$	&	$\alpha~[\GeV^{-1}]$	 \\ 
	\hline  
	 Extrap. ($\widetilde{y}=0$)				& 	907(37)		&	0.257(18) 	 \\
	 Extrap. ($\widetilde{y}=\widetilde{y}_{\exp}$)	& 	944(34)		&	0.243(18)\\
	\hline
	$C_{\widetilde{y}}$						& 	2.74(40)		& 	-1.03(16)\\
	$C_{a}$								&	-0.06(3) 		&	0.11(2)\\ 
	\hline 
	$\chi^2/{\dof}$    						& 	\multicolumn{2}{c}{2.94}\\
	\hline  
	\end{tabular}
\end{center}

\end{minipage}
\begin{minipage}{0.49\textwidth}

\begin{center}
	\begin{tabular}{lccc}
	\hline 
				&	\multicolumn{3}{c}{LMD}			 \\
	\cline{2-4} 
				& 	$M_V~[\MeV]$	& $\beta~[\MeV]$	&	$\alpha~[\GeV^{-1}]$	\\ 
	\hline  
	 Extrap. ($\widetilde{y}=0$)				&	669(30)	&	 -28(4)	&	0.289(19)	 \\
	 Extrap. ($\widetilde{y}=\widetilde{y}_{\exp}$)	&	705(24)	&	 -28(4)	&	0.275(18) \\
	\hline
	$C_{\widetilde{y}}$						& 	2.68(49)	& 	$\times$		&	 -1.14(19)	\\
	$C_{a}$								&	$\times$		& 	-0.006(5)	&	0.11(2) 	\\ 
	\hline 
	$\chi^2/{\dof}$    	& 	\multicolumn{3}{c}{1.30}	\\
	\hline  
	\end{tabular}
\end{center}

\end{minipage}

\begin{center}
	\begin{tabular}{lccccc}
	\hline 
				&	\multicolumn{5}{c}{LMD+V}			 \\
	\cline{2-6} 
				& 	$M_{V_1}~[\MeV]$	&	$\alpha~[\GeV^{-1}]$	&	$\widetilde{h}_0~[\GeV]$	& $\widetilde{h}_2~[\GeV^3]$	&	$\widetilde{h}_5~[\GeV]$	\\ 
	\hline  
	 Extrap. ($\widetilde{y}=0$)				& 	747(8)	&	0.287(25)		&	-0.031	&	0.345(167)	&	-0.182(74)		\\
	 Extrap. ($\widetilde{y}=\widetilde{y}_{\exp}$)	& 	775\,\, \, \,		&	0.273(24)		&	-0.030(5)	&	0.345(167)	&	-0.195(70)		\\
	\hline 	
	$C_{\widetilde{y}}$						& 	2.10(57)		& 	-1.04(22)		& 	-0.03(37)	& 	$\times$	&	 -0.98(45)	\\
	$C_{a}$								&	$\times$		&	0.10(2)		&	-0.03(4)	& 	0.09(19)	&	-0.08(6)	\\ 
	\hline 
	$\chi^2/{\dof}$    	& 	\multicolumn{5}{c}{1.36}		\\
	\hline  
	\end{tabular}
	
\end{center}

\end{table}

\subsection{Systematics errors : Finite-time extent}

Table~\ref{tab:glb_fit_fte} corresponds to Table~\ref{tab:glb_fit} but using the LMD model to fit the tail of $\widetilde{A}(\tau)$ rather than the VMD model. This is discussed in Sec.~\ref{subsubsec:FTE}.

\renewcommand{\arraystretch}{1.3}
\begin{table}[h!]

\caption{Study of finite-time effects. Same as Table~\ref{tab:glb_fit} but using the LMD model to fit the tail of $\widetilde{A}(\tau)$ rather than the VMD model.}
\label{tab:glb_fit_fte}
    \vskip 0.1in

\begin{minipage}{0.49\textwidth}

\begin{center}
	\begin{tabular}{lcc}
	\hline 
				&	\multicolumn{2}{c}{VMD} 		 \\
	\cline{2-3}
				& 	$M_V~[\MeV]$	&	$\alpha~[\GeV^{-1}]$	\\ 
	\hline 
	 Extrap. ($\widetilde{y}=0$)				& 	907(36)		&	0.257(19) 		 \\
	 Extrap. ($\widetilde{y}=\widetilde{y}_{\exp}$)	& 	943(33) 		&	0.243(18) 	\\
	\hline 
	$C_{\widetilde{y}}$			& 	2.66(38)		& 	-1.00(15)		\\
	$C_{a}$					&	-0.06(3) 		&	0.12(2)		\\ 
	\hline 
	$\chi^2/{\dof}$    	& 	\multicolumn{2}{c}{3.20}\\
	\hline 
	\end{tabular}
\end{center}

\end{minipage}
\begin{minipage}{0.49\textwidth}

\begin{center}
	\begin{tabular}{lccc}
	\hline 
				&	\multicolumn{3}{c}{LMD}			 \\
	\cline{2-4} 
				&	$M_V~[\MeV]$	& $\beta~[\MeV]$	&	$\alpha~[\GeV^{-1}]$	\\ 
	\hline 
	 Extrap. ($\widetilde{y}=0$)				&	651(29)	&	 -29(4)	&	0.292(20)	 \\
	 Extrap. ($\widetilde{y}=\widetilde{y}_{\exp}$)	&	686(23)	&	 -29(4)	&	0.277(19) \\
	\hline 
	$C_{\widetilde{y}}$			& 	2.55(47)	& 	$\times$		&	 -1.09(18)	\\
	$C_{a}$					&	$\times$		& 	-0.007(4)	&	0.11(2) 	\\ 
	\hline 
	$\chi^2/{\dof}$    	& 	\multicolumn{3}{c}{1.30}	\\
	\hline 
	\end{tabular}
\end{center}
	
\end{minipage}

\begin{center}
	\begin{tabular}{lccccc}
	\hline 
				&	\multicolumn{5}{c}{LMD+V}			 \\
	\cline{2-6} 
				& 	$M_{V_1}~[\MeV]$	&	$\alpha~[\GeV^{-1}]$	&	$\widetilde{h}_0~[\GeV]$	& $\widetilde{h}_2~[\GeV^3]$	&	$\widetilde{h}_5~[\GeV]$	\\ 
	\hline 
	 Extrap. ($\widetilde{y}=0$)				& 	759(7) 	&	0.283(28)		&	-0.031	&	0.279(136)	&	-0.200(70)		 \\
	 Extrap. ($\widetilde{y}=\widetilde{y}_{\exp}$)	& 	775		&	0.270(27)		&	 -0.029(4)	&	 0.279(136) 	&	-0.216(67) \\
	\hline 
	$C_{\widetilde{y}}$					& 	1.18(53)		& 	-0.94(21)		& 	0.10(30)	& 	$\times$		& 	-1.16(43)	\\
	$C_{a}$							&	$\times$		&	0.11(3)		&	-0.04(3)	& 	0.08(15)		&	0.07(6)	\\ 
	\hline
	$\chi^2/{\dof}$    	& 	\multicolumn{5}{c}{1.43}		\\
	\hline  
	\end{tabular}
	\end{center}

\end{table}

\subsection{Systematics errors : Sampling \label{app:sampling}}

\begin{figure}[h!]
	\begin{minipage}[c]{0.49\linewidth}
	\centering 
	\includegraphics*[width=\linewidth]{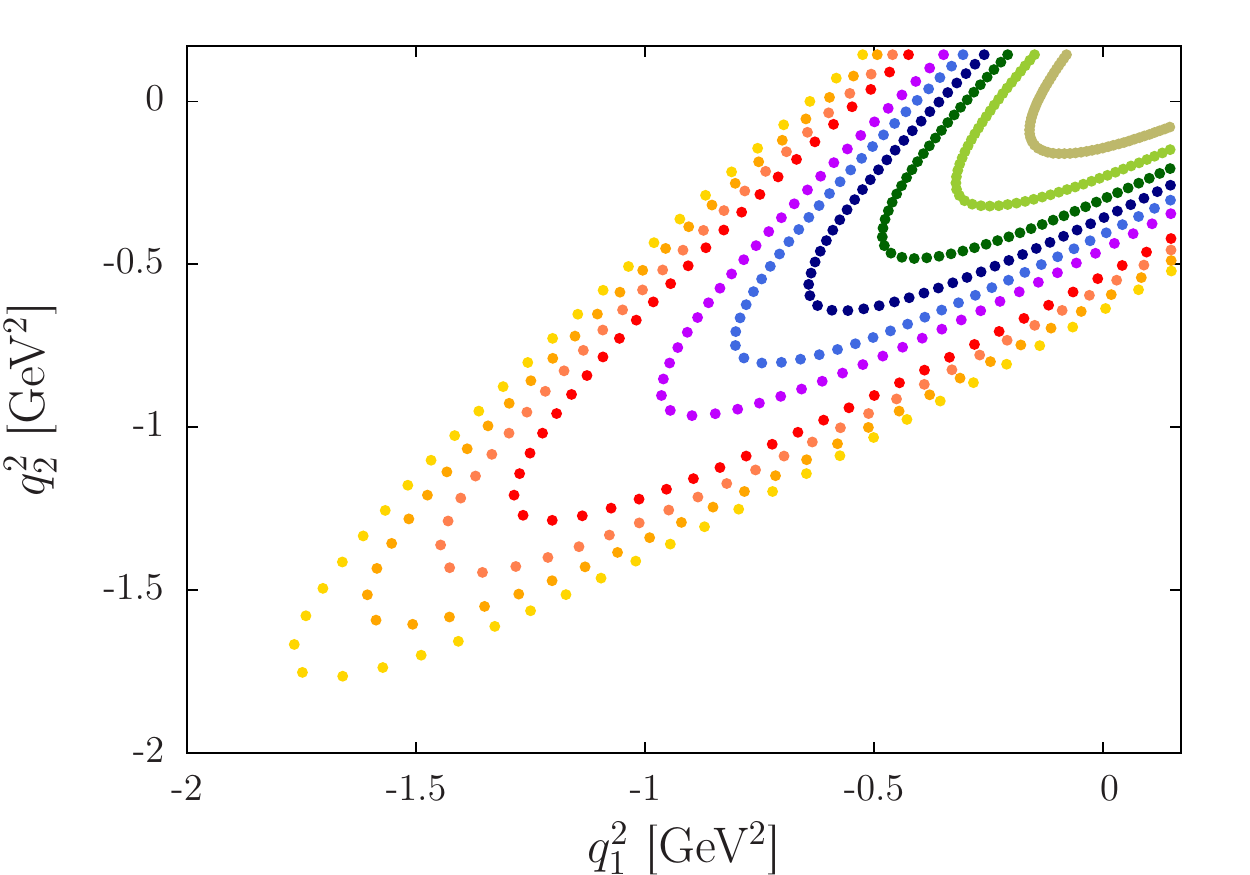}
	\end{minipage}
	\begin{minipage}[c]{0.49\linewidth}
	\centering 
	\includegraphics*[width=\linewidth]{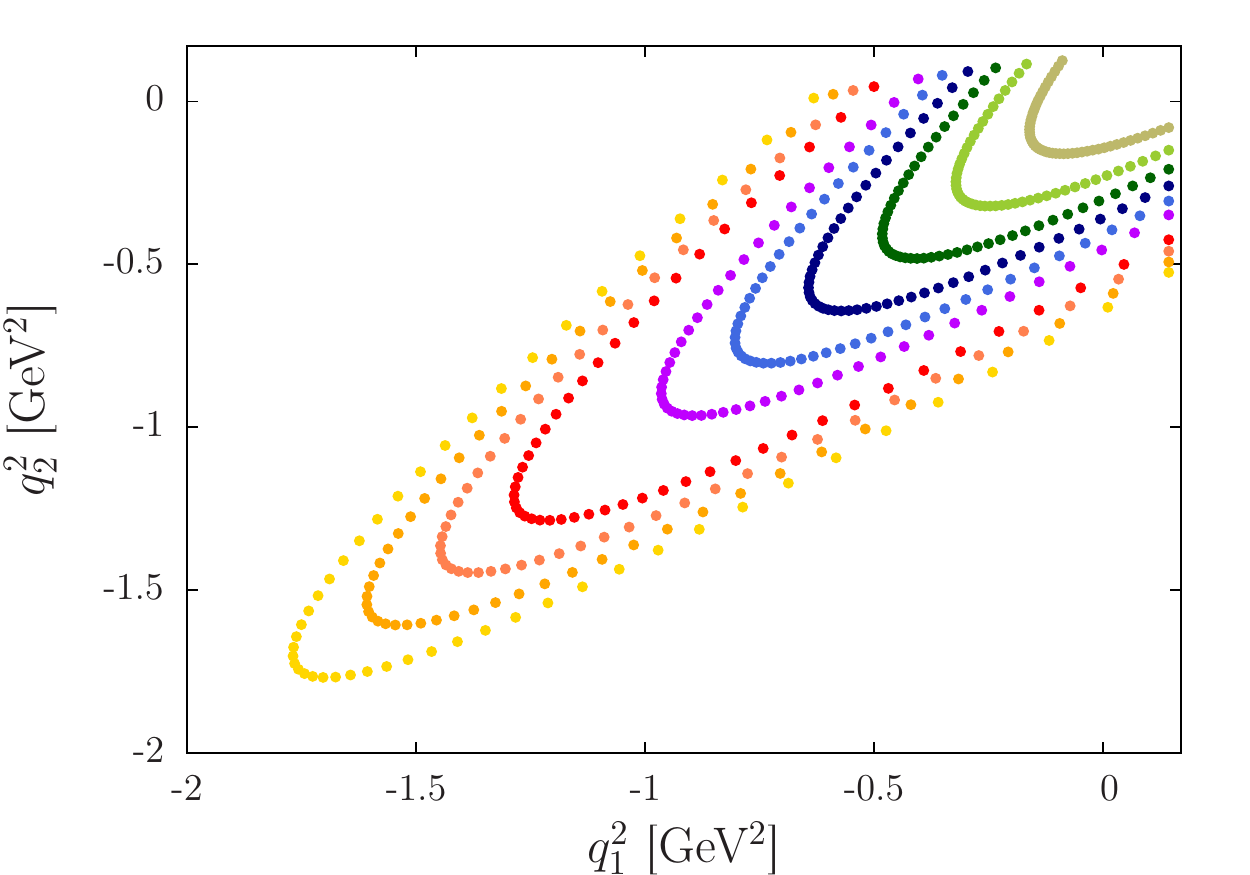}
	\end{minipage}

	\caption{Illustration of the two different samplings used for the lattice ensemble O7. Left: Sampling 1. Right: Sampling 2. }	
	\label{fig:sampling}
\end{figure}

Table~\ref{tab:glb_fit_s2} corresponds to Table~\ref{tab:glb_fit} but using a different method to sample our data as explained in Sec.~\ref{subsubsec:Sampling}. The two different samplings used in this work are illustrated in Fig.~\ref{fig:sampling} for the lattice ensemble~O7.
\begin{itemize}
\item Sampling 1 (Table~\ref{tab:glb_fit}) : Points are regularly distributed along each curve in the  ($q_1^2,q_2^2)$ plane. 
\item Sampling 2 (Table~\ref{tab:glb_fit_s2}) : Points are chosen using a constant step in $\omega_1$ in Eq.~(\ref{eq:kin}).
\end{itemize}

\renewcommand{\arraystretch}{1.3}
\begin{table}[h!]

\caption{Influence of the sampling used to discretize our data. Same as Table~\ref{tab:glb_fit} but using the second sampling to compute the form factor.}
\label{tab:glb_fit_s2}
    \vskip 0.1in

\begin{minipage}{0.49\textwidth}
	 
\begin{center}
	\begin{tabular}{lcc}
	\hline 
				&	\multicolumn{2}{c}{VMD} 		 \\
	\cline{2-3}
				& 	$M_V~[\MeV]$	&	$\alpha~[\GeV^{-1}]$	\\ 
	\hline 
	 Extrap. ($\widetilde{y}=0$)				& 	928(34)		&	0.248(16) 		 \\
	 Extrap. ($\widetilde{y}=\widetilde{y}_{\exp}$)	& 	968(32) 		&	0.233(16) 	\\
	\hline 
	$C_{\widetilde{y}}$				& 	2.90(37)		& 	-1.04(14)		\\
	$C_{a}$					&	-0.07(3) 		&	0.12(2)		\\ 
	\hline 
	$\chi^2/{\dof}$    	& 	\multicolumn{2}{c}{3.69}\\
	\hline 
	\end{tabular}
\end{center}

\end{minipage}
\begin{minipage}{0.49\textwidth}

\begin{center}
	\begin{tabular}{lccc}
	\hline 
				&	\multicolumn{3}{c}{LMD}			 \\
	\cline{2-4} 
				&	$M_V~[\MeV]$	& $\beta~[\MeV]$	&	$\alpha~[\GeV^{-1}]$	\\ 
	\hline 
	 Extrap. ($\widetilde{y}=0$)				&	659(29)	&	 -28(4)	&	0.291(17)	 \\
	 Extrap. ($\widetilde{y}=\widetilde{y}_{\exp}$)	&	696(22)	&	 -28(4)	&	0.276(16) \\
	\hline 
	$C_{\widetilde{y}}$			& 	2.76(48)	& 	$\times$		&	 -1.15(18)	\\
	$C_{a}$					&	$\times$		& 	-0.007(4)	&	0.11(2) 	\\ 
	\hline 
	$\chi^2/{\dof}$    	& 	\multicolumn{3}{c}{1.37}	\\
	\hline 
	\end{tabular}
\end{center}
	
\end{minipage}
	 
\begin{center}
	\begin{tabular}{lccccc}
	\hline 
				&	\multicolumn{5}{c}{LMD+V}			 \\
	\cline{2-6} 
				& 	$M_{V_1}~[\MeV]$	&	$\alpha~[\GeV^{-1}]$	&	$\widetilde{h}_0~[\GeV]$	& $\widetilde{h}_2~[\GeV^3]$	&	$\widetilde{h}_5~[\GeV]$	\\ 
	\hline 
	 Extrap. ($\widetilde{y}=0$)				& 	745(10) 	&	0.294(24)		&	-0.031	&	0.395(194)	&	-0.152(77)		 \\
	 Extrap. ($\widetilde{y}=\widetilde{y}_{\exp}$)	& 	775		&	 0.279(23)		&	 -0.030(4)	&	0.395(194)	&-0.168(74)  \\
	\hline 
	$C_{\widetilde{y}}$					& 	2.23(71)		& 	-1.09(21)		& 	0.02(36)	& 	$\times$		& 	-1.16(40)	\\
	$C_{a}$							&	$\times$		&	0.10(2)		&	-0.02(3)	& 	0.07(21)		&	0.09(6)	\\ 
	\hline
	$\chi^2/{\dof}$    	& 	\multicolumn{5}{c}{1.45}		\\
	\hline  
	\end{tabular}
\end{center}

\end{table}

\section{Results of the fits in the time-momentum representation \label{app:tables2}} 

In this appendix, we collect in Tables~\ref{tab:loc_fit_TMR} and \ref{tab:glb_fit_TMR} our fit results for the form factor in the time-momentum representation, presented in Sec.~\ref{subsec:fit_res2}. Similarly to the previous appendix, a cross indicates that the parameter is not fitted and explicitly set to zero and a number quoted without error indicates that the parameter is not fitted but set to a constant value.

\begin{table}[h!]

\caption{Fit results using the first method in the time-momentum representation with $\tau_{\rm min}/a = 3$. We collect the values of the fit parameters for each lattice ensemble and the associated $\chi_1^2/\dof$. Then the continuum and chiral extrapolation of each parameter is given in both the chiral limit ($\widetilde{y}=0$) and at the physical point ($\widetilde{y}=\widetilde{y}_{\exp}$). The last lines correspond to the slope for discretization effects and chiral corrections and the $\chi_2^2/\dof$ of the corresponding fit (see Eq.~(\ref{eq:fit_param})). } 
\label{tab:loc_fit_TMR}

\begin{center}
	\begin{tabular}{l S[table-format=3.2] S[table-format=3.2] S[table-format=0.6]@{\quad}c}
	\hline 
				&	\multicolumn{4}{c}{LMD ($\tau_{\rm min}/a=3$)}	 \\
	\cline{2-5} 
				&	$M_V~[\MeV]$	& $\beta~[\MeV]$	&	$\alpha~[\GeV^{-1}]$		&	$\chi_1^2/{\dof}$ \\ 
	\hline 
	A5    			&	805(35)	&	 -38(3)	&	0.333(12) 		&	0.93\\ 
	B6    			&	801(21)	& 	-32(3)	&	0.366(13)		& 	0.64\\ 
	E5			&	891(17)	& 	 -38(2)	&	0.292(4)		& 	0.26\\ 
	F6			&	760(22)	&	-38(2)	&	0.344(10) 		&	0.48\\ 
	F7    			&	731(22)	&	 -36(3)	&	0.361(11) 		&	0.62\\ 
	G8    		&	759(33)	&	-30(4)	&	0.345(13)		&	0.83\\ 
	N6    		&	794(19)	&	 -33(1)	&	0.299(9)		&	1.25\\ 
	O7    		&	771(25)	& 	-32(3)	& 	0.334(10)		&	1.17\\ 
	\hline 
	\hline 
	Extrap. ($\widetilde{y}=0$)  				& 	624(61) 	&	 -20(5)		&	0.318(24) 		&	\\
	Extrap. ($\widetilde{y}=\widetilde{y}_{\exp}$)	& 	651(57)	&	-22(6)		&	0.303(23)		&	\\
	\hline 
	$C_{\widetilde{y}}$						&	\mc{1.97(48)}	&	\mc{-0.12(4)}		&	\mc{-1.13(13)}		&	\\
	$C_{a}$								&	\mc{-0.061(48)}	&	\mc{-0.008(5)}		&	\mc{0.08(2)}	&\\	
	\hline 
	$\chi_2^2/{\dof}$    						&	\mc{1.41}		&		\mc{0.87}		&	\mc{1.14}		&	\\
	\hline  
	
	\end{tabular}
\end{center}

\end{table}

\begin{table}[h!]

\caption{Fit results using the second method in the time-momentum representation. The continuum and chiral extrapolation of each parameter is given both in the chiral limit ($\widetilde{y}=0$) and at the physical point ($\widetilde{y}=\widetilde{y}_{\exp}$). The last lines correspond to the slope for discretization effects and chiral corrections and the $\chi^2/\dof$ of the fit.}
\label{tab:glb_fit_TMR}

\begin{center}
	\begin{tabular}{l@{\quad}ccc@{\qquad}ccc}
	\hline 
				&	\multicolumn{3}{c}{LMD ($\tau_{\rm min}/a=2$)}	&	\multicolumn{3}{c}{LMD ($\tau_{\rm min}/a=3$)}			 \\
	\cline{2-4} \cline{5-7} 
				& 	$M_V~[\MeV]$	& $\beta~[\MeV]$	&	$\alpha~[\GeV^{-1}]$		& 	$M_V~[\MeV]$	& $\beta~[\MeV]$	&	$\alpha~[\GeV^{-1}]$	\\ 
	\hline  
	 Extrap. ($\widetilde{y}=0$)				&	631(24)	&	 -27(4)	&	0.316(17)	&	648(25)	&	 -25(4)	&	0.316(18)	 \\
	 Extrap. ($\widetilde{y}=\widetilde{y}_{\exp}$)	&	667(20)	&	 -27(4)	&	0.297(17) &	682(20)	&	 -25(4)	&	0.297(17) \\
	\hline
	$C_{\widetilde{y}}$			& 	2.76(39)	& 	$\times$		&	 -1.38(16)	& 	2.46(39)		& 	$\times$		&	 -1.33(15)	\\
	$C_{a}$					&	$\times$		& 	-0.009(5)	&	0.10(2) 	&	$\times$		& 	-0.009(5)	&	0.10(2) \\ 
	\hline 
	$\chi^2/{\dof}$    	& 	\multicolumn{3}{c}{1.35}	& 	\multicolumn{3}{c}{1.32}	\\
	\hline  
	\end{tabular}
\end{center}

\end{table}

\section{Operator-product expansion analysis of $\widetilde A_{\mu\nu}(\tau)$ \label{app:OPE}}

Consider the operator-product expansion (OPE) of $\psib_f(x)\gamma_\mu\psi_f(x)\;\psib_f(0)\gamma_\nu\psi_f(0)$.
At dimension three, only quark bilinears with no derivatives are candidate operators in the OPE. Of the five types of bilinears, only the pseudoscalar density and the axial current couple to the pion. Based on the Euclidean SO(4) symmetry group we have the possible terms
\begin{gather}
\nonumber \delta_{\mu\nu}\psib_f \gamma_5 \psi_f,\qquad (x_\mu x_\nu - \frac{x^2}{4}\delta_{\mu\nu})\psib_f \gamma_5 \psi_f \,,
\\
\psib_f (x_\mu\gamma_\nu - x_\nu \gamma_\mu)\gamma_5\psi_f\,,
\qquad
\psib_f(x_\mu\gamma_\nu + x_\nu\gamma_\mu - \frac{1}{2}\delta_{\mu\nu} x_\rho\gamma_\rho) \gamma_5\psi_f\,,
\qquad
\epsilon_{\mu\nu\rho\sigma} x_\rho \psib_f  \gamma_\sigma\gamma_5\psi_f\,.
\end{gather}
Parity now eliminates all of the candidates except the last one. Indeed, $V_i(x)V_j(0)$ does not acquire a minus sign under a parity transformation, whereas all but the last operator does acquire one (recall that the axial current does not receive a minus sign under parity).
The only dimension-three operator that can contribute is thus
\begin{equation}
\epsilon_{\mu\nu\rho\sigma} \; \frac{x_\rho}{(x^2)^2} \;A_\sigma\,.
\end{equation}

We also briefly consider dimension-four operators. Since we need an isovector
operator to couple to the pion, the only possibility is either
\begin{equation}
\label{eq:mA}
m \epsilon_{\mu\nu\rho\sigma} \; \frac{x_\rho}{(x^2)^2} \;A_\sigma\,,
\end{equation}
or a quark bilinear with one derivative.  Consider the correlation function of the product of vector currents with a pion interpolating operator with vanishing spatial momentum. The derivative inside the quark bilinear must be temporal, otherwise the operator will not overlap with the pion. Indeed, the option $\psib_f  \gamma_5 \gamma_i D_i\psi_f$ can be replaced by $-\psib_f\gamma_5 (\gamma_0 D_0+m)\psi_f$ using the equation of motion $\psib_f (\gamma_\mu D_\mu + m)\psi_f=0$. 
Therefore, taking into account the pseudoscalar quantum number of the pion, the only option is 
\begin{equation}
\label{eq:o2}
\epsilon_{\mu\nu\rho\sigma} x_\rho\; \psib_f \gamma_5 D_\sigma \psi_f\,.
\end{equation}
In the candidate $\epsilon_{\mu\nu\rho\sigma}\psib_f \gamma_5 \gamma_\rho D_\sigma \psi_f$, 
at least one of the spacetime indices inside the bilinear would have to be spatial, preventing an overlap with a pion at rest.

Thus at dimension four, we have the candidates (\ref{eq:mA}) and (\ref{eq:o2}). So far, we have not taken into account the constraints of chiral symmetry. Taking into account the latter, (\ref{eq:mA}) is forbidden, and so is (\ref{eq:o2}). Both must appear with an additional power of the quark mass, making them dimension-five in terms of the degree of singularity of the Wilson coefficient.

However, on the lattice with Wilson fermions, exact chiral symmetry is not realized exactly. Therefore operators (\ref{eq:mA}) and (\ref{eq:o2}) are not a priori excluded on the lattice.

\subsection{Wilson coefficient of the axial current and asymptotics of $\FF$}

We work in Euclidean notation. At tree level, performing single Wick contractions of the quark fields yields two terms,
\begin{equation}
\label{eq:wick}
\psib_f(x)\gamma_\mu \psi_f(x)\; \psib_f(0)\gamma_\nu \psi_f(0)
= \psib_f(x) \gamma_\mu G(x) \gamma_\nu \psi_f(0)
+ \psib_f(0) \gamma_\nu G(-x) \gamma_\mu \psi_f(x)\,,
\end{equation}
where $G(x)$ is the quark propagator. Now using the massless propagator
\begin{equation}
G(x) = \frac{ x_\sigma \gamma_\sigma}{2\pi^2 (x^2)^2}\,,\qquad (m=0)
\end{equation}
and 
\begin{equation}
\gamma_\mu\gamma_\sigma \gamma_\nu = \epsilon_{\mu\sigma\nu\rho}\gamma_5 \gamma_\rho + 
(\delta_{\mu\sigma}\gamma_\nu + \delta_{\sigma\nu} \gamma_\mu - \delta_{\mu\nu}\gamma_\sigma)\,,
\end{equation}
we obtain 
\begin{equation}
\psib_f(x)\gamma_\mu \psi_f(x)\; \psib_f(0)\gamma_\nu \psi_f(0)
\stackrel{x\to 0}{=} \frac{\epsilon_{\mu\nu\rho\sigma}}{\pi^2(x^2)^2}\; x_\rho\; \psib_f(0)\gamma_\sigma \gamma_5 \psi_f(0) + \dots
\end{equation}
This (tree-level) equality holds when inserted into a Euclidean correlation function, 
in particular in a three-point function with an interpolating operator for the pion.
Now using 
\begin{equation}
\label{eq:int1}
\int d^3x\; \frac{x_k}{(x_0^2+\vec x^2)^2}\; e^{-i\vec q\cdot \vec x} = -i\pi^2 \;\frac{q_k}{|\vec q|}\; e^{-|\vec q||x_0|}\,,
\end{equation}
we obtain 
\begin{equation}
\int d^3x\;e^{-i\vec q\cdot \vec x}\;\<0|{\rm T}\{V_i(x)V_j(0)\}|\pi,{\vec p=0}\> = i\epsilon_{ijk} \frac{q_k}{|\vec q|}e^{-|\vec q||x_0|} 
\<0|\psib_f(0)\gamma_0\gamma_5 \psi_f(0)|\pi,\vec p=0\>\,.
\label{eq:VVAres1}
\end{equation}
Note that Euclidean correlation functions automatically yield matrix elements of time-ordered products of fields.

For the up quark contribution, the matrix element is given by (see the text below Eq.~(\ref{eq:MFF}) for our convention
concerning the phase of the pion state) 
\begin{equation}
\<0|\overline{\psi}(0)\gamma_0\gamma_5 \psi(0)|\pi^0,\vec p=0\> = 
\frac{1}{2} \<0|\bar u(0)\gamma_0\gamma_5 u(0) - \bar d(0)\gamma_0\gamma_5 d(0)|\pi^0,\vec p=0\> 
=i F_\pi m_\pi\,.
\end{equation}
For $V_i$ and $V_j$ the electromagnetic currents, we must multiply (\ref{eq:VVAres1}) by the charge factor
\[
Q_u^2 \cdot 1 + Q_d^2 \cdot (-1) = \frac{1}{3}\,.
\]
If we define 
\begin{equation}
{\rm Disc}(f) \equiv f'(x_0=0^-) - f'(x_0=0^+)\,,
\end{equation}
we have
\begin{equation}
\label{eq:disc1}
{\rm Disc} \int d^3x\;e^{-i\vec q\cdot \vec x}\;\<0|{\rm T}\{V^{\rm em}_i(x)V^{\rm em}_j(0)\}|\pi,{\vec p=0}\>
= -\frac{2F_\pi}{3} \epsilon_{ijk}\; q_k  m_\pi \,.
\end{equation}

\subsection{Comparison with Appendix \ref{app:A}}

The generic connection between $\widetilde A_{ij}$ and the pion matrix element of the product
of two vector currents is given by Eqs.~(\ref{eq:Mminkowski}) and (\ref{eq:lat_M}),
\begin{equation}
\widetilde A_{\mu\nu}(x_0) = -\frac{Z_\pi}{2E_\pi} \int d^3x\; e^{-i\vec q\cdot \vec x}\;\<0|{\rm T}\{V_\mu(x)V_\nu(0)\}|\pi,\vec p\>\,.
\end{equation}
Thus, comparing the cusp of $\widetilde A_{ij}$ given for the LMD model in Eq.~(\ref{eq:discLMD}) with Eq.~(\ref{eq:disc1}), we obtain 
\begin{equation}
\beta= - \frac{F_\pi}{3}\,.
\end{equation}
This corresponds to an asymptotic behavior of the transition form factor consistent with 
Eq.~(\ref{eq:OPE}).

\subsection{Contribution of a dimension-four operator to $\widetilde A_{ij}(\tau)$}

Analogously to Eq.~(\ref{eq:int1}), the relevant integral for the contribution of a dimension-four operator to \\
$\int d^3x\; e^{-i\vec q\cdot \vec x}\<0|{\rm T}\{V_i(x)V_j(0) \}|\pi(\vec p=0)\>$ is then 
\begin{equation}
\int d^3x\; \frac{x_k}{(x_0^2+\vec x^2)^{3/2}}\; e^{-i\vec q\cdot \vec x} = -4\pi i\; \frac{q_k}{|\vec q|}|x_0| K_1(|\vec q||x_0|)\,.
\end{equation}
Expanding at small $x_0$, we get
\begin{equation}
\int d^3x\; \frac{x_k}{(x_0^2+\vec x^2)^{3/2}}\; e^{-i\vec q\cdot \vec x} = -4\pi i \frac{q_k}{|\vec q|} + {\mathcal O}(x_0^2\log(|\vec q||x_0|)\,.
\end{equation}
The leading term is analytic in $x_0$.
We conclude that dimension-four operators do not contribute to the discontinuity in the derivative at $x_0=0$.

\subsection{Wilson coefficient of the axial current on the lattice}

We  repeat the OPE calculation above on the lattice.
For two local vector currents on the lattice, the starting point is again Eq.~(\ref{eq:wick}), where now $G(x)$ must be replaced by the lattice Wilson propagator $G_{\rm w}(x)$, where in the time-momentum representation\footnote{$\hat p_\mu = \frac{2}{a} \sin\frac{ap_\mu}{2}$, $\circp_\mu = \frac{1}{a} \sin ap_\mu$}
\begin{gather}
G_{\rm w}(x_0,\vec q) = h(x_0,\vec q) + \sum_\mu \gamma_\mu\, g_\mu(x_0,\vec q)\,,
\\
g_k(x_0,\vec q) = -{i \circq_k}\; \frac{e^{-\omega_{\vec q}|x_0|}}{D_{\vec q}}\,,
\end{gather}
\begin{eqnarray}
A(\vec p) &=& 1 + am + \frac{1}{2} a^2 \hat{\vec p}^{\,2},
\\
 B(\vec p) &=& m^2 + (1+am) \hat{\vec p}^{\,2} + \frac{1}{2} a^2 \sum_{k<l} \hat p_k^2 \hat p_l^2\,.
\\
\omega_{\vec p} &=& \frac{2}{a} {\rm \,asinh\,}\left( \frac{a}{2}\sqrt{B(\vec p)/A(\vec p)}\right)\,,
 \\
{\cal D}_{\vec p} &=& 
\frac{2}{a}A(\vec p) \sinh(a\omega_{\vec p})
  = \sqrt{B(\vec p)\,(4A(\vec p) + a^2 B(\vec p))}\,.
\end{eqnarray}
Going to the time-momentum representation of the product of vector currents as in Eq.~(\ref{eq:int1}), we obtain 
\begin{equation}
a^3 \sum_{\vec x} e^{-i\vec q\cdot \vec x}\;\bar q(x)\gamma_i q(x)\; \bar q(0)\gamma_j q(0)
\stackrel{x\to 0}{=} -2\sum_{k=1}^3 \epsilon_{ijk}\; g_k(x_0,\vec q)   \;
 \bar q(0)\gamma_0 \gamma_5 q(0) + \dots
\end{equation}
Unlike in the continuum massless case, there is also a piece proportional to the unit Dirac matrix in $G_{\rm w}(x)$, however it does not
contribute to the Wilson coefficient of the axial current.  In the last equation, the dots stand for other operators contributing to the
OPE and for terms of order $a$.

Thus the result for the Wilson coefficient of the axial current depends on the bare quark mass. However, for the discontinuity we find
again Eq.~(\ref{eq:disc1}), up to ${\mathcal O}(a)$ effects, since ${\cal   D}_{\vec q} = 2\omega_{\vec q}(1+{\mathcal O}(a))$.  We do not expect
this agreement to persist at higher order in perturbation theory for $x_0$ of order the lattice spacing.

\section{Spectral representation of the three-point correlator in infinite volume \label{app:FSE}}

Consider the following Euclidean correlation function for $x_0>0$,
\be
\widetilde A_{ij}^{\rm r}(x_0,\vec P,\vec p)\equiv -\frac{2E_\pi}{Z_\pi} \widetilde A_{ij}(x_0) 
 = \int d^3x\; e^{-i\vec P\cdot \vec x} \,\<0|V_i(x)\,V_j(0)| \pi,\vec p\>.
\ee
In this appendix we use the notation $\vec P$ instead of $\vec q_1$, which is more natural in the following dispersive representation.
We insert a complete set of (outgoing) two-pion states, which dominate the correlation function for $x_0\to +\infty$,
\be
1 = \int \frac{d^3k}{(2\pi)^3 2E_{\vec k}}\int \frac{d^3k'}{(2\pi)^3 2E_{\vec k'}}\; |\pi_{\vec k}\pi_{\vec k'}\>_{\rm out}~
{}_{\rm out}\<\pi_{\vec k}\pi_{\vec k'}| + \dots
\ee
and obtain, using 
\be
\<0| V_i(x)|\pi_{\vec k}\pi_{\vec k'}\>_{\rm out}
 = e^{-(E_{\vec k}+E_{\vec k'}) x_0 + i(\vec k+\vec k')\vec x}
~ \<0| V_i(0)|\pi_{\vec k}\pi_{\vec k'}\>_{\rm out}
\ee
the following expression
\be
\widetilde A_{ij}^{\rm r}(x_0,\vec P,\vec p)
 = \int \frac{d^3k\; e^{-(E_{\vec k}+E_{\vec P-\vec k})x_0}}{(2\pi)^3 (2E_{\vec k})(2E_{\vec P-\vec k})}\;
\<0| V_i(0)|\pi_{\vec k}\pi_{\vec P-\vec k}\>_{\rm out}\; {}_{\rm out}\<\pi_{\vec k}\pi_{\vec P-\vec k}|V_j(0)| \pi,\vec p \>.
\ee
Now the matrix elements are parametrized in terms of form factors (see \cite{Meyer:2011um} Eq.~(12) and \cite{Hoferichter:2012pm} Eq.~(4))
\begin{eqnarray}
\<0|V_i(0)|\pi_{\vec k}\pi_{\vec k'}\>_{\rm out} &=& -i(\vec k'-\vec k)_i\; F_\pi^V(E^*)^*
\\
_{\rm out}\<\pi_{\vec k}\pi_{\vec k'}|V_j(0)|\pi_{\vec p}\> &=&
 i \epsilon_{j\nu\alpha\beta}\, p^\nu k^\alpha k'{}^\beta {\cal F}(E^{*\,2},t,q_\gamma^2).
\end{eqnarray}
Here $E^*$ is the center-of-mass energy of the $\pi\pi$ system, $t$ is the Mandelstam variable
 and $F_\pi^V$ the pion vector form factor.

The second matrix element, between the one-pion state and the outgoing two-pion state,
determines the invariant amplitude ${\cal M}$ of the process $\pi(p) \gamma^*(q_\gamma) \to \pi(k)\pi(k')$.
We only keep the $p$-wave component of ${\cal F}(E^{*\,2},t,q_\gamma^2)$,
$f_1(E^*,q_\gamma^2)$.  The latter  depends on the center-of-mass energy and the virtuality of the photon;
we have removed the factor $e$ (electromagnetic coupling constant) from $f_1$ as compared to \cite{Hoferichter:2012pm}.
Energy-momentum conservation leads to 
\be
\vec q_\gamma + \vec p = \vec P, \quad E_\gamma+E_{\vec p} = E_{\vec k} + E_{\vec P-\vec k},
\qquad q_\gamma^2 = (E_{\vec k}+E_{\vec P-\vec k} -E_{\vec p})^2- (\vec P - \vec p)^2.
\ee

We now choose the rest frame of the $(\pi\pi)$ system, where the amplitude simplifies to
\be
\widetilde A_{ij}^{\rm r}(x_0,\vec P=0,\vec p) = \int \frac{d^3k\;e^{-2E_{\vec k}x_0}}{(2\pi)^3 (2E_{\vec k})^2}\; 
\Big[i (2k_i) F_\pi^V(2E_{\vec k})^*\Big]\;
\Big[ i \epsilon_{j\nu\alpha\beta}\; p^\nu k^\alpha k'{}^\beta\; 
f_1(E,(2E_{\vec k} -E_{\vec p})^2- \vec p^2)\Big]_{k'=(E_{\vec k},-\vec k)}.
\ee
We note $ \epsilon_{j\nu\alpha\beta}\; p^\nu k^\alpha k'{}^\beta = -2E_{\vec k} \;\epsilon_{jlm}\, p^l  k^m$, so that 
\be
\widetilde A_{ij}^{\rm r}(x_0,\vec P=0,\vec p) = \frac{1}{3}\epsilon_{ijl} p^l \int \frac{d^3k\; \vec k^2\;e^{-2E_{\vec k}x_0}}{(2\pi)^3E_{\vec k}}
F_\pi^V(2E_{\vec k})^*\; f_1(2E_{\vec k},(2E_{\vec k} -E_{\vec p})^2- \vec p^2).
\ee
We have made use of rotation symmetry, 
which implies $\int d^3k f(|\vec k|) k^i k^j = \frac{1}{3}\delta_{ij} \int d^3k f(|\vec k|)\; \vec k^2$.
Performing the trivial angular integrations and inserting $1=\int_0^\infty d\omega \delta(\omega-2E_{\vec k})$,
we obtain the spectral representation
\begin{eqnarray}
\widetilde A_{ij}^{\rm r}(x_0,\vec P=0,\vec p) &=& \epsilon_{ijl}\,p^l \int_0^\infty d\omega\,\rho(\omega,|\vec p|) \; e^{-\omega x_0}, \qquad (x_0>0)
\\
\rho(\omega,|\vec p|) &= & 
\frac{1}{12\pi^2}\Big(\frac{\omega^2}{4}-m_\pi^2\Big)^{3/2}\,F_\pi^V(\omega)^*\,f_1(\omega,(\omega-E_{\vec p})^2-\vec p^2).
\end{eqnarray}
We see that for our purposes, unlike in \cite{Hoferichter:2012pm}, the
dispersion relation is not in one virtuality, with the other photon
virtuality fixed, but rather in the energy of the $\pi\pi$ system, at
fixed spatial momentum $\vec P$.

The same dispersion relation can be set up in finite volume, where the
energy eigenstates are discrete.  Using the relations in~\cite{Luscher:1991cf} and
\cite{Meyer:2011um,Meyer:2012wk,Briceno:2014uqa}, the spectrum and the
finite-volume matrix elements can be related to their infinite-volume
counterpart, so that the finite-size effects can be evaluated once
$F_\pi^V$ and $f_1$ have been specified.


\end{document}